\documentclass[prb,twocolumn,amssymb,floatfix,amsmath,showpacs,superscriptaddress]{revtex4}
\usepackage{epsfig}

\usepackage{amssymb}
\usepackage{amsmath}
\usepackage{amsfonts}
\usepackage{bbold}
\usepackage{bm}
\usepackage{graphicx}
\usepackage{braket}
\usepackage{color}

\begin{document}
\title{A brief review on $\mu$SR studies of unconventional Fe and Cr-based superconductors}
\author{A. Bhattacharyya}
\email{amitava.bhattacharyya@rkmvu.ac.in}
\affiliation{ISIS Facility, Rutherford Appleton Laboratory, Chilton, Didcot, Oxon, OX11 0QX, United Kingdom} 
\affiliation{Department of Physics, Ramakrishna Mission Vivekananda Educational and Research Institute, Belur Math, Howrah 711202,West Bengal, India} 
\author{D. T. Adroja} 
\email{devashibhai.adroja@stfc.ac.uk}
\affiliation{ISIS Facility, Rutherford Appleton Laboratory, Chilton, Didcot, Oxon, OX11 0QX, United Kingdom} 
\affiliation{Highly Correlated Matter Research Group, Physics Department, University of Johannesburg, Auckland Park 2006, South Africa}
\author{M. Smidman}
\affiliation{Center for Correlated Matter and Department of Physics, Zhejiang University, Hangzhou 310058, China}
\author{V. K. Anand}
\affiliation{Helmholtz-Zentrum Berlin f\"{u}r Materialien und Energie GmbH, Hahn-Meitner Platz 1, D-14109 Berlin, Germany}

\begin{abstract}
Muon spin relaxation/rotation ($\mu$SR) is a very useful technique for probing the superconducting gap structure, pairing symmetry and time reversal symmetry breaking, enabling an understanding of the mechanisms behind the unconventional superconductivity of cuprates and Fe-based high-temperature superconductors, which remain a puzzle. Very recently double layered Fe-based superconductors having quasi-2D crystal structures and Cr-based superconductors with a quasi-1D structure have drawn considerable attention. Here we present a brief review of the characteristics of a few selected Fe- and Cr-based superconducting materials and highlight some of the major unsolved problems, with an emphasis on the superconducting pairing symmetries of these materials. We focus on $\mu$SR studies of the newly discovered superconductors $A$Ca$_2$Fe$_4$As$_4$F$_2$ ($A$ = K, Rb, and Cs), ThFeAsN, and  $A_2$Cr$_3$As$_3$ ($A$ = K, Cs), which were used to determine the superconducting gap structures, the presence of spin fluctuations, and to search for time reversal symmetry breaking in the superconducting states.  We also briefly discuss the results of $\mu$SR investigations of the superconductivity in hole and electron doped BaFe$_2$As$_2$.
\end{abstract}
\date{\today} 

\maketitle

\section{Introduction}
\noindent In 1911, Kamerlingh Onnes measured the electrical conductivity of numerous metals and discovered the abrupt disappearance of  the resistance of a solid mercury wire in liquid helium~\cite{Onnes}. Subsequently a wide range of superconductors have been found with increasing high transition temperatures ($T_{c}$), and it is hoped that a room-temperature superconductor can ultimately be realized. Following the discovery of superconductivity it took  a long time for a microscopic theoretical understanding of the phenomenon. Such a microscopic theory (BCS theory) was proposed in 1957 by John Bardeen, Leon Cooper, and John Robert Schrieffer \cite{Bardeen,Cooper}, where electrons with opposite momenta condense into Cooper pairs, which are bound via an attractive interaction mediated by lattice vibrations (phonons).  Superconductors which are adequately characterized by BCS theory, where the Cooper pairs are bound by the electron-phonon interaction, are known as conventional superconductors, which make up the vast majority of known superconducting materials to date. In these systems, the Cooper pairs condense in a $s$-wave pairing state with zero orbital angular momentum, which leads to a gap in the single particle excitation spectrum across the whole Fermi surface~\cite{Poole}. For phonon mediated BCS superconductivity, the transition temperature is comparatively low, with an expected maximum of 30--40~K with strong coupling (at ambient pressure) ~\cite{Bardeen,Cooper,McMillan}. A dramatic breakthrough in the field of superconductivity came from the discovery of a La-Ba-Cu-O system with a $T_c$ of around 30~K  by Bednorz and M$\ddot{\rm u}$ller \cite{Bednorz}, for which they were awarded the Nobel Prize in Physics 1987. The cuprates materials are quite distinct from previously known superconductors, consisting of  doped CuO$_2$ layers. Soon after, cuprates with significantly higher values of $T_c$ were found, including the first material to become superconducting above the boiling point of liquid nitrogen, YBa$_2$Cu$_3$O$_{7-\delta}$, with a $T_c$ of about 93~K \cite{Wu1987}. The largest $T_c$ reported at ambient pressure is at around 135~K in Hg$_2$Ba$_2$Ca$_2$Cu$_3$O$_{10+\delta}$ \cite{Schilling1993}, which can be increased even further under high applied pressures \cite{Chu1993}.
\begin{figure*}[t]
\centering
    \includegraphics[width=0.47\linewidth]{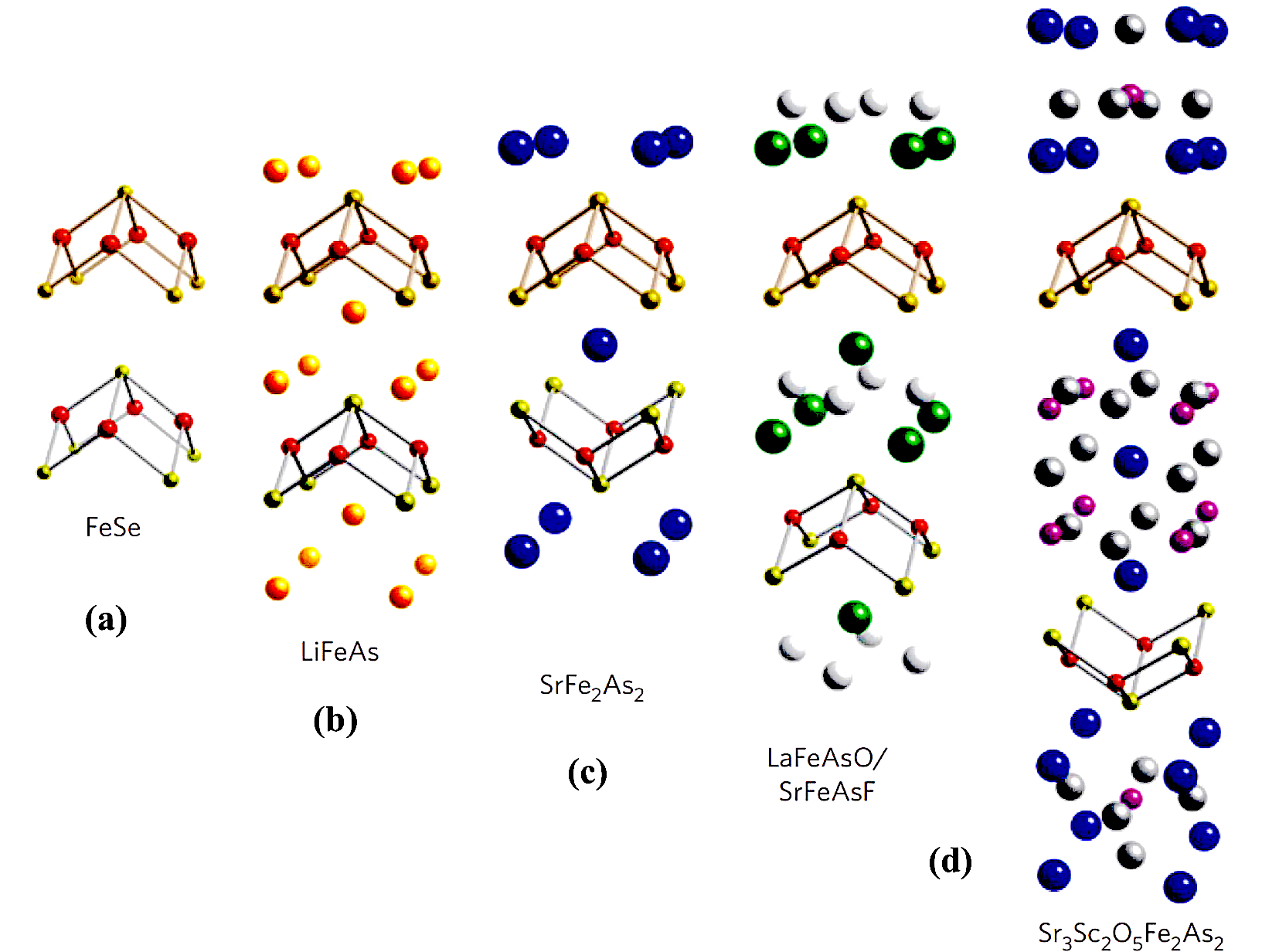}\hfil
    \includegraphics[width=0.47\linewidth]{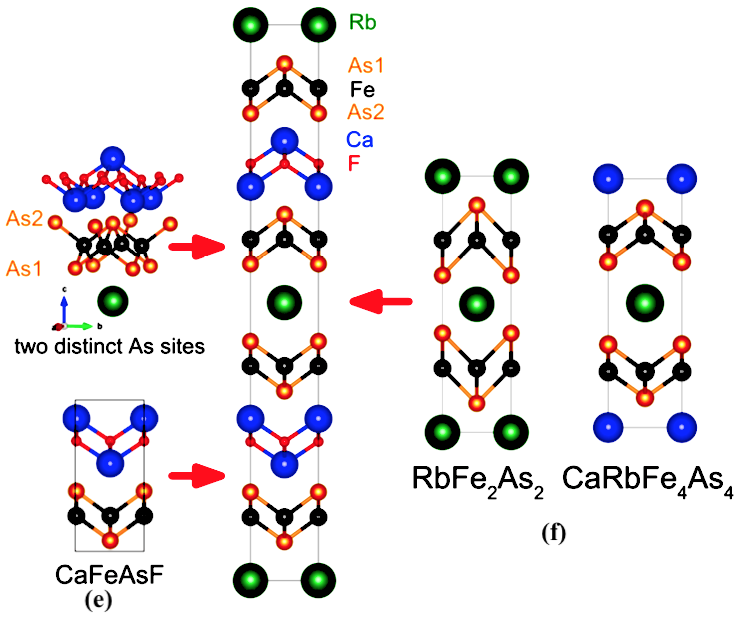}\par\medskip
\caption{Illustration of the different crystal structures of the parent compounds of various FeSCs, (a) 11-type, (b) 111-type, (c) 122-type and (d) 1111-type and 32522-type materials. Reprinted with permission from Paglione {\ et al.} Nature Physics {\bf 6}, 645 (2010) \cite{Paglione}. Copyright 2010 by Springer Nature. (e) and (f) represent the tetragonal crystal structure of newly discovered $A$Ca$_2$Fe$_4$As$_4$F$_2$ ($A$ = K, Rb and Cs) compounds along with those for (f) RbFe$_2$As$_2$, CaRbFe$_4$As$_4$  and (e) CaFeAsF from Adroja {\ et al.} arXiv:1802.07334~\cite{Adroja2}.}
\label{Crystal structure}
\end{figure*}

\noindent The iron pnictide superconductors  (FeSCs) were the second family of high$-T_{c}$ superconductors to be discovered, by Kamihara {\it et al.}  in 2006~\cite{Kamihara2}, which was even more exciting given the long standing notion that ferromagnetic elements such as iron are detrimental to superconductivity. Prior to this finding, superconductivity had already been discovered with $T_c= 4$~K in LaOFeP \cite{Kamihara1}, but the subsequent study of La(O$_{1-x}$F$_x$)FeAs revealed a clearly enhanced value of $T_c= 26$~K \cite{Kamihara2}. It was further found that upon switching La with other rare earth elements and with either fluorine doping or introducing oxygen deficiencies, $T_{c}$ could be further increased~\cite{Chen2008,Takahashi2}. Much like the cuprates, it was quickly apparent that these classes of materials are also unconventional superconductors, with a pairing mechanism not mediated by the electron-phonon interaction of BCS theory ~\cite{Mazin,Johnston,Paglione,Wen,Stewart,Hirschfeld}. Here the essential component for superconductivity appears to be the FeAs layers, as opposed to the Cu-O planes of the cuprates. One similarity to the cuprates is the tunability of the properties upon carrier doping, starting from a non-superconducting parent compound. In the cuprates this parent compound corresponds to a Mott insulator, whereas for the Fe-based superconductors these  are generally metallic or semimetallic with a spin-density wave ground state~\cite{Paglione,J. Schmalian1,Y. S. Lee1}. The crystal structures typically consist of arrangements of  layers of FeAs and spacer layers, and a number of varieties are found. The structures of the 11, 111, 122,1111 and 32522-type materials are shown in Figs.~\ref{Crystal structure}(a-f) \cite{Paglione}, along with that of the  recently discovered $A$Ca$_2$Fe$_4$As$_4$F$_2$ ($A$ = K, Rb and Cs, 122442-family) superconductors.
\begin{figure*}[t]
\centering
    \includegraphics[width=0.5\linewidth]{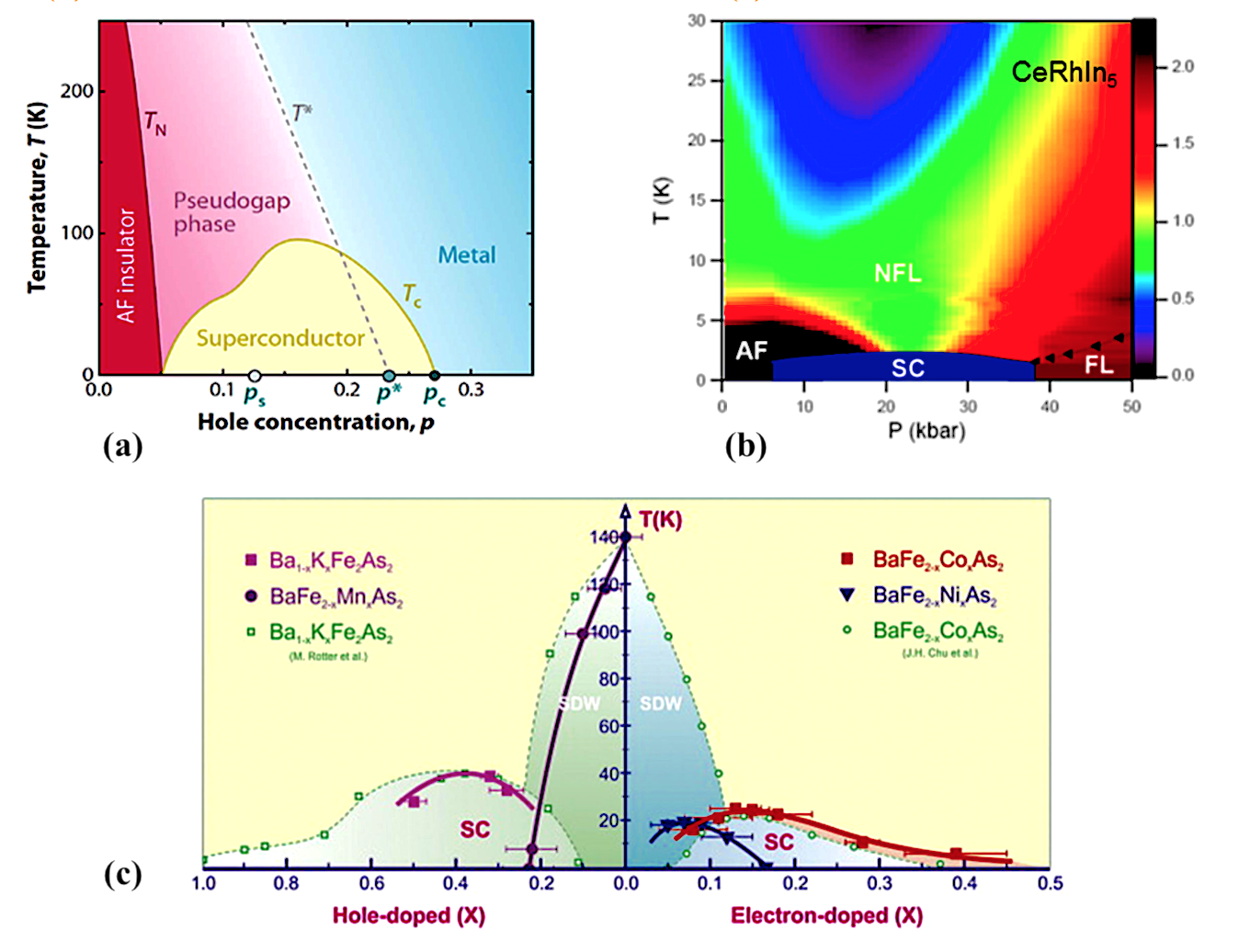}\hfil
    \includegraphics[width=0.5\linewidth]{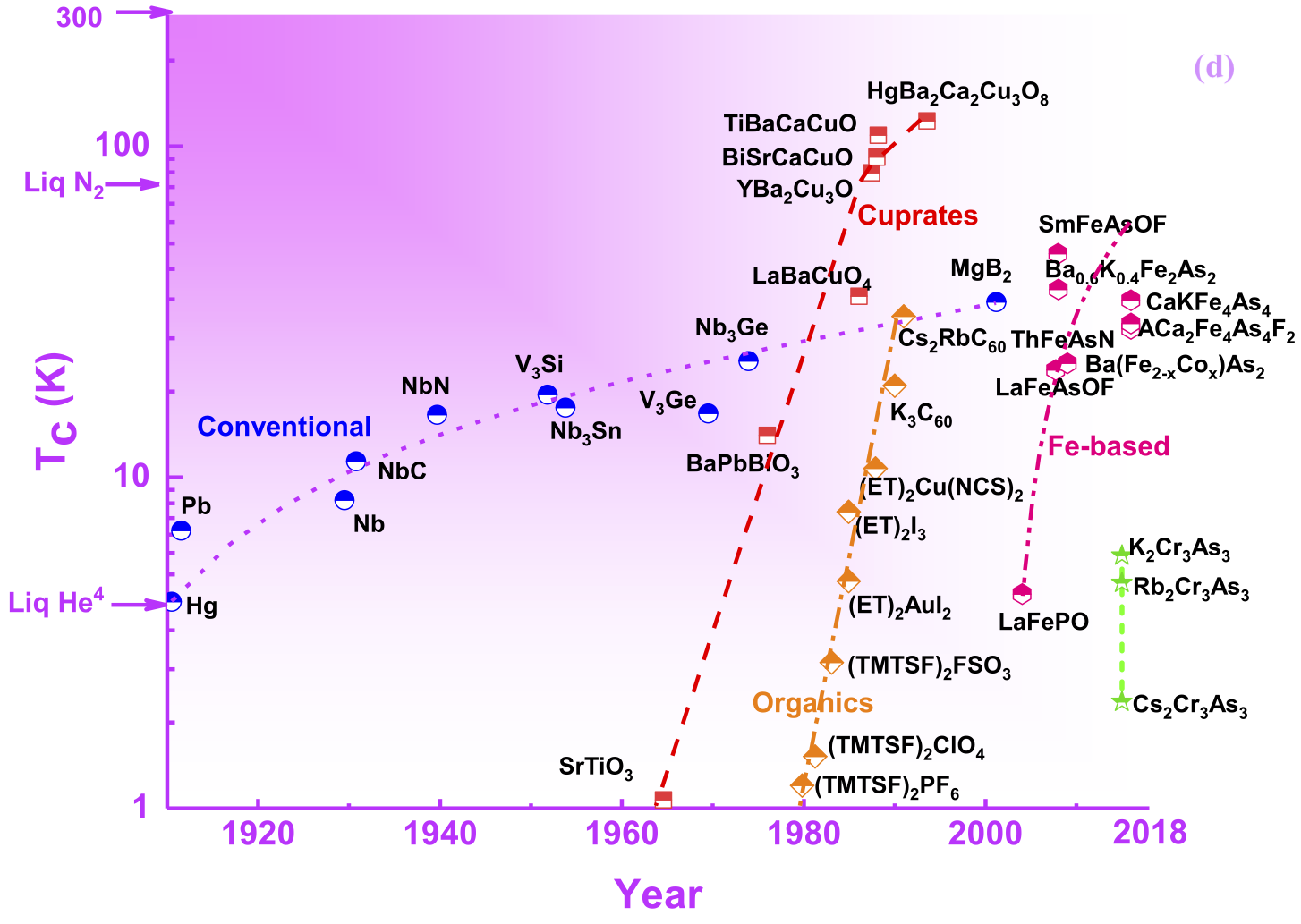}\par\medskip
\caption{Prototypical examples of the phase diagrams of three major classes of unconventional superconductors, following Bang {\ et al.} J. Phys.: Condens. Matter {\bf 29}, 123003 (2017)~\cite{Y. Bang}. (a) Cuprate superconductor (YBCO) reprinted with permission from Leyraud {\ et al.} Nature  {\bf 447}, 565-568 (2007)~\cite{Nicolas Doiron-Leyraud}. Copyright 2007 by Springer Nature.  (b) Heavy fermion superconductor (CeRhIn$_5$) reprinted with permission from Park {\ et al.} Nature {\bf 456}, 366-368 (2008)~\cite{T. Park}. Copyright 2008 by Springer Nature. (c) Fe-based 122 superconductors  reprinted with permission from Liu {\ et al.} Physica C: Superconductivity and its Applications, {\bf 470}, S513-S515 (2010)~\cite{Y. Liu1}. Copyright 2010 by Elsevier.  (d) A plot showing the historical progression of the $T_{c}$ of superconductors.}
\label{phasediagram}
\end{figure*}
\noindent One of the striking features found in some iron based superconductors is the observation of electronic nematicity below a structural transition, \cite{Fisher1,Fernandes1,Avci1} the origin of which  is not yet resolved. In the nematic state the four fold rotational symmetry of the high temperature tetragonal structure is broken, and this appears to be electronically driven, corresponding to either spin  or orbital nematic ordering \cite{Fernandes2,Fernandes3,Fang4,Xu1,J. Dai,S. Onari,S. Onari1,S. Kumar,W. Lv,C.-C. Lee}. Whether the nematic order in all the Fe-based systems has a common origin is another open question, particularly in the case of FeSe where there appears to be a lack of magnetic order in the nematic state, which has been accounted for by orbital ordering  originating from the Aslamazov-Larkin vertex correction~\cite{Yamakawa1}. In general,  nematic order appears to compete with superconductivity \cite{Moon2012}, although when the $s$-wave and $d$-wave states are nearly degenerate, it has been suggested that the nematic order can lift the frustration and hence lead to an increase of $T_c$ \cite{Fernandes1,R. M. Fernandes1,Fan Yang1}.
  
\noindent Another example of unconventional superconductivity is in the heavy fermion superconductors~\cite{F Steglichhfs}, such as CeCu$_2$Si$_2$~\cite{F. Steglich}, CeCoIn$_5$~\cite{C Petrovic1}, CePt$_3$Si~\cite{Bauer}, UPt$_3$~\cite{G. R. Stewart1} and URu$_2$Si$_2$~\cite{Premala Chandra}, which in spite of the much smaller values of $T_c$, show a number of similarities to the high temperature superconductors. These similarities can be seen more clearly in Figs. ~\ref{phasediagram}(a)-(c), which show the schematic phase diagrams of three of the most investigated classes of unconventional superconductors, including  (a), a cuprate superconductor (YBCO) from Leyraud {\ et al.}~\cite{Nicolas Doiron-Leyraud}, (b), a heavy fermion superconductor (CeRhIn$_5$) from Park {\ et al.}~\cite{T. Park}, and (c) Fe-based (doped Ba-122) superconductors from Liu {\ et al.}~\cite{Y. Liu1} A common feature is that the application of a non-thermal tuning parameter such as doping or pressure suppresses a magnetically ordered phase, but leads to a superconducting dome, which may encompass a quantum critical point near to where $T_c$ is largest. Meanwhile, a plot showing the historical progression of realized values of $T_{c}$   is shown in Fig.~\ref{phasediagram}(d).
\noindent

\begin{figure*}[t]
\centering
    \includegraphics[width = 0.8\linewidth]{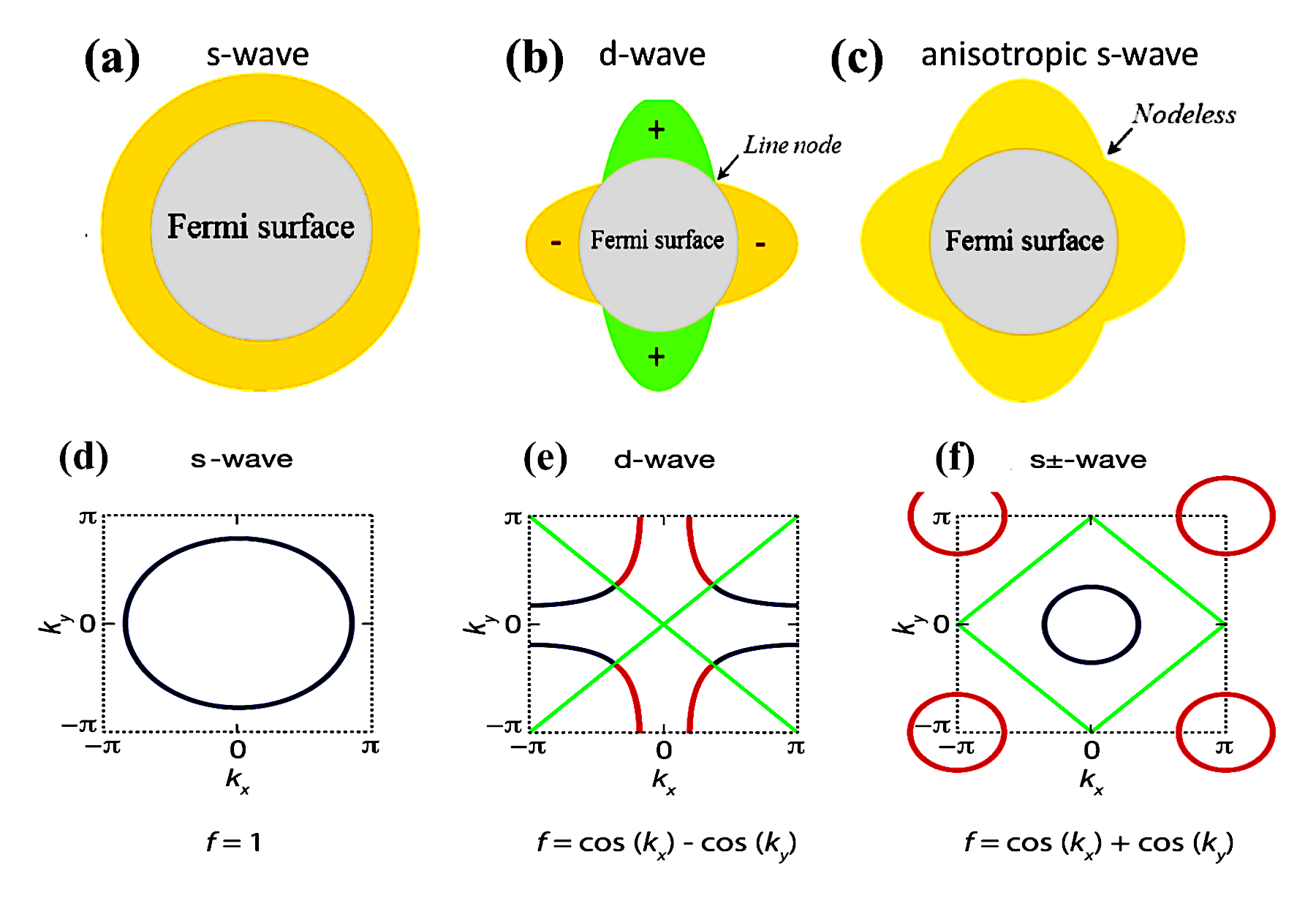}\hfil
\caption{ (a-c)  Illustrations of the superconducting gap structures corresponding to various gap symmetries, from P.~K. Biswas (unpublished). (d-f)  Schematic drawings of two dimensional Fermi surfaces and the Brillouin zones from Tegel ~\cite{Tegel}, where the green lines indicate the nodes in the superconducting order parameter and the blue/red circles highlight the different signs of the superconducting order parameter.}
\label{Fermi surfaces}
\end{figure*}
\noindent Cooper pairs can be described by a wave function, consisting of spatial and spin parts, corresponding to particular orbital anglar momentum $L$ and spin $S$.  Spin-singlet wave functions have $S=0$ and $L=0, 2, ... $, whereas the spin-triplet wave functions have $S=1$ and $L=1, 3, ...$. One salient feature of unconventional  pairing states is that the Cooper pairs do not only form in the  $s$-wave state of BCS theory, but can condense in higher angular momentum states, and the gap can be highly anisotropic, disappearing at points or lines on the Fermi surface. Figure~\ref{Fermi surfaces} depicts the gap structures of several pairing states. Conventional superconductors have the most symmetric orbital part, $L=0$, which corresponds to $s$-wave pairing   [see Fig.~\ref{Fermi surfaces}(a)]. A less symmetric orbital part, $L>0$, is a feature of many unconventional superconductors, where $L=1$ and $L=2$ correspond to $p$-wave and $d$-wave, respectively [see Fig.~\ref{Fermi surfaces}(b)]. A notable example of this is the cuprate superconductors where the experimental results are most consistent with $d$-wave symmetry \cite{Hardy1993,Annett1990}, and the Cooper pairs are thought to mediated by antiferromagnetic spin fluctuations \cite{Scalapino2012}. The gap corresponding to such a $d$-wave state is highly anisotropic, and it changes sign and has line nodes on the Fermi surface, as shown in Fig.~\ref{Fermi surfaces}(b).  In fact a sign changing superconducting order parameter is generally required for the realization of spin-fluctuation mediated superconductivity \cite{Scalapino2012}, and therefore in analogy with the cuprates, one may also have anticipated  $d$-wave symmetry in iron-arsenide superconductors. However,  experimental probes on iron-based materials have in most cases revealed nodeless superconductivity, and this has often been accounted for by an $s_{\pm}-$state~\cite{Mazin2008} similarly mediated by antiferromagnetic spin fluctuations, where there is a different sign of the order parameter between disconnected Fermi surface pockets, as shown in Fig.~\ref{Fermi surfaces}(f). Similar to the cuprates, Fe-based superconductors have also been much studied with inelastic neutron scattering (INS). These studies have revealed a spin resonance peak below $T_{c}$ in the 1111, 122, 111, and 11 families of superconductors, where  the energy of the resonance peak has a linear relationship with $T_{c}$~\cite{Dai1}.  The presence of such a spin resonance suggests a change of sign of the superconducting pairing function between regions of the Fermi surface separated by the ordering wave vector of the antiferromagnetic parent compound. Therefore in many cases, the observation of a spin resonance is in good agreement with that expected for the $s_{\pm}$ state, since the AFM wave vector often connects the hole pocket at the zone center, and the electron pockets at the zone edge \cite{Dai1}.

\noindent There has been considerable debate over to what extent the superconductivity of iron pnictides can  be universally described by an $s_{\pm}$ state. This is particularly the case for overdoped 122 FeAs-based systems such as KFe$_2$As$_2$ which lack  electron pockets at the zone edge \cite{Sato1}, and FeSe-based materials such as K$_x$Fe$_{2-y}$Se$_2$ \cite{Qian2011} and (Li$_{1-x}$Fe$_x$)OHFeSe \cite{Niu2015,Zhao2016}, where the hole pockets at the zone center are absent.   For KFe$_2$As$_2$ it was suggested that there is a crossover from a nodeless $s_{\pm}$ state in optimally doped Ba$_{1-x}$K$_x$Fe$_2$As$_2$ to a nodal $d$-wave state in KFe$_2$As$_2$ upon hole doping \cite{J.-Ph. Reid,Maiti2011,Thomale2011,Reid2012}. Before reaching the $d$-wave phase, the system would be expected to pass through an $s+id$ state at intermediate doping, in which time reversal symmetry is broken \cite{Lee2009}. On the other hand, other results suggest that even in the heavily hole doped region, the superconducting state remains $s$-wave \cite{Cho2016,KFe2As2ARPES}.  ARPES measurements report evidence that KFe$_2$As$_2$ has $s$-wave symmetry with a nodeless gap on the innermost hole Fermi surface, an octet line node structure on the middle surface,  and a very small gap on the outer one \cite{KFe2As2ARPES}. For the FeSe based materials with no hole pockets at the zone center, there have been proposals for a different $s_{\pm}$ state compatible with the electronic structure \cite{Lee2009,Khodas2012}. Another  proposed alternative is  an orbital selective $s\times \tau_3$ state, where the interband and intraband pairing functions have different $d$-wave symmetries, resulting in a sign change of the intraband pairing function between the electron pockets, but the superconducting gap remains fully open \cite{Nica2017}.
\begin{figure*}[t]
\centering
    \includegraphics[width=0.6\linewidth]{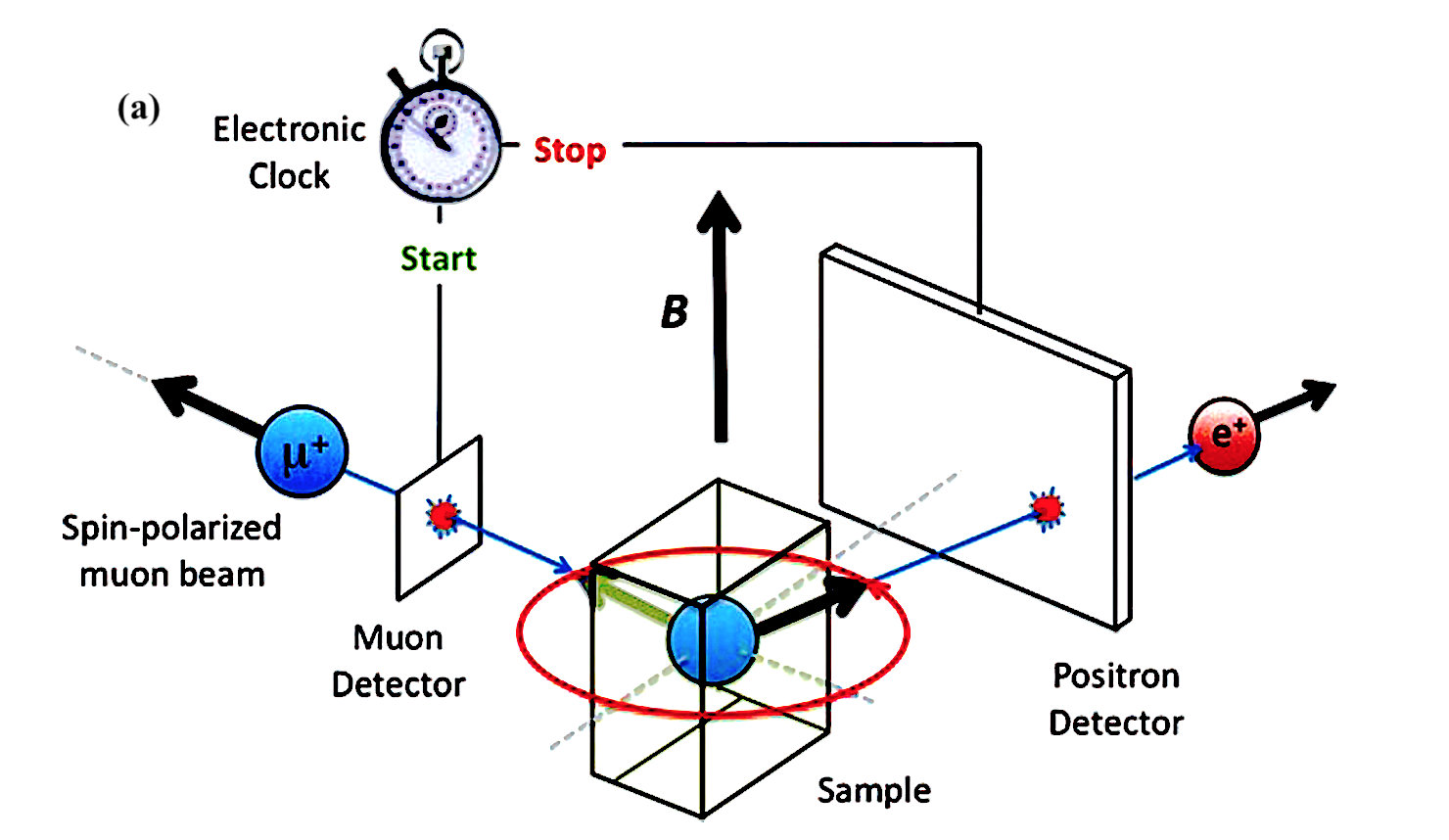}\hfil
    \includegraphics[width=0.4\linewidth]{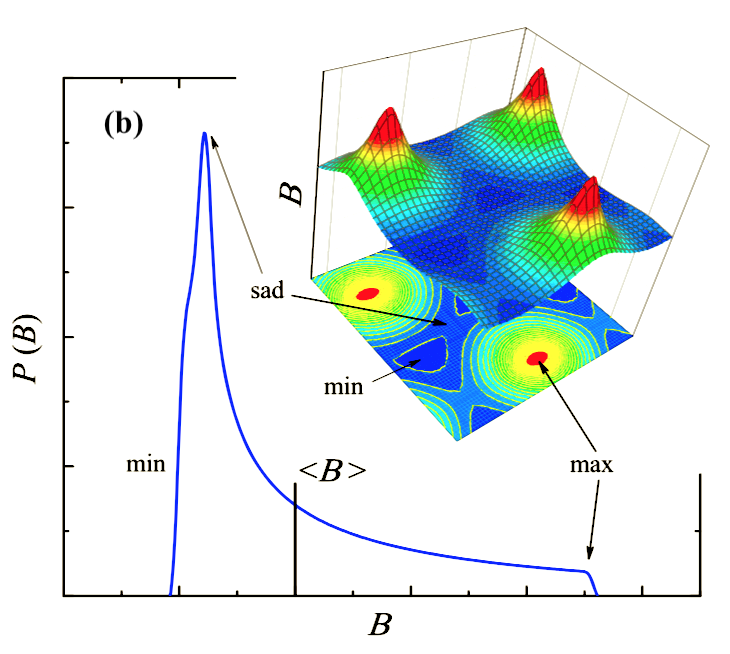}\par\medskip
\caption{Schematic representation of a $\mu$SR experiment. (a) Illustration of the transverse field $\mu$SR experimental setup. After the spin-polarized  muons are implanted in the sample,  the number of decay positrons detected in the forward and backward detectors is counted as a function of time from Iain McKenzie, Annu. Rep. Prog. Chem., Sect. C: Phys. Chem. {\bf 109}, 65--112 (2013)~\cite{McKenzie1}. (b) Field distribution of a regular triangular flux-line lattice, with the minimum ($B_{min}$), maximum ($B_{max}$), and saddle point ($B_{sad}$) fields indicated by the arrows. Both panels are reprinted with permission   from Khasanov {\ et al.} Supercond. Sci. Technol. {\bf 28} 034003 (2015).~\cite{Khasanov}.  Copyright 2015, the Institute of Physics.}
\label{musr experiment}
\end{figure*}

\noindent Unconventional superconductors often exhibit the breaking of additional symmetries in the superconducting state, as well as the broken global gauge symmetry common to all superconductors. One possibility is that time reversal symmetry (TRS) may be broken, which leads to the spontaneous appearance of magnetic fields below $T_c$. Two representative examples  showing such behavior are the  unconventional superconductors Sr$_2$RuO$_4$ and UPt$_3$, where evidence for TRS breaking was initially found from zero-field $\mu$SR \cite{Luke1998,Luke1993}, which were later corroborated by optical Kerr rotation experiments \cite{Xia2006,Schemm2014} (although in Ref.\cite{Reotier1995} conflicting results were found for $\mu$SR measurements of UPt$_3$). The broken TRS is well accounted for by the leading theories of triplet superconductivity for these systems, where there are two-component order parameters corresponding to  two-dimensional irreducible representations of the crystallographic point group \cite{Mackenzie2003,Joynt2002}. However, in other cases such as LaNiC$_2$ and LaNiGa$_2$, the broken TRS is ascribed to non-unitary triplet pairing \cite{Hillier2009,Hillier2012,Weng2016}, where the Cooper pairs have an intrinsic spin polarization \cite{Sigrist2005}. Meanwhile, it has also been proposed that loop Josephson-currents can account for the TRS breaking signals in some otherwise apparently conventional superconductors \cite{Ghosh2018}. In the case of iron pnictides, $\mu$SR measurements were performed on Ba$_{1-x}$K$_x$Fe$_2$As$_2$ to look for the evidence of TRS breaking $s+id$ state predicted to occur between the $s_{\pm}$ and $d$-wave phases \cite{Lee2009}. However, these yielded conflicting results, with one study finding no signatures of broken TRS \cite{Mahyari2017}, while evidence was found from measurements of ion irradiated samples \cite{Grinenko2017}. As such, performing $\mu$SR measurements to look for the presence of broken TRS is an important method for identifying the pairing state of new unconventional superconducting materials.

\noindent Recent discoveries have led to a particular interest in superconductivity of Cr-based materials. For example, CrAs, exhibits helimagnetic order at $T_N$ = 273 K at ambient pressure and becomes superconducting with $T_{c}$ = 1.5 K in an applied pressure of 1.09 GPa~\cite{CrAsSC1,CrAsSC2}. This spurred the synthesis of new Cr-based superconductors, which resulted in the discovery of superconductivity in $A_2$Cr$_3$As$_3$ ($A$ = Na, K, Rb, and Cs), which have a quasi-1D crystal structure~\cite{J. Bao,Z. Tang1,Z. Tang,Mu1}. Subsequently, superconductivity was also observed in isostructural $A_2$Mo$_3$As$_3$ ($A$ = K, Rb and Cs)~\cite{Mu,Zhao2}. As the elements Cr and Mo belong to the same group of transition metals with a similar electronic configuration, this poses the question of whether these two families of superconductors share some common origins for the superconductivity~\cite{Mu,Zhao2}. The  $A_2$Cr$_3$As$_3$ superconductors have been intensively investigated both experimentally and theoretically~\cite{J. Bao,T. Kong,Balakirev,Z. Tang1,Z. Tang,Mu1,Pang2,X. Wu,Zhou,Wu2,H. Zhong,Zhi,Adroja1,Watson,Zhi3,Yang2015,Adroja5}, as they are strong candidates for multiband triplet pairing. Proposals of either triplet $p_z$ or $f$-wave pairing states arise from theoretical calculations \cite{Wu2,Zhou}, which may also account for the evidence of line nodes in the superconducting gap \cite{Pang2,Adroja1,Adroja5}.
\noindent 
 Various techniques can be used to investigate the gap structure of superconductors including measuring  the temperature and field dependence of thermodynamic properties, angle-resolved measurements, and spectroscopic techniques. To completely understand the gap structure and pairing state, it is often necessary to consider a variety of measurements, and here we primarily discuss $\mu$SR investigations of unconventional superconductors. The temperature and field dependence of TF-$\mu$SR measurements yields information about the  flux-line lattice and magnetic penetration depth, which can reveal the structure of the superconducting gap. Furthermore, TF-$\mu$SR measurements on single crystals can allow for the gap anisotropy to be directly probed \cite{Khasanov2007,Ohishi2012}. $\mu$SR in zero-field can be used to determine whether time reversal symmetry is broken in the superconducting state, as well as the magnetic properties of the system. Detailed reviews about $\mu$SR measurements on superconductors and other condensed matter systems can also be found in Refs~\cite{Amato,Blundell,Feyerherm,Sonier,Khasanov}. In this review, we principally concentrate on using $\mu$SR to probe the superconducting pairing states of newly discovered double layered FeSCs, as well as a family of Cr-based one-dimensional materials. In the first part, a brief introduction to the muon-spin rotation/relaxation techniques is given, while the second part reviews the superfluid density of  $A$Ca$_2$Fe$_4$As$_4$F$_2$ ($A$ = K, Cs, and Rb), ThFeAsN, doped BaFe$_2$As$_2$ and  $A_2$Cr$_3$As$_3$ (A = K, and Cs). Finally, we close with concluding remarks where we discuss the results and suggest future directions. We note that this review only includes a small subsection of the extensive research on high-temperature superconductors, and further detailed reviews can be found in Refs.~\cite{Dai1,Stewart,Sonier1,Y. Jie1,Paglione,PCarretta,Khasanov}.

\section{Introduction to the $\mu$SR technique}
\noindent Muon spin rotation and relaxation are very sensitive local probes which can be used to investigate magnetic and superconducting materials~(for further details see Refs.~\cite{Amato,Blundell,Feyerherm,Sonier,Khasanov}).  The muons utilized in $\mu$SR facilities are generated by colliding a high energy proton beam with a graphite target, which produces pions, and each positive (negative) pion decays into a positive (negative) muon and a (anti)neutrino. As a result of parity violation, the muon beam is spin-polarized, and these muons are implanted into the sample and come to rest via electrostatic interactions which preserve the muon spin direction. For  $\mu$SR measurements, positive muons $\mu^+$  are usually utilized, which decay with a half-life of 2.2~$\mu$s into a pair of neutrinos and a positron. Since the positrons are more likely to be emitted along the axis of the muon spin, the angular distribution of the emitted positrons reflects the distribution of the spins of the muons. In the presence of any local magnetic fields at the muon stopping site $B_{loc}$, the $\mu^+$ spins precess at the Larmor frequency $\omega_\mu = \gamma_\mu B_{loc}$, where $\gamma_\mu/2\pi$ = 135.5~MHz T$^{-1}$ is the muon gyromagnetic ratio. The time dependence of the polarization of the implanted muons is given by $P_\mu (t)$ = $G(t)P_\mu(0)$, where the function $G(t)$ corresponds to the $\mu^{+}$ spin autocorrelation function, which is determined by the internal magnetic field distribution \cite{Amato}. A schematic illustration of a $\mu$SR experiment (adopted from Ref.~\cite{Khasanov}) is shown in Fig.~\ref{musr experiment}(a).  A commonly measured quantity in $\mu$SR experiments is the time dependent asymmetry $A(t)$ which is proportional to  $P_\mu (t)$ and is given by  $A(t)=\frac{N_F(t)-\alpha N_B(t)}{N_F(t)+\alpha N_B(t)}$, where $N_F(t)$ and $N_B(t)$ are the number of positrons detected in the forward and  backward positions respectively, and $\alpha$ is a calibration constant. 
\noindent  $\mu$SR measurements performed in zero external magnetic field are very sensitive to weak internal fields in a sample, down to the order of 0.1~G. In studies of superconductors this can be utilized to look for time reversal symmetry (TRS) breaking in the superconducting state. TRS breaking corresponds to the spontaneous appearance of magnetic fields below $T_c$, which in turn leads to a more rapid muon spin depolarization. While a number of expressions can be utilized to analyze ZF-field $\mu$SR spectra, the data are often analyzed using \cite{Luke1998,Hillier2009}

\begin{equation}
A(t) = A_0G_{KT}(t)e^{-\lambda t} + A_{bg},
\end{equation}

\noindent where $\lambda$ is the Lorentzian relaxation rate, $A_0$ is the initial asymmetry of the relaxing component, and $A_{bg}$ is the background term. $G_{KT}(t)$ is the Kubo-Toyabe function given by

\begin{equation}
G_{KT}=\frac{1}{3}+\frac{2}{3}(1-\sigma_{KT}^2t^2)e^{-\frac{\sigma_{KT}^2t^2}{2}},
\end{equation}

\noindent where $\sigma_{KT}$ is the Gaussian relaxation rate. Here $G_{KT}$ characterizes the decay of the asymmetry due to a Gaussian distribution of magnetic fields which are static on the timescale of the muon lifetime (including those arising from nuclear moments), while the Lorentzian decay can correspond to either a Lorentzian distribution of static fields or rapidly fluctuating fields, and hence is often termed the electronic relaxation rate. The breaking of TRS in the superconducting state can be inferred from either an increase of $\lambda$  \cite{Luke1998,Hillier2009}, or $\sigma_{KT}$ \cite{Hillier2012} below $T_c$. The application of a longitudinal field (LF) parallel to the initial muon-spin serves to decouple the muon from the local magnetic fields in the sample. This decoupling occurs more rapidly for static fields than for dynamic fields, and therefore LF-$\mu$SR reveals information about both the magnitude and dynamics of the internal fields \cite{Blundell}.
On the other hand, $\mu$SR measurements in applied transverse magnetics field (TF), perpendicular to the initial muon spin direction, are important for probing the flux line lattice \cite{Sonier}. In TF-$\mu$SR, the precession of the muon spins leads to oscillations in $A(t)$, which decay due to the presence of a broadened field distribution $P(B)$. The shape of the field distribution corresponding to the flux line lattice is shown in Fig.~\ref{musr experiment}(b) \cite{Khasanov}, and $P(B)$ can be quantitatively utilized to extract the magnetic penetration depth $\lambda$ and the coherence length $\xi$ \cite{Sonier,W. Barford1,E. H. Brandt}. The decay of the oscillations in $A(t)$ is often modelled using a Gaussian function and therefore $A(t)$ can be analyzed using

\begin{align}
\begin{split}
A(t) = A_0\cos(2\pi \nu_s t+\phi)\exp\left({-\frac{\sigma^2t^2}{2}}\right)\\ + A_{bg}\cos(2\pi \nu_{bg} t+\phi)
\end{split}
\end{align}

where $\nu_s$ and $\nu_{bg}$ are the frequencies of the muon precession from the sample and background, respectively with a phase offset $\phi$. The Gaussian decay rate in the first term $\sigma$ has two contributions below $T_c$,  where $\sigma$ = $\sqrt{(\sigma_{\rm sc}^2+\sigma_{\rm nm}^2)}$, and $\sigma_{\rm sc}$ is the contribution from the vortex lattice and $\sigma_{\rm nm}$ corresponds to the nuclear dipolar relaxation rate, which can be obtained from spectra measured in the normal state. $\sigma_{\rm sc}$ can be directly related to the effective penetration depth $\lambda_{eff}$ via $\sigma_{\rm sc}/\gamma_{\mu}=0.0609\Phi_0/\lambda^2_{eff}$, where $\Phi_0$  is the magnetic flux quantum. This relation between $\sigma_{\rm sc}$ and $\lambda_{eff}$ is valid for $0.13/\kappa^{2}\ll(H/H_{c2})\ll1$, for a Ginzburg-Landau parameter $\kappa$=$\lambda$/$\xi$$\gg$70 \cite{E. H. Brandt}.

\noindent Since $\lambda(T)$ is sensitive to low energy excitations, this in turn can provide  information about the superconducting gap structure. For instance, if the superconducting gap is fully open across the whole Fermi surface, $\lambda(T)$ shows exponentially activated behaviour, but if there are line or point nodes in the gap, $\lambda(T)$ shows a linear or quadratic temperature dependence in the clean and dirty limits, respectively \cite{Prozorov2006}. In particular, since $\sigma_{\rm sc}$ is proportional to the superfluid density $n_s$, it can be modelled using \cite{Bhattacharyya1,Bhattacharyya2,Adroja1},

\begin{multline}
\frac{\sigma_{\rm sc}(T)}{\sigma_{\rm sc}(0)}
 = \frac{\lambda^{-2}(T,\Delta_{0,i})}{\lambda^{-2}(0,\Delta_{0,i})}\\
 = 1+\frac{1}{\pi}\int_{0}^{2\pi}\int_{\Delta(T)}^{\infty}\left (\frac{\delta f}{\delta E}\right) \frac{EdEd\phi}{\sqrt{E^2-\Delta(T,\Delta_{i})^2}}
\end{multline}

where $f = [1+{\rm exp}(-E/K_B T)]^{-1}$ is the Fermi function, $\phi$ is the azimuthal angle across the Fermi surface, and $\Delta_i(T,\phi) = \Delta_{0,i}\delta(T/T_c)g(\phi)$. The temperature dependence of the superconducting gap function is approximated by the expression $\delta(T/T_c)$ = tanh$\{1.82[1.018(T_c/T-1)]^{0.51}\}$ while $g(\phi)$ describes the azimuthal angle dependence of the gap, given by 1 and $|cos(2\phi)|$ for an isotropic $s-$wave gap and a  $d-$wave gap with line nodes, respectively~\cite{Annett2,Carrington2003}. A range of other potential gap structures have been used to analyze the results from iron-based superconductors, including  $g(\phi)=1+\alpha{\rm cos}2\phi$ for an extended $s$-wave state \cite{Lin2011}, and  $\Delta(T,\phi)$=[($\Delta_1(0)$)$^2$+($\Delta_2(0)$sin($\phi$))$^2$]$^{1/2}$ for an $s\times \tau_3$ state, where components corresponding to two different $d$-wave symmetries are summed in quadrature \cite{Nica2017,Smidman2017}. Therefore by analyzing $\sigma_{\rm sc}$ with various gap models, the magnitude and structure of the superconducting gap can be probed, which in turn allows for the identification of the superconducting pairing symmetry.
 
\section{Multiband superconductivity in $\rm {ACa_2Fe_4As_4F_2}$}

\subsection{Crystal structure and superconductivity}
\noindent Recently, Wang {\it et al.} discovered superconductivity at 29-33 K in $A$Ca$_2$Fe$_4$As$_4$F$_2$ (A = K, Rb, Cs, 12442-family) ~\cite{Wang1,Wang2}. Figures~\ref{Crystal structure} (e)-(f) show the tetragonal crystal structure with space group $I4/mmm$ (No. 139, Z = 2) in which the Fe$_2$As$_2$ layers are sandwiched between A atoms on one side and Ca$_2$F$_2$ on the other side, leading to two distinct As sites above and below the Fe plane. This crystal structure is similar to the Ca$A$Fe$_4$As$_4$-family ($A$ = K, Rb, Cs, 1144 family) and since the superconducting properties are closely associated with the nature of the spacer layers between adjacent conducting layers~\cite{Wang1,Wang2}, it is of great interest to study the superconductivity of compounds with these double Fe$_2$As$_2-$layers. Furthermore, it was noted that for the 12442-family, $T_{c}$ has an inverse correlation   with the lattice parameters $a$ and $c$ \cite{Wang1,Wang2}. This is in contrast to the case of optimally hole-doped $A$Fe$_2$As$_2$ systems,  where the maximum $T_c$  increases with the lattice parameters~\cite{Wang1,Wang2}. More specifically, for the 12242 materials $T_{c}$ decreases with increasing $d_{intra}$ (the spacing of Fe planes within the double Fe$_2$As$_2-$layers), yet $T_{c}$ increases almost linearly with the distance between the Fe planes separated by the Ca$_2$F$_2$ layers, $d_{inter}$. In FeSCs, $T_c$ has often been found to be related to both the As-Fe-As bond angle, and the As height from the Fe plane~\cite{Johnston2010}, and for the 12442-family there are two sets of these parameters due to the asymmetric Fe$_2$As$_2$ layers leading to inequivalent As sites \cite{Wang1,Wang2}. In addition, unlike many other FeSCs, no structural phase transitions or SDW transitions have been reported at low temperatures~\cite{Wang1,Wang2}. These materials are intrinsically near  optimal doping with hole conduction (0.25 holes/Fe$^{2+}$)~\cite{Ishida}. Electronic structure calculations for KCa$_2$Fe$_4$As$_4$F$_2$ reveal a multiband character and suggest that the system is close to a stripe antiferromagnetic ground state \cite{Wang7}. It is suggested that the magnetic ground state is suppressed by the self-doping, leading to the observed high temperature superconductivity.
\begin{figure*}[t]
\centering
    \includegraphics[width=0.2\linewidth]{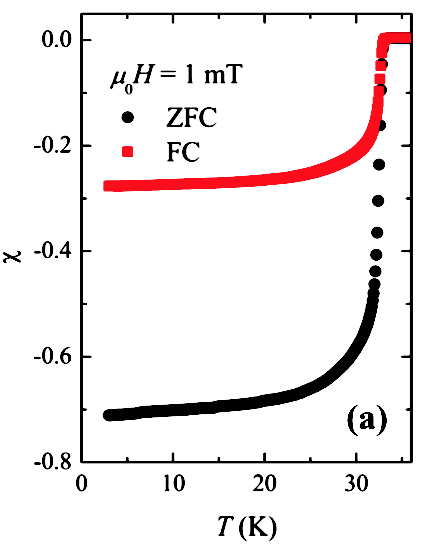}
     \includegraphics[width=0.35\linewidth]{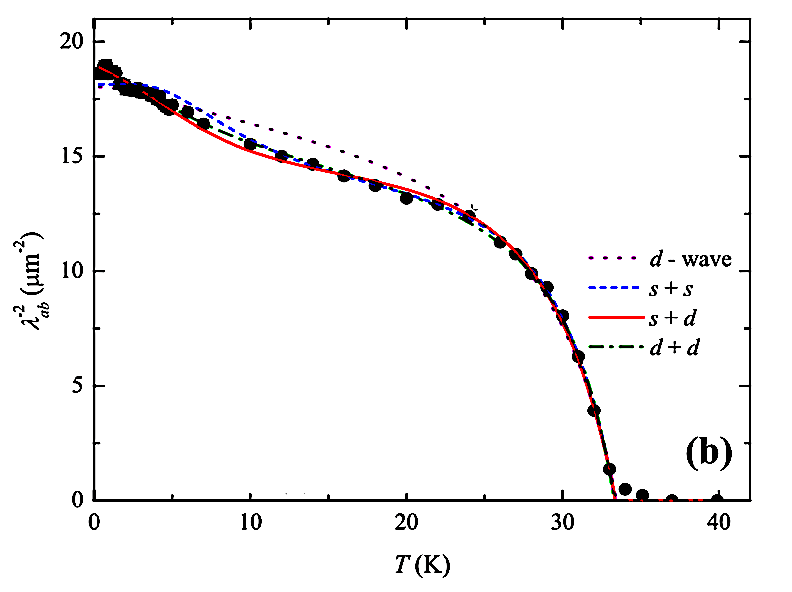}
    \includegraphics[width=0.35\linewidth]{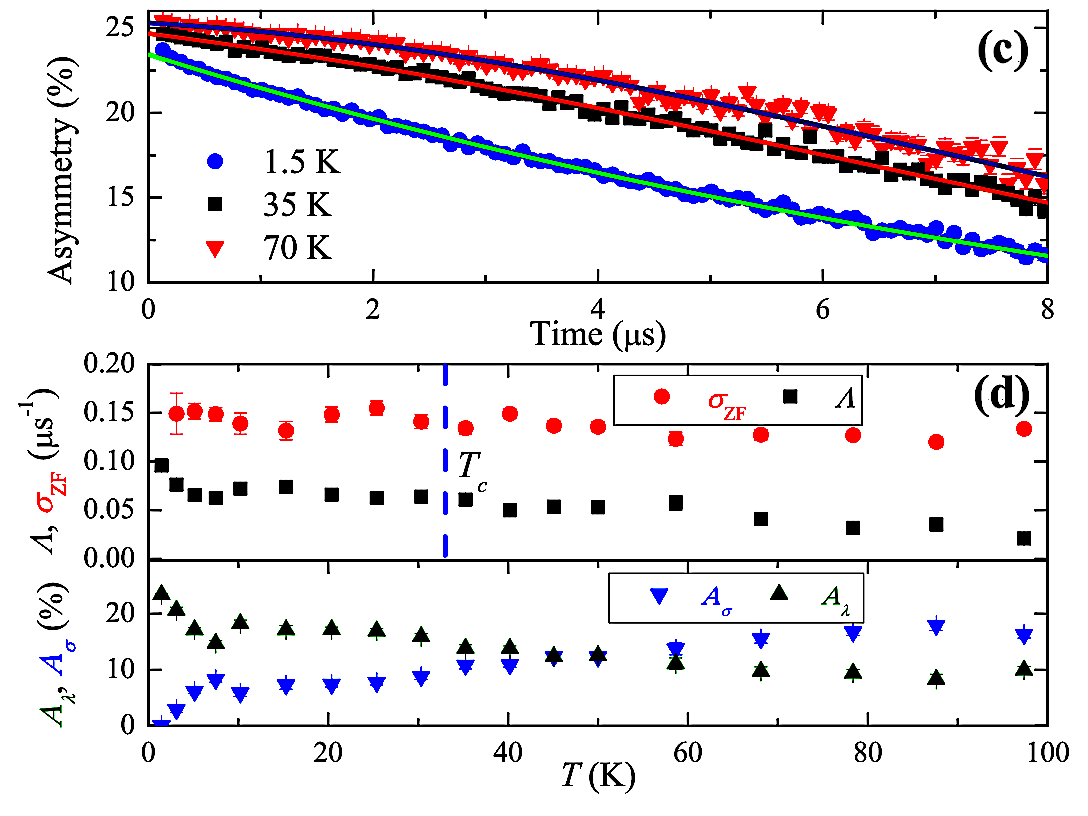}
\caption{(a) Magnetization as a function of temperature for KCa$_2$Fe$_4$As$_4$F$_2$ in zero-field cooled (ZFC) and field-cooled (FC) modes. (b) Variation of the square of the inverse in-plane penetration depth with temperature $\lambda_{ab}^{-2}(T)$ for KCa$_2$Fe$_4$As$_4$F$_2$. The lines represent  fits using different models for the superconducting gap structures. The single gap $d-$wave and fully gapped $s + s$ models do not fit the data well, but good fits are obtained using two-gap nodal $s + d$ and $d + d$ models. (c) Time dependence of zero field muon spin relaxation spectra at different temperatures for KCa$_2$Fe$_4$As$_4$F$_2$, where the solid lines show fits consisting of sum of a Gaussian and a Lorentzian decay. (d) Variation of the Lorentzian ($\lambda$) and Gaussian ($\sigma_{ZF}$) relaxation rates with temperature, along with the corresponding amplitudes of each component. All panels are reprinted  with permission from Smidman {\it et al.} Phys. Rev. B {\bf 97}, 060509(R) (2018)~\cite{Smidman1}. Copyright 2018 by the American Physical Society.}
\label{KCa$_2$Fe$_4$As$_4$F$_2$sigmasc}
\end{figure*}
\subsection{Superconducting gap structure and time reversal symmetry}

\subsubsection{KCa$_2$Fe$_4$As$_4$F$_2$}

\noindent KCa$_2$Fe$_4$As$_4$F$_2$ with $T_{c}$ = 33.36 K exhibits multigap superconductivity with line nodes as reported by Smidman {\it et al.}~\cite{Smidman1}. TF-$\mu$SR data reveal that the temperature dependence of the in-plane penetration depth $\lambda_{ab}^{-2}$ does not have the form of a $s$-wave superconductor, since it does not become flat and saturate at low temperatures as shown in Fig.~\ref{KCa$_2$Fe$_4$As$_4$F$_2$sigmasc}(b). Furthermore, $\lambda_{ab}^{-2}$  cannot be described by either single band models or a multigap $s$-wave expression, but are well described by two-gap models with line nodes. Here good fits are obtained for both a $s+d$ model where one gap is fully open and the other is nodal, as well as a $d+d$ scenario where both gaps have line nodes. The respective fitted values of the larger of the two gaps are $3.52\,k_BT_{c}$ and $5.08\,k_BT_{c}$, which are significantly higher than the weakly-coupled gap values ($1.76\,k_BT_{c}$ and $2.14\,k_BT_{c}$ for $s$- and $d$-wave, respectively), suggesting strongly coupled superconductivity. An interesting comparison can be made with the ARPES studies of two other nodal iron-pnictide superconductors BaFe$_2$(As$_{0.7}$P$_{0.3}$)$_2$ and  KFe$_2$As$_2$  \cite{Zhang6n,KFe2As2ARPES}. For BaFe$_2$(As$_{0.7}$P$_{0.3}$)$_2$ it was found that the superconducting gap nodes only exist on a small region of the Fermi surface \cite{Zhang6n}, and it was suggested that this relatively limited nodal area can account for the relatively high $T_c$ of 30~K. On the other hand, the ARPES study of KFe$_2$As$_2$ revealed a much more extended nodal area of the Fermi surface with an octet line node structure, and in turn this compound has a much lower $T_c$ of around 3~K \cite{KFe2As2ARPES}. In KCa$_2$Fe$_4$As$_4$F$_2$ the superfluid density does not drop as dramatically with increasing temperature as would for instance be expected for a weakly coupled $d$-wave superconductor, which also suggests a relatively small nodal region of the Fermi surface as in BaFe$_2$(As$_{0.7}$P$_{0.3}$)$_2$, which is consistent with the similarly large $T_c$.

\noindent Figures~\ref{KCa$_2$Fe$_4$As$_4$F$_2$sigmasc}(c)-(d) show the results from ZF-$\mu$SR, which indicate that TRS is preserved in the superconducting state. In addition, a temperature dependence of the relaxation rate well above $T_{c}$ suggests the presence of spin fluctuations in this system. It is of interest to understand the origin of nodal superconductivity in this system and in particular how changing the spacer layers in these asymmetric FeSC tunes the properties of the pairing state. In comparison, the 1144 superconductor CaKFe$_4$As$_4$  exhibits two-gap nodeless superconductivity \cite{Biswas,Cho,Mou}, and therefore the exchange of a Ca layer for Ca$_2$F$_2$ can tune the pairing state from nodeless to nodal in these materials~\cite{Cui}. Since it is difficult to account for this change along similar lines to other nodal FeSCs, this suggests that this material represent a different means of realizing nodal iron-based superconductors.  Furthermore, from the TF$-\mu$SR study superconducting parameters of an in-plane penetration depth $\lambda_{ab}(0)$ = 229.5~nm, superconducting carrier density $n_s$ = 1.39 $\times$ 10$^{27}$ m$^{-3}$, and carrier's effective-mass $m^{*}$ = 2.59 $m_e$ are estimated using $\Theta_D$ = 366 K, and $\lambda_{e-ph}$ = 1.59. 
\begin{figure*}[htbp]
\centering
    \includegraphics[width=0.487\linewidth]{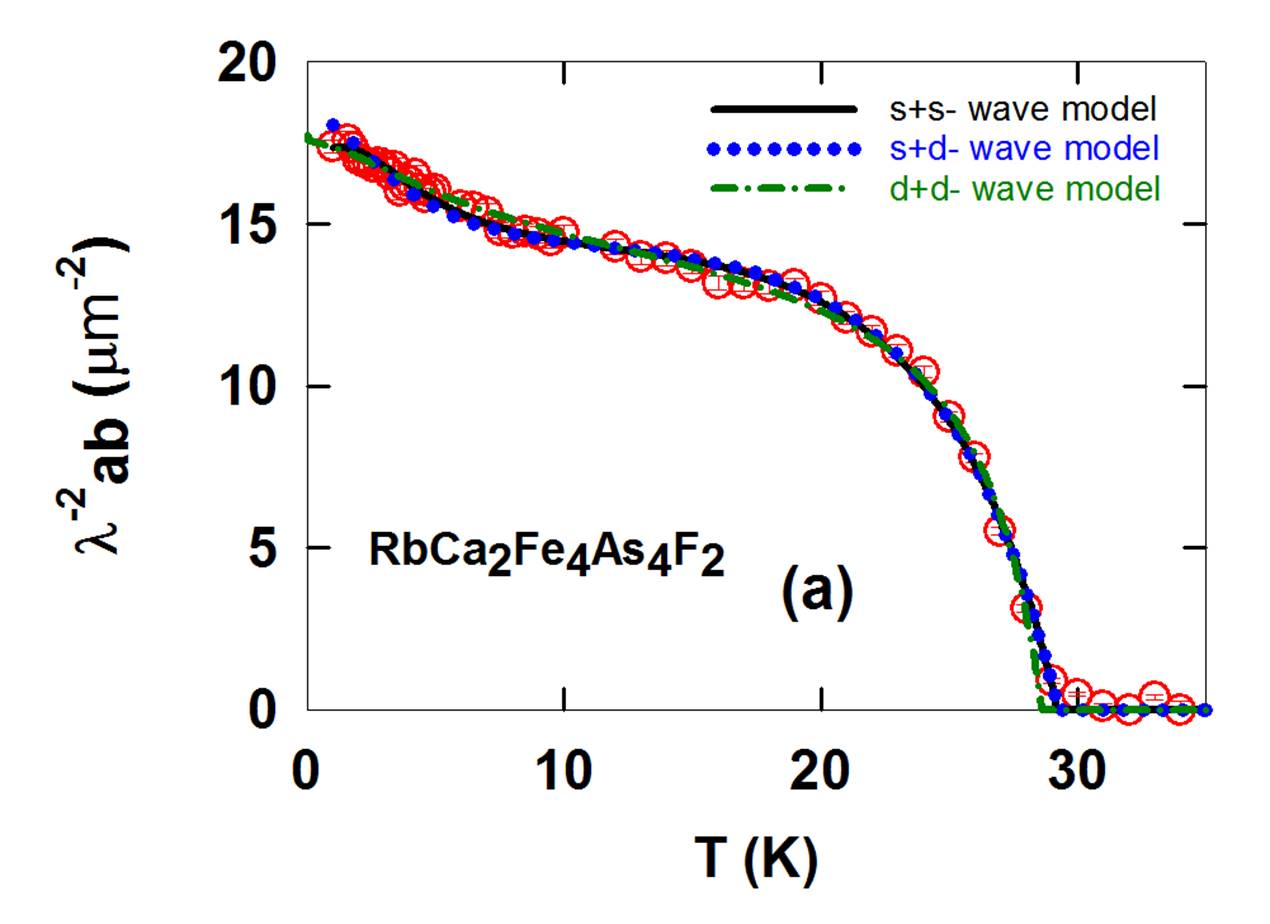}\hfil
     \includegraphics[width=0.35\linewidth]{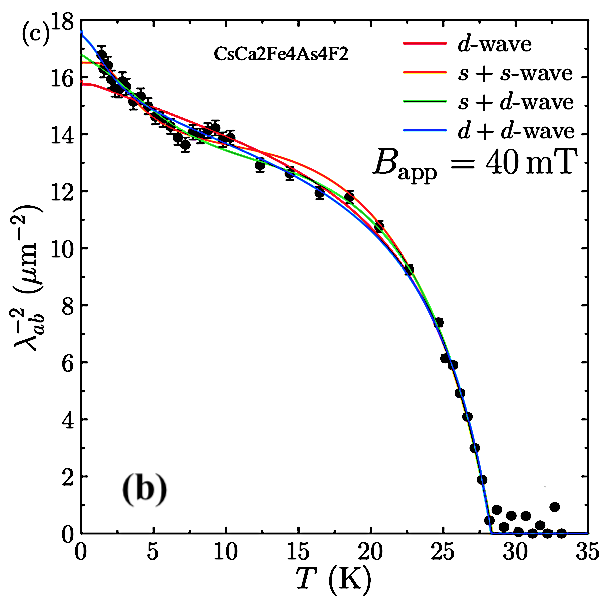}\hfil
\caption{(a) Variation of the square of the inverse in-plane penetration depth $\lambda_{ab}^{-2}(T)$ with temperature for RbCa$_2$Fe$_4$As$_4$F$_2$ from Adroja {\ et al.} arXiv:1802.07334 ~\cite{Adroja2}. The solid and dotted lines represent the fits using different models of the superconducting gap. (b) $\lambda_{ab}^{-2}(T)$ of CsCa$_2$Fe$_4$As$_4$F$_2$ reprinted  with permission from Kirschner {\ et al.} Phys. Rev. B {\bf 97}, 060506(R) (2018)~\cite{Kirschner}. Copyright 2018 by the American Physical Society. Fits using different models with  $d-$, $(s + s)-$, $(s + d)-$, and $(d + d)-$wave superconducting gap structures are also shown.}
\label{Rb/CsCa$_2$Fe$_4$As$_4$F$_2$sigmasc}
\end{figure*}
\subsubsection{RbCa$_2$Fe$_4$As$_4$F$_2$}

\noindent Another member of the 12442 family RbCa$_2$Fe$_4$As$_4$F$_2$ with $T_{c}$ = 29.19 K was investigated by Adroja {\ et al.}~\cite{Adroja2}. For this compound, the magnetic susceptibility and specific heat reveal the presence of bulk superconductivity with a $T_{c}$ of around 29~K. Figure~\ref{Rb/CsCa$_2$Fe$_4$As$_4$F$_2$sigmasc}(a) displays the TF$-\mu$SR results which show that $\lambda_{ab}^{-2}$ cannot be accounted for by either nodal or nodeless single-gap functions. The $\lambda_{ab}^{-2}$ data are well described by both an $s+s$ model with two nodeless gaps, as well as an $s+d$ model where one of the gaps has line nodes. From the TF$-\mu$SR study, superconducting parameters of an in-plane penetration depth $\lambda_{ab}(0)$ = 231.5 nm, superconducting carrier density $n_s$ = 1.29 $\times$ 10$^{27}$ m$^{-3}$, and carrier's effective-mass $m^{*}$ = 2.44 $m_e$ are estimated using $\Theta_D$ = 352 K, and $\lambda_{e-ph}$ = 1.45. Additionally, the ZF$-\mu$SR study does not reveal a clear sign of TRS breaking below $T_{c}$, but the temperature dependence of the relaxation rate of the asymmetry suggests the occurrence of spin-fluctuations (not shown here)~\cite{Adroja2}.  

\subsubsection{CsCa$_2$Fe$_4$As$_4$F$_2$}

\noindent Kirschner {\ et al.}~\cite{Kirschner} reported a TF- and ZF-$\mu$SR study on CsCa$_2$Fe$_4$As$_4$F$_2$ with $T_{c}$ = 28.3 K to understand its superconducting and magnetic properties. They find that $\lambda_{ab}(T)^{-2}$ of CsCa$_2$Fe$_4$As$_4$F$_2$ does not plateau at low temperature, indicating nodal superconductivity as shown in Fig.~\ref{Rb/CsCa$_2$Fe$_4$As$_4$F$_2$sigmasc}(b). At intermediate temperatures, there is an inflection point in the temperature dependence of the superfluid density from which it is concluded that CsCa$_2$Fe$_4$As$_4$F$_2$ is a nodal multigap superconductor, which can be fitted  with a $s + d$-wave model.  ZF$-\mu$SR measurements again show  two components in the relaxation of the asymmetry, and both the Gaussian and a Lorentzian relaxation rates show a weak temperature dependence between 2~K and 50~K, without any clear sign of TRS breaking at $T_{c}$, which is very similar to the results for   KCa$_2$Fe$_4$As$_4$F$_2$ and RbCa$_2$Fe$_4$As$_4$F$_2$ \cite{Smidman1,Adroja2}. This compound contains the largest alkali atom in this family of superconductors, and  $\lambda_{ab}(0)$ = 244 nm,  $n_s$ = 1.16 $\times$ 10$^{27}$ m$^{-3}$, and $m^{*}$ = 2.44$m_e$ are estimated using $\Theta_D$ = 344 K, and $\lambda_{e-ph}$ = 1.44.

\begin{figure*}[t]
\centering
    \includegraphics[width=0.6\linewidth]{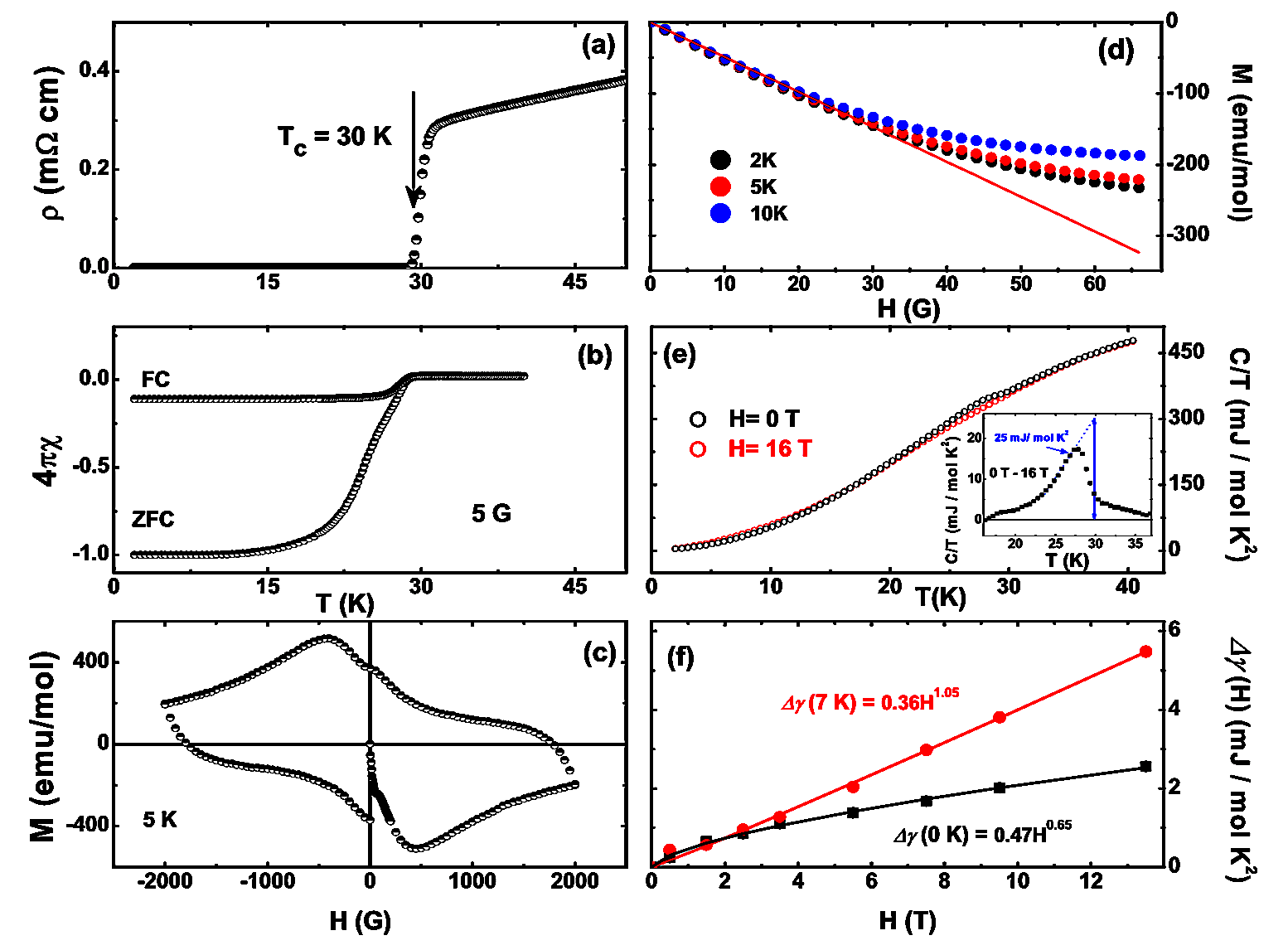}\hfil
     \includegraphics[width=0.39\linewidth]{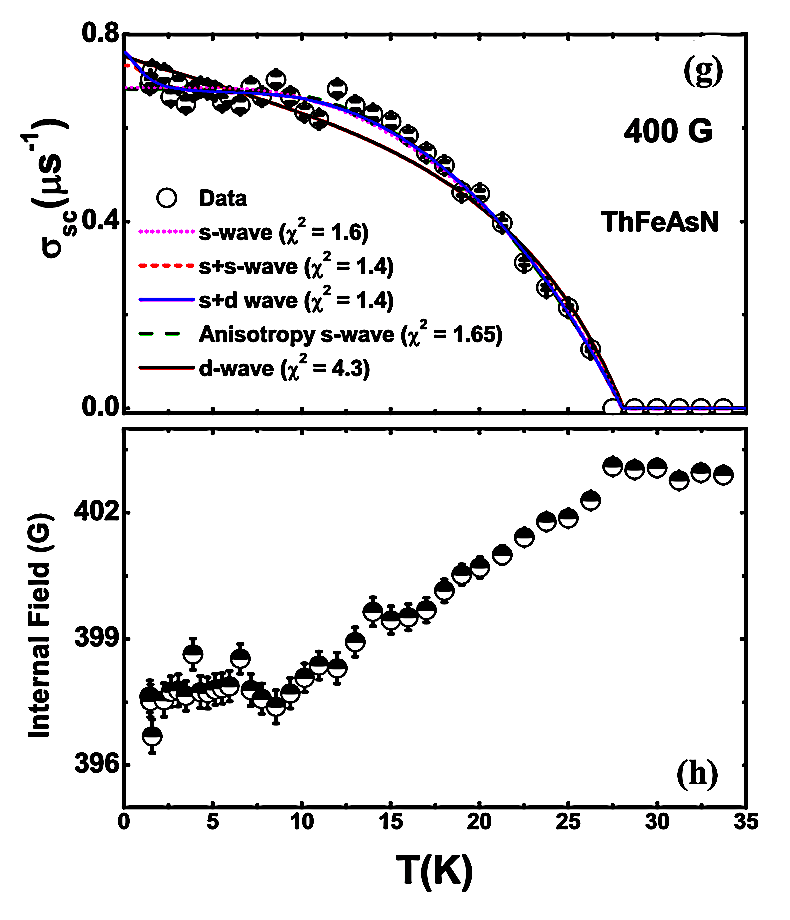}\hfil
\caption{(a) Resistivity of ThFeAsN as a function of temperature. (b) $\chi(T)$ vs. $T$ in zero-field cooled (ZFC) and field cooled (FC) modes with an external applied field of 5 G. (c) Isothermal field dependence of the magnetization at 5 K performed upon sweeping the field to positive and negative fields. (d) Isothermal $M(H)$  at 2, 5 and 10 K. (e)  Temperature dependence of the heat capacity in zero-field and 16 T. The inset of (e) shows the difference of the heat capacity data between the two datasets plotted as $C/T$ vs $T$. The blue vertical arrow shows the heat capacity jump. (f) Variation of the electronic specific heat coefficient $\Delta$$\gamma$ (=$\gamma$(H)-$\gamma$(0)) with field  extrapolated to T$\sim$0 K and 7 K. The solid red line shows a power law fit, $\gamma$(H)$\sim$H$^{n}$. (g) Muon depolarization rate  $\sigma_{\rm sc}(T)$ as function of temperature for ThFeAsN cooled in a field of 400~G, after subtracting the nuclear contribution. The lines correspond to fits of the data using various models of the superconducting gap. The dotted magenta line shows the fit using an isotropic single-gap $s-$wave model, the dashed red line and blue solid line show the fits to the two-gap models $s + s$ wave and $s + d$ wave, respectively. The green long-dashed line represents the fit using an anisotropic $s-$wave model, and the solid purple line shows the fit using the $d-$wave model. (h) Temperature dependence of the internal field of  ThFeAsN from TF-$\mu$SR. All panels reprinted  with permission from Adroja {\ et al.} Phys. Rev. B {\bf 96}, 144502 (2017)~\cite{Adroja3}. Copyright 2017 by the American Physical Society.}
\label{ThFeAsNsigmasc}
\end{figure*}

\noindent It is interesting to note that the multigap behavior observed in $A$Ca$_2$Fe$_4$As$_4$F$_2$ ($A$ = K, Rb, and Cs) is very common in other FeSC materials  (see Table-I). For example, recent $\mu$SR measurements on CaKFe$_4$As$_4$ revealed nodeless two-gap behavior, with gap sizes of 2.5 and 8.6 meV~\cite{Biswas}, which is consistent with the results from tunnel-diode resonator and scanning tunneling microscopy measurements~\cite{Cho}, as well as a high-resolution ARPES study \cite{Mou}. Meanwhile NMR results  reveal spin singlet superconductivity from the decrease of the Knight shift below $T_c$, as well as nearly temperature independent behavior in the normal state \cite{Cui}. Polarized Raman spectroscopy reveals no evidence for Pomeranchuk-like electronic nematic fluctuations \cite{Zhang}, suggesting that the occurence of electronic nematicity is not a requirement for realizing high$-T_{c}$ superconductivity in FeSCs. On the other hand, the analysis of the superfluid density of KCa$_2$Fe$_4$As$_4$F$_2$ and CsCa$_2$Fe$_4$As$_4$F$_2$ indicate that at least one gap has line nodes, which is very different to the fully gapped behavior of CaKFe$_4$As$_4$, while the analysis for RbCa$_2$Fe$_4$As$_4$F$_2$ cannot distinguish between nodeless and nodal two-gap models. Why swapping the Ca layer in CaKFe$_4$As$_4$ for Ca$_2$F$_2$ induces nodal behavior needs to be determined by future studies since it does not correspond to a pressure effect, reduced height of the As atoms, or hole doping~\cite{Smidman1}.
\section{Multiband superconductivity in $\rm {ThFeAsN}$}

\noindent The study of stoichiometric iron pnictide superconductors has been of particular interest, due to the advantage of having reduced disorder, which can be useful for the development of realistic theoretical models for understanding the unconventional superconductivity in the Fe-based materials. Recently, Wang {\ et al.}~\cite{Wang3,Wang4} discovered the first nitride iron pnictide superconductor ThFeAsN with $T_{c}$ = 30 K (the highest of the 1111-type series in the absence of chemical doping), which contains layers with nominal compositions [Th$_2$N$_2$] and [Fe$_2$As$_2$]. The $T_{c}$ of ThFeAsN is as high as the electron-doped 1111-based superconductors, as well as the $A$Ca$_2$Fe$_4$As$_4$F$_2$ ($A$ = K, Rb, and Cs, $T_{c}\sim$ 30 K) and CaAFe$_4$As$_4$ (A = K, Rb and Cs) families.  The first-principles calculations suggest that the lowest-energy ground state corresponds to the stripe-type antiferromagnetic phase~\cite{Singh,Wang5}. However, the normal-state resistivity is metallic~\cite{Wang3}, and a lack of long range magnetic order is confirmed by both neutron diffraction~\cite{Mao} and  $^{57}$Fe M$\ddot{\rm o}$ssbauer spectroscopy~\cite{Albedah}. Furthermore, the application of pressure as well as both electron doping via substitution of N with O and hole doping by swapping Th with Y  suppresses the superconducting $T_{c}$ \cite{Barbero,Bai-Zhuo Li}. Electronic structure calculations show the presence of nested hole and electron Fermi surfaces \cite{Singh,Wang5}, and as a result this material presents the  scenario of a stoichiometric material  near optimal $T_c$, which has an electronic structure much like most other FeSCs.

\subsection{Crystal structure and $T_{c}$}

\noindent Wang {\ et al.}~\cite{Wang3} reported that ThFeAsN  crystallizes in the ZrCuSiAs-type tetragonal crystal structure (space-group $P4/nmm$, No. 129, $Z $= 2) with lattice parameters $a$ = 4.0367 and $c$=8.5262 \AA. In this  structure there are separate layers of Th and N perpendicular to $c$-axis. The As and Fe layers are halfway along the $c$-axis, and the Fe and As ions form tetrahedrons with two As-Fe-As bond angles $\alpha\sim107.0^\circ$ and $\beta\sim 114.5^\circ$ at 300 K. The layered structure of ThFeAsN is very similar to others in the 1111 family of iron pnictide superconductors, such as LaFeAsO~\cite{Paglione}. The $\rho(T)$ data  shown in Fig.~\ref{ThFeAsNsigmasc}(a) shows a superconducting transition onsetting below $T_{c}$ = 30 K.  As shown in Figs.~\ref{ThFeAsNsigmasc} (b) and (e), bulk  superconductivity below 30~K is revealed by measurements of the magnetic susceptibility and specific heat.
\begin{figure*}[t]
\includegraphics[width=0.36\linewidth]{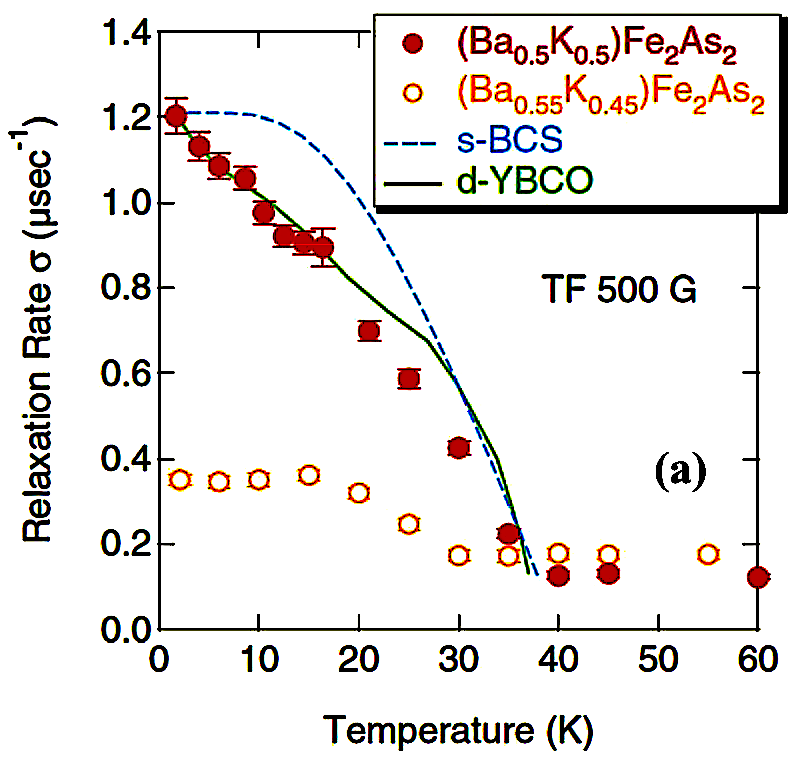} 
\includegraphics[width=0.25\linewidth]{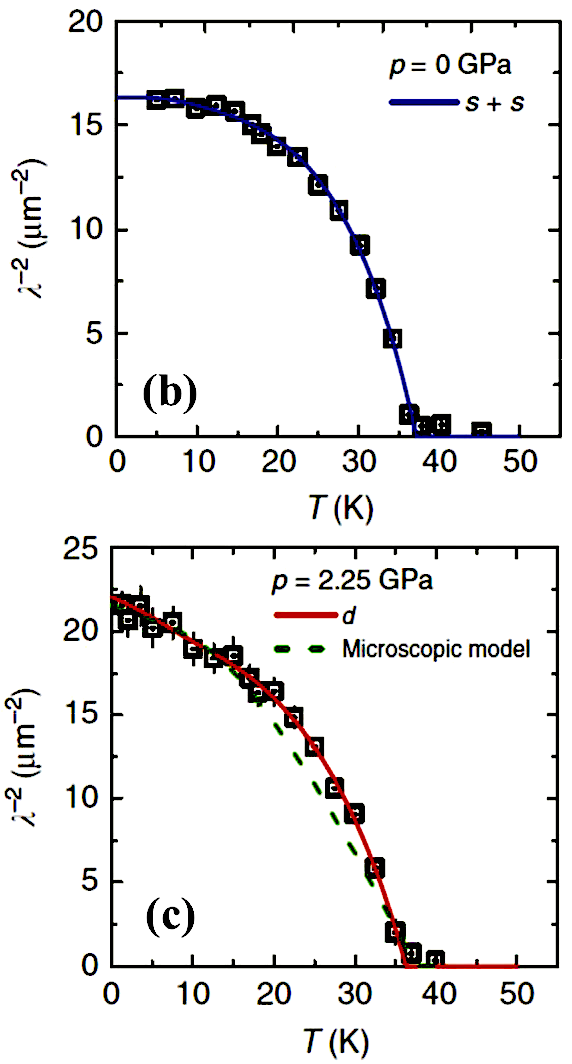} 
\includegraphics[width=0.35\linewidth]{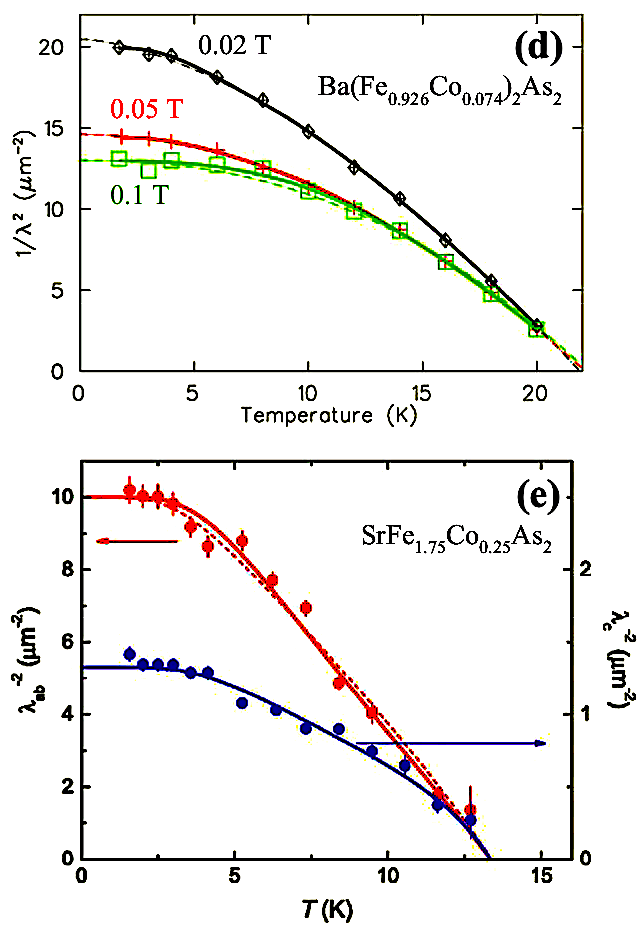} 
\caption{(a) Temperature  dependence of the muon spin relaxation rate $\sigma$ for  Ba$_{0.5}$K$_{0.5}$Fe$_2$As$_2$ and Ba$_{0.55}$K$_{0.45}$Fe$_2$As$_2$.  The dashed curve represents the isotropic gap $s$-wave pairing BCS superconductivity and the solid curve represents the scaled results for YBCO. Reprinted  with permission from Goko {\ et al.} Phys. Rev. B {\bf 80}, 024508 (2009)~\cite{Goko2009}. Copyright 2009 by the American Physical Society. (b,c) Penetration depth $\lambda$ (plotted as $1/\lambda^2$ vs. $T$) obtained from the transverse field muon spin relaxation measurements on Ba$_{0.65}$Rb$_{0.35}$Fe$_2$As$_2$ at (b) ambient pressure and (c) 2.25 GPa. The solid curve in (b) represents the fit with two-gap $s$-wave model and in (c) the fit according to a multiband $d$-wave model. The dashed line in (c) represents the fit by a microscopic model. Both (b) and (c) are reprinted from Guguchia {\ et al.} Nat. Commun. {\bf 6}, 8863 (2015)~\cite{Guguchia2015}. Available under a Creative Commons Attribution 4.0 International License. (d) $1/\lambda^2$ vs. $T$ obtained from the transverse field muon spin relaxation measurements on Ba(Fe$_{0.926}$Co$_{0.074}$)$_2$As$_2$. The solid curve represents the fit with two gaps ($s-$wave) and the dashed line to the fit by a power-law.  Reprinted  with permission from Williams {\ et al.} Phys. Rev. B {\bf 80}, 094501 (2009).~\cite{Williams2009}. Copyright 2009 by the American Physical Society. (e) $1/\lambda^2$ vs. $T$ obtained from the transverse field muon spin relaxation measurements on SrFe$_{1.75}$Co$_{0.25}$As$_2$.  The solid curve represents the fit with two gaps ($s-$wave) in the clean-limit and the dashed shows the fit in the dirty-limit. Reprinted  with permission from Khasanov {\ et al.} Phys. Rev. Lett. {\bf 103}, 067010 (2009)~\cite{Khasanov2009}. Copyright 2009 by the American Physical Society.}
\label{fig:BaKFe2As2_MuSR}
\end{figure*}
\subsection{Superconducting gap structure and presence of time reversal symmetry}

\noindent Adroja {\it et al.}~\cite{Adroja3} used $\mu$SR to investigate the gap symmetry of ThFeAsN. Figures~\ref{ThFeAsNsigmasc}(g) and (h) show the TF-$\mu$SR results, where the temperature  dependence of $\sigma_{\rm sc}$ in  Fig.~\ref{ThFeAsNsigmasc}(g) is fitted better with models with two gaps, than either the  single-gap isotropic $s$-wave or $d$-wave models~\cite{Adroja3}. It can be seen that similarly good fits are obtained with either an isotropic $s + s$ wave model, or an $s + d$ wave model which has one nodal gap as as well as a nodeless one. However  as shown in Fig.~\ref{ThFeAsNsigmasc}(f),   a nonlinear field-dependence of  the specific heat can be observed in the low temperature  limit. This is evidence for  a nodal superconducting gap, suggesting that the $(s + d)-$wave model best explains the gap structure of ThFeAsN~\cite{Adroja3}. Furthermore, the estimated superconducting parameters were $\lambda_L(0)$ = 375 nm, $n_s$ = 4.97$\times$10$^{27}$ m$^{-3}$, and $m^{*}$ = 2.48 $m_e$~\cite{Adroja3}.   Shiroka {\it et al.} ~\cite{Shiroka}, also investigated ThFeAsN using TF- and ZF-$\mu$SR, as well as NMR measurements, which detect the presence of magnetic fluctuations, which weaken upon entering the superconducting state. This was taken as evidence for competition between superconductivity and magnetism, which is in contrast to the often observed scenario of microscopic coexistence \cite{Sanna2010,Dai1}. They also  report that the superconductivity is best accounted for by pairing states without gap nodes, with both a  two-gap $s$-wave and anisotropic $s$-wave models accounting for the data. Moreover, their ZF-$\mu$SR study suggests that time reversal symmetry is preserved in the superconducting state, since there is only a very weak change of the relaxation rate with decreasing temperature.

\subsection{Physical and chemical pressure effects}

\noindent Very recently Bai-Zhuo Li {\ et al.} have investigated  the superconducting properties of ThFeAsN$_{1-x}$O$_x$  ($x$ = 0-0.6) \cite{Bai-Zhuo Li}.   It was found that  increasing the O concentration (electron doping) very quickly suppresses $T_c$  and the system is not superconducting for 0.1$\le x \le$0.2 down to at least 2~K. However further doping leads to remergence of superconductivity, reaching a maximum $T_c$ of 17.5~K for $x=0.3$. As a result there are two superconducting domes in the temperature-doping phase diagram separated by a non-superconducting region, where  the second dome corresponding to heavier doping has a lower maximum $T_c$. Two domes of superconductivity have also been observed under pressure  in the  heavy fermion superconductor CeCu$_2$Si$_2$~\cite{Yuan2003} and FeS~\cite{Jun Zhang1}, where in both cases  it is the second dome at high pressures which has a higher $T_c$. On the other hand the $T$--$P$ phase diagram of FeSe only has one superconducting dome with a maximum $T_c$ of around 37~K at 8.9 GPa~\cite{Medvedev1}.  Two superconducting domes are also observed in LaFeAs$_{1-x}$P$_x$O~\cite{Shen C,Mukuda H}, but in this case  there is an antiferromagnetic phase in the intermediate non-superconducting region~\cite{Mukuda H}. Second dome is also found in the phase diagram of LaFeAsO$_{1-x}$H$_x$ ($x \le$ 0.53) reaching a maximum of $T_c$ = 36 K~\cite{Soshi Iimura1}, in addition to the  dome with $T_c$ = 26 K corresponding to LaFeAsO$_{1-x}$F$_x$~\cite{Yang J}.  It would be interesting to perform $\mu$SR studies on the materials corresponding to the second dome, to examine whether there is change of gap structure and pairing symmetry between the separated phases. 

\noindent The above findings from ThFeAsN can also be compared with $\mu$SR studies on others in the 1111-family of materials.  R. Khasanov {\ et al.,}~\cite{Rustem Khasanov} reported ZF and TF$-\mu$SR measurements on SmFeAsO$_{0.85}$ ($T_c$ = 52 K) and NdFeAsO$_{0.85}$ ($T_c$ = 51~K). From the TF$-\mu$SR measurements,  absolute values of the penetration depth of $\lambda_{ab}(0)$ = 189 and 195 nm are reported for the Sm and Nd samples, respectively. Considering the Uemura classification scheme, both  materials  were designated as unconventional superconductors. The ZF$-\mu$SR data of SmFeAsO$_{0.85}$ shows both slow and fast components to the relaxation, which increase with decreasing temperature, and in particular there is a rapid increase of the fast component below around 10~K. An analysis  of $\sigma_{\rm sc}(T)$ using $\sigma_{\rm sc}(T)$/$\sigma_{\rm sc}(0)$ = (1-$T/T_c$)$^n$ was reported to yield exponents close to $n = 4$ and $n = 2$
for SmFeAsO$_{0.85}$ and NdFeAsO$_{0.85}$, respectively, where the former is close to the universal two fluid value while the latter is expected for $d$-wave superconductors in the dirty limit \cite{Rustem Khasanov}.

\section{\label{Sec:Fe122} Multiband Superconductivity in doped AF\lowercase{e}$_2$A\lowercase{s}$_2$}

\noindent The observation of high-temperature superconductivity in doped $A{\rm Fe_2As_2}$ (A = Ba, Sr, Ca) with $T_{\rm c}$ as high as 38~K in Ba$_{0.6}$K$_{0.4}$Fe$_2$As$_2$ attracted significant interests in these materials~\cite{Johnston2010, Canfield2010, Mandrus2010, Stewart2011}. In particular, the research activities were sparked due to their simple ${\rm ThCr_2Si_2}$-type crystal structure, the easy growth of large single crystals, and the similarity of their generic phase diagram with that of the high-$T_{\rm c}$ cuprates. With a semimetallic ground state the parent $A{\rm Fe_2As_2}$ compounds exhibit an itinerant antiferromagnetic spin density wave transition accompanied by a tetragonal to orthorhombic structural distortion and become superconducting upon suppression of the SDW transition by partial chemical substitutions at either the A, Fe or As sites or by application of external pressure. The emergence of superconductivity upon the suppression of long-range antiferromagnetic ordering is qualitatively similar to the observation of superconductivity in high-$T_{\rm c}$ cuprates where the parent compounds are antiferromagnetic insulators~\cite{Johnston2010, Canfield2010, Mandrus2010, Stewart2011, Damascelli2003, Lee2006}. 
\noindent Understanding the superconducting gap structure and the microscopic pairing mechanism is a key problem for FeSCs, and in general for the understanding of high-$T_{\rm c}$ superconductivity. While two-gap superconductivity (with one small gap and one large gap) is generally believed to occur for the doped-122 superconductors, different superconducting gap structures have been proposed. For optimally doped Ba$_{1-x}$K$_x$Fe$_2$As$_2$, Ba$_{1-x}$Rb$_x$Fe$_2$As$_2$, Ba(Fe$_{1-x}$Co$_x$)$_2$As$_2$ and Ba(Fe$_{1-x}$Ni$_x$)$_2$As$_2$ a nodeless isotropic gap structure seems more appropriate \cite{Hiraishi2009, Goko2009, Ding, Zhao2008, Nakayama2009, Guguchia2011, Williams2009, Williams2010}. On the other hand signatures of nodal superconducting gap structures have been reported for BaFe$_{2}$(As$_{1-x}$P$_{x}$)$_{2}$ and Ba(Fe$_{1-x}$Ru$_x$)$_2$As$_2$ \cite{Hashimoto2010,Yamashita2011,Zhang6,Qiu2012}, as well as in overdoped Ba$_{1-x}$K$_x$Fe$_2$As$_2$ and Ba$_{1-x}$Rb$_x$Fe$_2$As$_2$ \cite{Guguchia2016,Cho2016}. The Fermi surface structure plays an important role in determining the superconducting gap structure. In optimally doped Ba$_{1-x}$K$_x$Fe$_2$As$_2$, the Fermi surface consists of both electron and hole pockets, whereas the Fermi surface of the end point compound KFe$_2$As$_2$ consists of only hole pockets \cite{Cho2016, Xu2013, Hodovanets2014}.
\subsection{Crystal structure and superconductivity}

\noindent The parent $A{\rm Fe_2As_2}$ compounds form in the familiar ${\rm ThCr_2Si_2}$-type  body-centered tetragonal (bct) structure (space group $I4/mmm$, No. 139, Z = 2) which consists of alternating layers of $A^{+2}$ and $[{\rm Fe_2As_2}]^{-2}$ slabs stacked along the $c$-axis.  In this structure each of the $A$, Fe and As atoms occupy only one crystallographic site.  The $A$ atoms form a bct sublattice and the Fe atoms form square-planar layers. Fe and As form ${\rm FeAs_4}$ tetrahedra which are edge shared to form the three-dimensional networks of $[{\rm Fe_2As_2}]^{-2}$; four of the six edges of ${\rm FeAs_4}$ are shared. In doped-$A{\rm Fe_2As_2}$ the $T_{\rm c}$ is suggested to be related to both the As-Fe-As bond angle, and the distance (height) between the Fe layer and the adjacent pnictogen layer~\cite{Johnston2010}. For the K-doped ${\rm BaFe_2As_2}$, the highest $T_{\rm c}$ appears for the As-Fe-As bond angle of $\approx 109.47^\circ$ within an undistorted  ${\rm FeAs_4}$ tetrahedron ~\cite{Johnston2010}. 

\subsection{Superconductivity in hole doped Ba$_{1-x}$K$_x$Fe$_2$As$_2$ and Ba$_{1-x}$Rb$_x$Fe$_2$As$_2$} 

\noindent The hole doping induced by partial substitution of Ba by K in ${\rm BaFe_2As_2}$, which exhibits structural and SDW  transitions near 140~K~\cite{Rotter1, Huang2008}, suppresses the SDW transition leading to the development of superconductivity in Ba$_{1-x}$K$_x$Fe$_2$As$_2$, with a maximum $T_{\rm c}\approx 38$~K for $x=0.4$, i.e. in Ba$_{0.6}$K$_{0.4}$Fe$_2$As$_2$ ~\cite{Rotter}. The $\mu$SR investigations on Ba$_{1-x}$K$_x$Fe$_2$As$_2$ have suggested the coexistence of magnetism and superconductivity as well as phase separation of superconducting (paramagnetic) and antiferromagnetic regions~\cite{Hiraishi2009, Aczel2008, Goko2009, Wiesenmayer2011,Mallett2017}. The $\mu$SR study on Ba$_{0.6}$K$_{0.4}$Fe$_2$As$_2$ by  Hiraishi {\it et al}.~\cite{Hiraishi2009} shows evidence for $s$-wave superconductivity where the $\sigma_{\rm sc}(T)$ data in the superconducting state could be fitted either by a single-gap structure with $2\Delta(0)/k_{\rm B}T_{\rm c} = 5.09(4)$ or by a two-gap structure with $2\Delta_1(0)/k_{\rm B}T_{\rm c} = 7.3$  and $2\Delta_2(0)/k_{\rm B}T_{\rm c} = 4.1(2)$. The large value of $2\Delta(0)/k_{\rm B}T_{\rm c}$ compared to the weak coupling BCS value of 3.53 clearly demonstrates strong coupling superconductivity in Ba$_{0.6}$K$_{0.4}$Fe$_2$As$_2$. 

\noindent A $\mu$SR study on Ba$_{0.5}$K$_{0.5}$Fe$_2$As$_2$ by Goko {\it et al}. ~\cite{Goko2009} found a nearly linear temperature dependence of $\sigma$ in the superconducting state [see Fig.~\ref{fig:BaKFe2As2_MuSR}(a), figure from Ref.~\cite{Goko2009}] which could not be explained by a single isotropic energy gap. Instead the result was suggested to represent either an anisotropic superconducting gap with line nodes or a multi-gap structure.  The ARPES study on Ba$_{0.6}$K$_{0.4}$Fe$_2$As$_2$ favours  two band superconductivity with  strong coupling~\cite{Ding, Zhao2008, Nakayama2009}. Based on ARPES data, the presence of two isotropic and nodeless superconducting gaps and an inter-band scattering scenario for the pairing mechanism have been suggested for superconductivity in Ba$_{0.6}$K$_{0.4}$Fe$_2$As$_2$. 

\noindent Attempts were also made to probe the time reversal symmetry breaking in  Ba$_{1-x}$K$_x$Fe$_2$As$_2$, however the $\mu$SR investigations on polycrystalline Ba$_{1-x}$K$_x$Fe$_2$As$_2$ ($0.5\leq x \leq 0.9$) could not detect any evidence for spontaneous magnetic fields associated with  TRS breaking~\cite{Mahyari2017}.  A recent  $\mu$SR  study on high quality ion-irradiated Ba$_{0.27}$K$_{0.73}$Fe$_2$As$_2$ single crystals reported evidence for TRS breaking~\cite{Grinenko2017}. It has been suggested that the broken TRS state exists only for a very narrow region of $x$ from 0.7 to 0.8 in the phase diagram of Ba$_{1-x}$K$_x$Fe$_2$As$_2$~\cite{Grinenko2017}. The end point compound KFe$_2$As$_2$ was proposed to exhibit a nodal $d$-wave pairing~\cite{Reid2012}, and a crossover from a nodeless gap structure to a nodal state upon heavy hole doping is expected to occur near $x \approx 0.8$~\cite{Cho2016}. The change from nodeless to nodal gap structure is believed to be related to a Lifshitz transition i.e., the disappearance of the electron pockets in the Fermi surface, since while the Fermi surface of optimally doped Ba$_{0.6}$K$_{0.4}$Fe$_2$As$_2$ possesses both electron and hole pockets, the Fermi surface of KFe$_2$As$_2$ has only hole pockets~\cite{Cho2016, Xu2013, Hodovanets2014}. 

\noindent The partial substitution of Ba by Rb in ${\rm BaFe_2As_2}$ also leads to superconductivity in Ba$_{1-x}$Rb$_x$Fe$_2$As$_2$. It is observed that with Rb in place of K, the Fe-sublattice AFM can be completely suppressed, consequently, there is no phase separation of superconducting and antiferromagnetic regions in optimally doped Ba$_{1-x}$Rb$_x$Fe$_2$As$_2$. The $\mu$SR study on Ba$_{1-x}$Rb$_x$Fe$_2$As$_2$ also finds evidence for two-gap superconductivity  ~\cite{Guguchia2011}. The  temperature dependence of the penetration depth $\lambda$ obtained from the $\mu$SR measurements on Ba$_{0.65}$Rb$_{0.35}$Fe$_2$As$_2$ is shown in Fig.~\ref{fig:BaKFe2As2_MuSR}(b) (figure from Ref.~\cite{Guguchia2015}) as $1/\lambda^2$ vs. $T$ along with the fit with two-gap nodeless $s$-wave model. Interestingly, $\mu$SR measurements under the pressure reveal that at pressure of about 2.25~GPa the nodeless gap structure of Ba$_{0.65}$Rb$_{0.35}$Fe$_2$As$_2$ becomes a nodal gap structure and pairing also changes from $s$-wave at ambient pressure to a $d$-wave paring at 2.25~GPa [Fig.~\ref{fig:BaKFe2As2_MuSR}(c)]~\cite{Guguchia2011, Guguchia2015}. A crossover from nodeless $s$-wave gap in optimally doped Ba$_{0.65}$Rb$_{0.35}$Fe$_2$As$_2$~\cite{Guguchia2011} to nodal $d$-wave gap is observed in the $\mu$SR study on overdoped Ba$_{0.35}$Rb$_{0.65}$Fe$_2$As$_2$~\cite{Guguchia2016}. These changes in gap structure by pressure and heavy hole doping is interpreted as the manifestation of Lifshitz transition due to the disappearance of the electron pocket from the Fermi surface at high hole concentration as also noticed in heavily hole doped Ba$_{1-x}$K$_x$Fe$_2$As$_2$~\cite{Guguchia2016,Cho2016}.
\begin{figure*}[t]
\centering
    \includegraphics[width=0.4\linewidth]{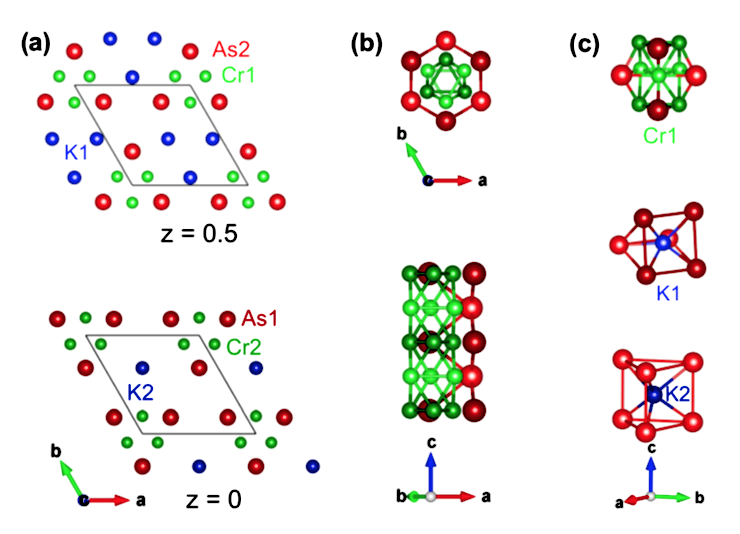}\hfil
     \includegraphics[width=0.23\linewidth]{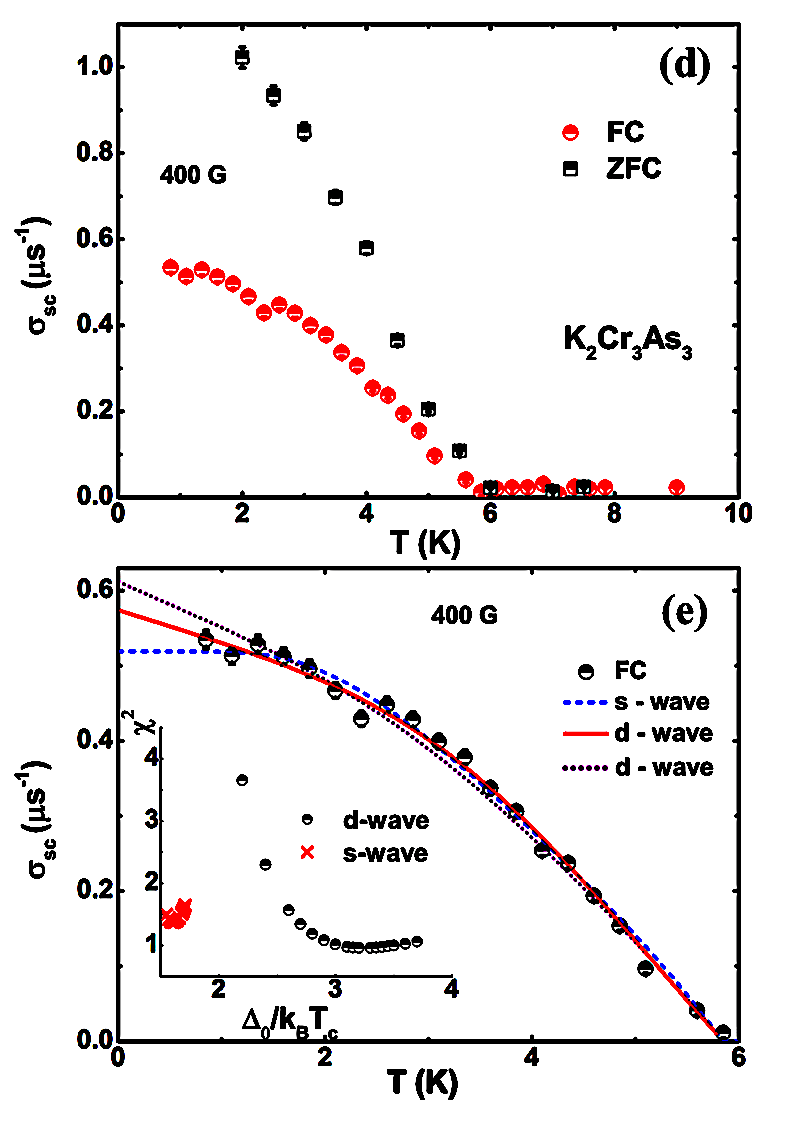}\hfil
     \includegraphics[width=0.25\linewidth]{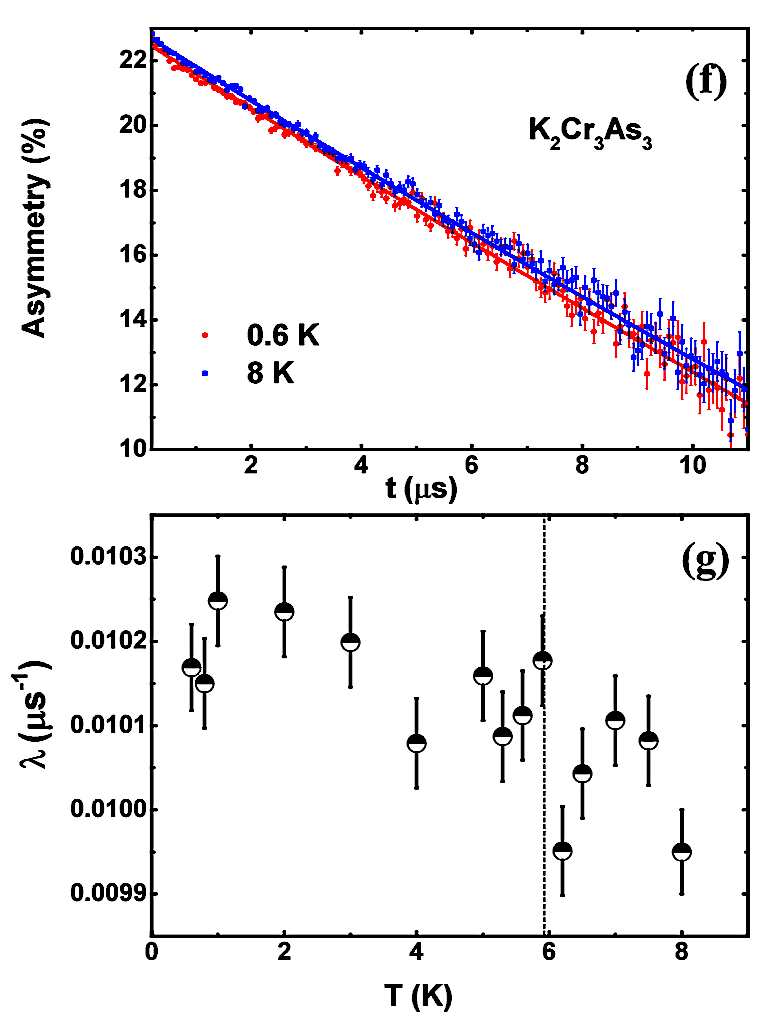}\hfil
\caption{(a) Illustration of the crystal structure of K$_2$Cr$_3$As$_3$.  (b) Crystal structure of the  [(Cr$_3$As$_3$)$^{2-}$]$_\infty$ tubes, which are major structural units in the $A_2$Cr$_3$As$_3$ materials. (c) Illustration showing the chemical bonding of Cr1 (top), K1 (middle), and K2 (bottom). (a)-(c) are reprinted from J. Bao {\ et al. } Phys. Rev. X {\bf 5}, 011013 (2015)~\cite{J. Bao}. Available under a Creative Commons Attribution 3.0 License. (d) Muon depolarization rate $\sigma_{\rm sc}(T)$ as function of temperature for K$_2$Cr$_3$As$_3$ for a field of 400 G with both zero-field and field-cooling. (e) $\sigma_{\rm sc}(T)$ for field-cooled measurements where the lines display fits for different gap models. The short-dashed (blue) line shows the fit using an isotropic single-gap $s-$wave model and the solid (red) line and dotted (purple) lines show fits to a $d$-wave model. The inset shows $\chi^2$ vs $\Delta (0)/k_B T_{c}$. (f) ZF$-\mu$SR asymmetry spectra for K$_2$Cr$_3$As$_3$ at 0.6 K (red) and 8~ K (blue), where the lines show fits to the data. (g) Electronic relaxation rate as a function of temperature, where $T_{c}$ = 5.8 K is shown by the dotted vertical line. (d)-(g) are reprinted  with permission from Adroja {\ et al. } Phys. Rev. B {\bf 92}, 134505 (2015)~\cite{Adroja1}. Copyright 2015 by the American Physical Society.}
\label{K$_2$Cr$_3$As$_3$sigmasc}
\end{figure*}

\subsection{Superconductivity in electron doped Ba(Fe$_{1-x}$Co$_x$)$_2$As$_2$}

\noindent The partial substitution of Fe by Co in ${\rm BaFe_2As_2}$ also suppresses the SDW transition and leads to the emergence of superconductivity in Ba(Fe$_{1-x}$Co$_x$)$_2$As$_2$ as a result of effective electron doping, with a maximum $T_{\rm c}\approx 23$~K for $x\approx 0.07$~\cite{Sefat2008, Ni2008, Wang2009}. The $\mu$SR investigations on Ba(Fe$_{1-x}$Co$_x$)$_2$As$_2$ report the coexistence of superconductivity and magnetism/SDW fluctuations even in the optimally doped case~\cite{Bernhard2009, Williams2009, Marsik2010, Williams2010, Sonier2011, Bernhard2012, Larsen2015}. A $\mu$SR study on Ba(Fe$_{0.926}$Co$_{0.074}$)$_2$As$_2$   showed evidence for two-gap superconductivity with $2\Delta_1(0)/k_{\rm B}T_{\rm c} = 3.77$  and $2\Delta_2(0)/k_{\rm B}T_{\rm c} = 1.57$~\cite{Williams2009, Williams2010}. The  temperature dependence of the penetration depth $\lambda$ is shown in Fig.~\ref{fig:BaKFe2As2_MuSR}(d) (figure from Ref.~\cite{Williams2009}) as $1/\lambda^2$ vs. $T$ along with the fit with two-gap model. The parameter $2\Delta(0)/k_{\rm B}T_{\rm c}$ suggests that the superconductivity in Ba(Fe$_{0.926}$Co$_{0.074}$)$_2$As$_2$ is not in a strong coupling limit as was the case for Ba$_{0.6}$K$_{0.4}$Fe$_2$As$_2$, though they both show a two-gap superconductivity. On the other hand $\mu$SR investigations on SrFe$_{1.75}$Co$_{0.25}$As$_2$ reveal the two gaps to be  $2\Delta_1(0)/k_{\rm B}T_{\rm c} = 7.2$  and $2\Delta_2(0)/k_{\rm B}T_{\rm c} = 2.7$  [Fig.~\ref{fig:BaKFe2As2_MuSR}(e)] \cite{Khasanov2009}. The value of large gap may suggest a strong coupling superconductivity in Sr(Fe$_{1-x}$Co$_x$)$_2$As$_2$, however Williams {\it et al}. ~\cite{Williams2010} report a much smaller value of large gap of  $2\Delta_1(0)/k_{\rm B}T_{\rm c} = 2.7$ in Sr(Fe$_{0.87}$Co$_{0.13}$)$_2$As$_2$, which is smaller than the BCS value of 3.53. Because of the contrasting values of reported large gap no consensus can be reached on the strong vs. weak coupling in Co-doped ${\rm (Ba,Sr)Fe_2As_2}$ .

\begin{figure*}[t]
\centering
    \includegraphics[width=0.3\linewidth]{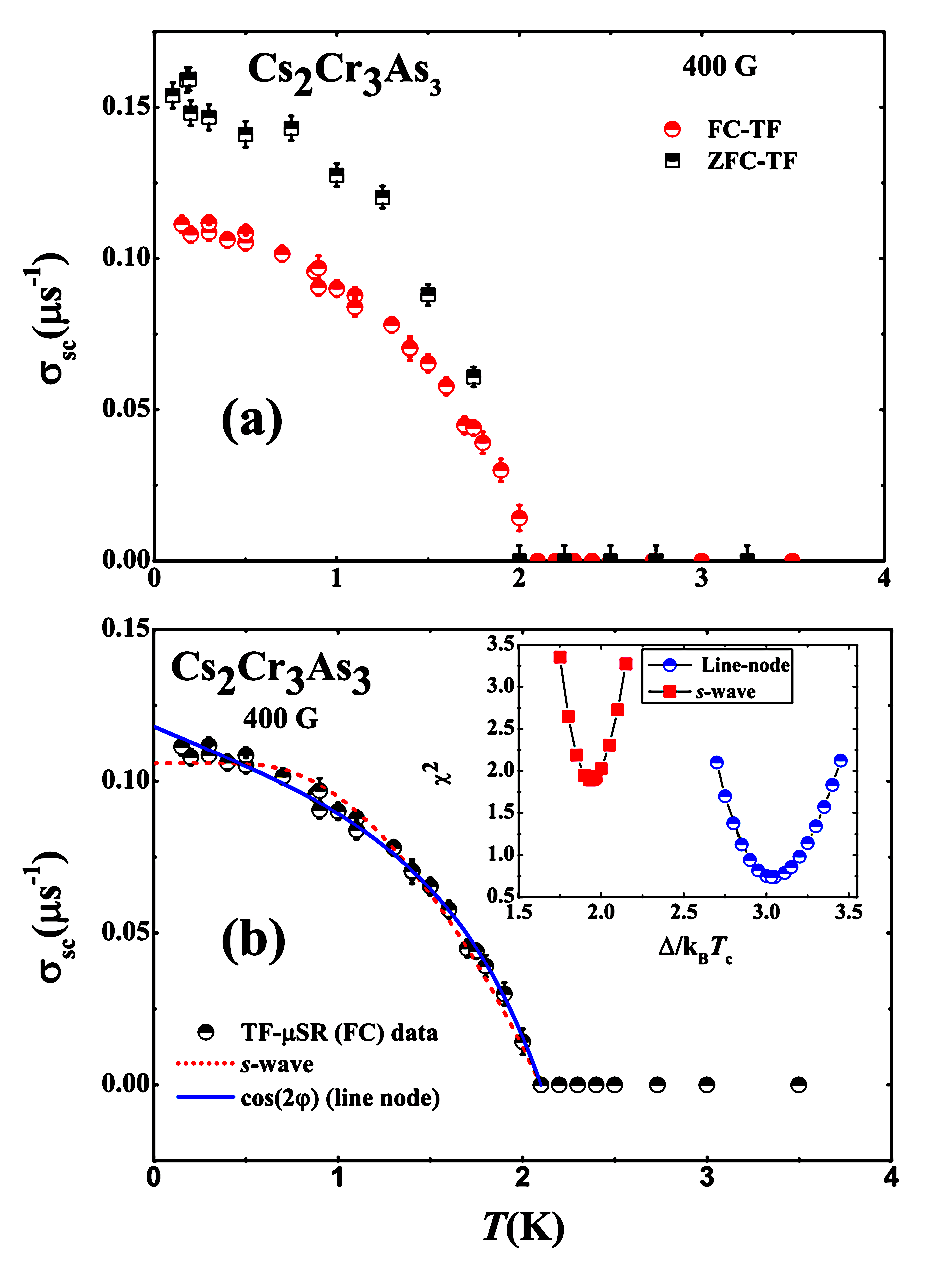}\hfil
     \includegraphics[width=0.43\linewidth]{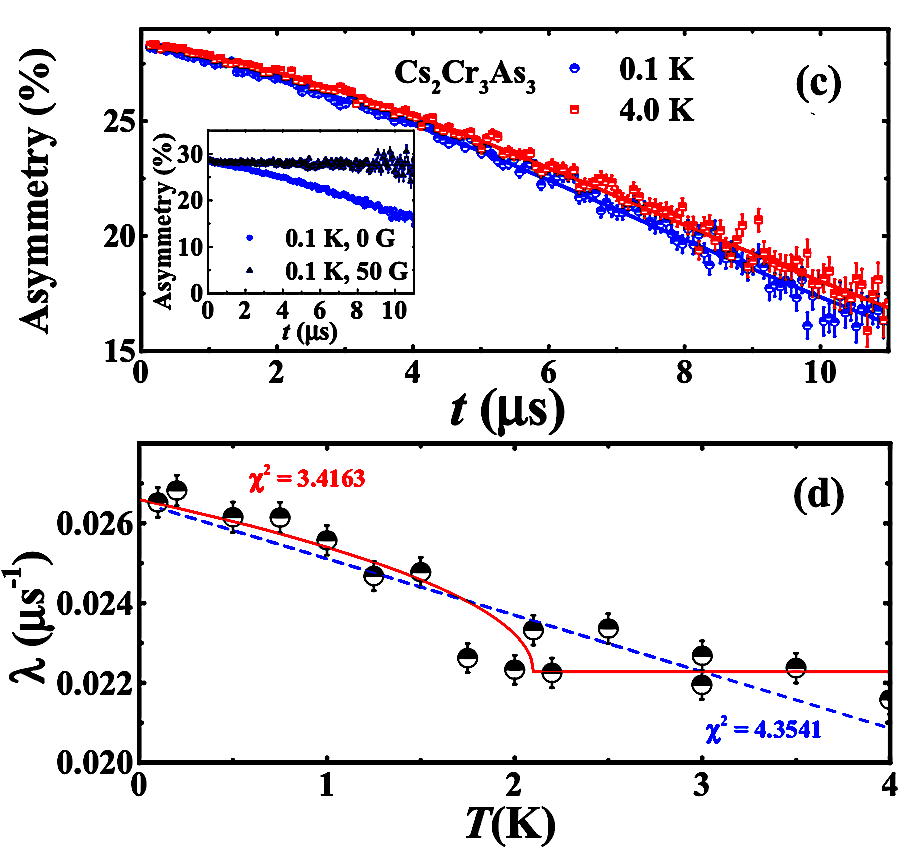}\hfil
\caption{(a) Temperature dependence of the superconducting component of the Gaussian relaxation rate from TF-$\mu$SR measurements of Cs$_2$Cr$_3$As$_3$ in 400~G, for both zero-field and field cooling. (b) shows the field-cooled data, which has been fitted with both  $s$-wave and nodal models for the superconducting gap. (c) ZF$-\mu$SR asymmetry spectra collected at 0.1 K (circles) and 4.0 K (red squares)  together with lines representing least squares fits to the data. The inset  shows the asymmetry spectra in a 50 G longitudinal field together with ZF field results for comparison.  (d) Temperature dependence of the electronic relaxation rate , where the dotted line marks $T_{c}$ = 2.1 K. The blue dotted line shows a linear fit and the solid red line shows a BCS-type function  with $T_{c}$ = 2.1 K fixed from the $\chi(T)$ data. All panels reprinted from Adroja {\ et al.} J. Phys. Soc. Jpn. {\bf 86}, 044710 (2017)~\cite{Adroja5}. Copyright 2017 by the Physical Society of Japan.}
\label{Cs$_2$Cr$_3$As$_3$sigmasc}
\end{figure*}

\section{Nodal superconductivity of quasi-one-dimensional $\rm {A_2Cr_3As_3 (A = K, Rb~and~Cs)}$  superconductors}
\subsection{Crystal structure and $T_{c}$}

\noindent Following the discovery of superconductivity under pressure in CrAs \cite{CrAsSC1,CrAsSC2}, a new family of isostructural superconductors   $A_2$Cr$_3$As$_3$  ($A$ = Na, K, Rb, and Cs) were discovered \cite{J. Bao,Z. Tang1,Z. Tang,Mu1}. These were found to exhibit superconductivity at ambient pressure, with higher values of $T_c$ of 8.6, 6.1, 4.8 and 2.2~K for $A$= Na, K, Rb, and Cs respectively. K$_2$Cr$_3$As$_3$ crystallizes in a hexagonal crystal structure [space group P$\bar{6}$m2 (No. 187)] with lattice parameters $a$ = 9.9832 \AA, $c$ = 4.2304 \AA. \cite{J. Bao}. Here the atoms do not form a planar hexagonal arrangement, but instead lie within two planes offset along the $c$-axis by half a unit cell, as illustrated in Figs.~\ref{K$_2$Cr$_3$As$_3$sigmasc}(a). The Cr and As atoms form quasi-one dimensional double-walled chains of [(Cr$_3$As$_3$)$^{2-}$]$_\infty$, consisting of face sharing octahedra with the outer wall being formed from As and with Cr atoms on the inside, as shown in Figs.~\ref{K$_2$Cr$_3$As$_3$sigmasc}(b)-(c). Within the Cr octahedra, the bond lengths lie between 2.61 and 2.69 \AA, suggesting that metallic bonding between Cr atoms is expected. The double-walled chains run along the crystallographic $c$-axis, and are separated by the K$^{+}$ cations, and hence giving rise to a quasi-one-dimensional crystal structure.

\subsection{Physical properties: K$_2$Cr$_3$As$_3$ and Cs$_2$Cr$_3$As$_3$}

\noindent Specific heat measurements of  K$_2$Cr$_3$As$_3$  reveal the presence of bulk superconductivity, where the large jump in the specific heat at the transition of $2.2\, \gamma T_{c}$ is considerably greater than that expected from weak coupling theory, indicating that the superconductivity is strongly coupled \cite{J. Bao}. From probing the normal state, strong electronic correlations were revealed by a large electronic specific heat coefficient and non-Fermi liquid transport behavior~\cite{J. Bao,T. Kong},  which are a common feature of unconventional superconductors. A lack of BCS pairing was also suggested from the very large and anisotropic upper critical fields ($H_{c2}$), where $H_{c2}$ perpendicular to the Cr$-$chains is significantly larger than that corresponding to the parallel direction. Moreover, paramagnetic pair breaking appears to be absent for the perpendicular direction, but parallel to the Cr$-$chain $H_{c2}$ appears still to have paramagnetic limiting, but  with an enhanced Pauli field~\cite{T. Kong, Balakirev}. 

\noindent Measurements of the magnetic susceptibility show that superconductivity occurs below 2.1 K and the isothermal $M(H)$ curve suggests  type-II superconductivity in Cs$_2$Cr$_3$As$_3$~\cite{Z. Tang,Adroja5}. The reported value of the upper critical field $H_{c2}$ = 6.45~T is higher than the Pauli limit, $H_P = 1.84 T_{c}$ = 3.86~T,  much like the other isostructural materials, which again indicates that the superconductivity is unconventional. In zero field,  $\rho(T)$ shows metallic behavior with a linear temperature dependence from 50 K down to just above $T_{c}$, indicating non-Fermi-liquid behavior, suggesting the importance of spin fluctuations~\cite{Z. Tang,Adroja5}.

\subsection{Superconducting gap structure and time reversal symmetry}

\subsubsection{K$_2$Cr$_3$As$_3$}

\noindent Adroja {\it et al.}~\cite{Adroja1}  reported a TF- and ZF-$\mu$SR study on K$_2$Cr$_3$As$_3$. The temperature dependence of  $\sigma_{\rm sc}$ obtained from the TF-$\mu$SR measurements could be fitted  to both an isotropic $s$-wave model, as well as a $d$-wave model with line nodes~\cite{Adroja1}. However, the goodness-of-fit values indicate that the data fit better to the $d$-wave model than the $s$-wave model [see Fig.~\ref{K$_2$Cr$_3$As$_3$sigmasc}(e)]. Therefore the $\mu$SR results are more consistent with having line nodes than a fully open superconducting gap, which is consistent with the findings of penetration depth measurements using a tunnel diode oscillator technique, where a linear temperature dependence is revealed at low temperatures in the highest quality samples~\cite{Pang2}. The ZF-$\mu$SR study reveals some evidence for the appearance of weak internal  fields below $T_{c}$, suggesting that the superconducting state may be  unconventional, but these results cannot ascertain whether there is spontaneous broken time reversal symmetry. Moreover, from the TF-$\mu$SR  study the estimated superconducting parameters are $\lambda_L(0)$ = 432(4) nm, $n_s$ = 2.7 $\times$ 10$^{27}$ m$^{-3}$, and $m^* = 1.75\,m_{\rm e}$, respectively.  The NMR study by Zhi {\it et al.}~\cite{Zhi} on K$_2$Cr$_3$As$_3$ also supported the conclusion of unconventional non $s$-wave superconductivity from the absence of a Hebel-Slichter coherence peak in the spin-lattice relaxation rate $1/T_1$ just below $T_c$. The measurements of $1/T_1$ also revealed enhanced Cr spin fluctuations in the normal state above $T_c$, where the power law dependence of $1/T T_1\sim T^{-0.25}$ was suggested to be compatible with a Tomonaga-Luttinger liquid. Signatures of Tomonoga-Luttinger liquid physics were also found from ARPES measurements by Watson {\it et al.,}~\cite{Watson}, and the relationship between the low dimensional spin fluctuations and the unconventional superconductivity requires further exploration.

\begin{figure*}[t]
\centering
    \includegraphics[width=0.6\linewidth]{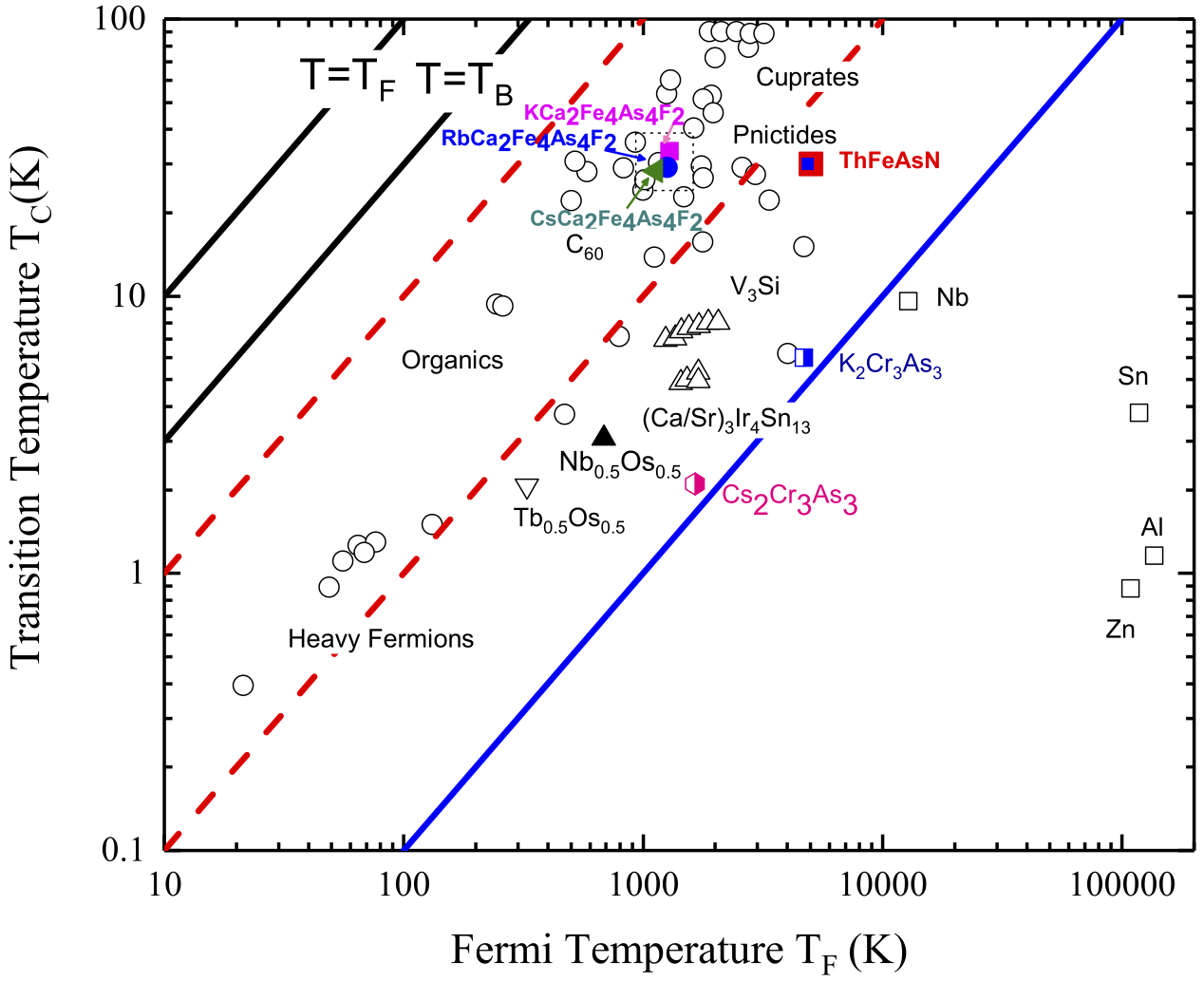}\hfil
\caption{Uemura plot of $T_{c}$ vs. effective Fermi temperature $T_{F}$. The ``exotic" superconductors lie within a region bounded by 1/100$\le T_{c}$/$T_{F}\le$1/10,as indicated by the red dashed lines from Adroja {\ et al.} Phys. Rev. B {\bf 96}, 144502 (2017)~\cite{Adroja3}. The solid black line correspond to the Bose-Einstein condensation temperature ($T_{B}$). The positions of $A$Ca$_2$Fe$_4$As$_4$F$_2$ ($A$ = K, Rb and Cs) and ThFeAsN on the plot place these materials as being exotic superconductors~\cite{Adroja3}.}
\label{fig:Uemura}
\end{figure*}

\subsubsection{Cs$_2$Cr$_3$As$_3$}

\noindent  Figures~\ref{Cs$_2$Cr$_3$As$_3$sigmasc} (a)-(b) show the TF-$\mu$SR results, and the temperature dependence of $\sigma_{\rm sc}$ is better accounted for by a nodal gap structure than an isotropic $s$-wave model \cite{Adroja5}. The observation of a nodal gap in Cs$_2$Cr$_3$As$_3$ is consistent with that observed in the isostructural K$_2$Cr$_3$As$_3$ compound via TF$-\mu$SR and tunnel diode oscillator based measurements \cite{Adroja1,Pang2}. Moreover, from the TF-$\mu$SR study the estimated values of the superconducting parameters are $\lambda_L(0)$ = 954 nm,  $n_s$ = 4.98 $\times$ 1026 m$^{-3}$, and  $m^* = 1.61\,m_e$.  ZF-$\mu$SR results are displayed in Figs.~\ref{Cs$_2$Cr$_3$As$_3$sigmasc} (c)-(d) , which reveal the existence of spin fluctuations below 4 K and the ZF relaxation rate $\lambda$  shows an enhancement below $T_{c}$ = 2.1 K, suggesting the slowing down of electronic spin fluctuations. To verify whether  the increase in $\lambda$ is correlated with $T_c$, two fits to the temperature dependence were performed; (1) a linear function across the whole temperature range, and (2) a BCS-type function onsetting at 2.1~K [see blue dashed and solid red lines in Fig. \ref{Cs$_2$Cr$_3$As$_3$sigmasc}(d), respectively]. The value of the goodness of fit for (1) is $\chi^2$ = 4.35, while for (2) is $\chi^2$ = 3.42. This results indicates that $\lambda$ shows an increase below $T_c$, suggesting that the superconductivity of Cs$_2$Cr$_3$As$_3$  is associated with the spin fluctuation mechanism, and is hence unconventional.

Zhi {\it et al.}~\cite{Zhi3} studied the NMR and $^{75}$As nuclear quadrupole resonance to probe the normal state of Cs$_2$Cr$_3$As$_3$. Their Knight shift  measurements corresponding to the Cs1 site show an increase of the uniform spin susceptibility upon decreasing the temperature from 295~K to around 60~K. Upon further lowering the temperature there is a slight decrease before flattening out below  about 10~K. On the other hand, the Knight shift for the Cs2 is very small, indicating little contribution to the exchange interactions. From a comparison of the low temperature spin-lattice relaxation rates with  K$_2$Cr$_3$As$_3$ and Rb$_2$Cr$_3$As$_3$, it is suggested that substituting increasingly large alkali atoms both suppresses the Cr spin fluctuations and the superconducting  $T_{c}$~\cite{Zhi,Zhi3,Yang2015}. This suggests a relationship between spin fluctuations and the superconductivity in $A$Cr$_3$As$_3$, again pointing towards an unconventional pairing mechanism.

\section{Uemura classification scheme:}

\noindent In this section we focus upon the Uemura classification scheme~\cite{Uemura1,Hillier1} which is based on the correlation between the superconducting $T_{c}$ and the effective Fermi temperature, $T_{F}$, determined from $\mu$SR measurements of the superconducting penetration depth. Within this scheme strongly correlated ``exotic" superconductors, i.e. high $T_{c}$ cuprates, heavy fermions and organic materials, all lie within a particular part of the diagram, which is indicative of a degree of universal scaling~\cite{Hillier1} of $T_{c}$ with $T_{F}$ such that $1/10\ge (T_{c} /T_{F}) \ge 1/100$. For conventional BCS superconductors $1/1000 \ge T_{c} / T_{F}$. Considering that the value of $T_{c}/T_{F}$ = 30/4969.4 = 0.006 for ThFeAsN [see Fig.11], this material can be classified as lying close to the proposed limit~\cite{Uemura1,Hillier1}. Taking into account the value of $T_{c}/T_{F}$ = 33.36/1285 =  0.026 for KCa$_2$Fe$_4$As$_4$F$_2$, $T_{c}/T_{F}$ = 29.19/1260 =  0.023 for RbCa$_2$Fe$_4$As$_4$F$_2$, $T_{c}/T_{F}$ = 28.31/1130 =   0.025 for CsCa$_2$Fe$_4$As$_4$F$_2$ (see Fig. 11), these materials can be designated as  exotic superconductors. For K$_2$Cr$_3$As$_3$, $T_{c}/T_{F}$ = 6.0/4689 = 0.0013 and Cs$_2$Cr$_3$As$_3$ = $T_{c}/T_{F}$ = 2.1/1651 = 0.0012. Moreover, we have also included the data of $A$Ca$_2$Fe$_4$As$_4$F$_2$ ($A$ = K, Rb and Cs) and $A_2$Cr$_3$As$_3$ ($A$ = K and Cs) to Uemura plot as shown in Fig. 11, which shows that the former compounds belong to the exotic class, while the latter are away from this region.

\section{Concluding remarks}

\noindent Despite extensive research activities worldwide following the discovery of FeSCs, the pairing states and mechanisms of superconductivity in these materials are still not well understood. Nevertheless, $\mu$SR is a powerful technique for elucidating these mysteries, in particular, due to the ability to characterize the superconducting gap structure, which is vital for revealing the underlying pairing interactions.  Although conventional phonon-mediated superconductors exhibit an isotropic $s$-wave gap with a 2$\Delta/k_BT_c \approx 3.53$ for weak-coupling, so far no consensus has been reached on the gap character of FeSCs. In this review, we discuss TF and ZF-$\mu$SR measurements in the normal and superconducting states of $A$Ca$_2$Fe$_4$As$_4$F$_2$ ($A$ = K, Rb and Cs), ThFeAsN,  $A_2$Cr$_3$As$_3$ (A = K and Cs) and in doped $A$F\lowercase{e}$_2$A\lowercase{s}$_2$.  

\noindent The superfluid densities of $A$Ca$_2$Fe$_4$As$_4$F$_2$ ($A$ = K, Rb and Cs) can be modelled using two-gap models, which better accounts for the data than single gap $s$-wave or $d$-wave models~ \cite{Smidman1,Kirschner,Adroja2}. Here an $s+d$ model can describe the data for all three members of this family of materials, while $A$ = K can also be accounted for by a $d+d$ model and $A$ = Rb can also be fitted with a $s+s$ model.  A larger value of 2$\Delta/k_BT_c$ ($\sim$ 6.5) than the 3.53 expected from BCS theory have also been obtained from the  $\mu$SR investigations of all these compounds, indicating  strongly coupled superconductivity. The observation of two gaps is very similar to what is often found in FeSCs (see Table-I).

\begin{table*}[tb]

\caption{Superconducting state properties of iron and chromium based superconductors probed by various techniques. Theoretically in the weak coupling limit $2\Delta/k_{\rm B}T_{\rm c}\approx 3.53$ for $s$-wave gap, and $2\Delta/k_{\rm B}T_{\rm c}\approx 4.28$ for $d$-wave. Note on abbreviations used for techniques in Table: TDO stands for tunnel diode oscillator, TC for thermal conductivity, HC for heat capacity, STM for scanning tunneling microscopy, NQR for nuclear quadrupole resonance and QPI for quasiparticle interference.}

\label{FeAsTable}
\begin{tabular}{l c c c c c r }

Compounds & $T_c$ (K) & $2\Delta/k_{\rm B}T_{\rm c}$ & Pairing state & Nodeless/nodal & Techniques & References \\


\hline

LaFePO &6, 7.5 & & $s_{+-}$ & Nodal &TC/TDO & \cite{J. D. Fletcher},\cite{Mike Sutherland}\\

LiFeAs &18 & & Anis. $s$-wave & Nodeless &ARPES/QPI & \cite{Allan2012, K. Umezawa}\\

LiFeP & 4.5 & & & Nodal &TDO & \cite{Hashimoto2012}\\

LiFeAsO$_{1-x}$F$_x$ &18 &4.2, 1.1 & $s_{+-}$ & Two-gap & $\mu$SR & \cite{Takeshita2009}\\

FeS & 4.04 & 4.54, 2.47 & $s+d$-wave & Nodal, two-gap & $\mu$SR &\cite{FeSCTan}\\

FeSe & 8.5&3.69, 1.64 & Iso. \& Anis. $s$-wave & Nodeless, two-gap & HC/STM &\cite{Hsu}\\

FeSe$_{0.85}$&8.3 & 4.49, 1.07 &$s+s$ &Nodeless &$\mu$SR & \cite{Khasanov2008}\\

& & 6.27 &Anis. $s$-wave &Nodeless &$\mu$SR & \cite{Khasanov2008}\\

FeTe$_{0.55}$Se$_{0.45}$&14.5 & &$s$-wave &Nodeless &ARPES & \cite{Miao}\\

KFe$_2$As$_2$ & 3.8 & & $d$-wave & Nodal &TC/TDO/$\mu$SR & \cite{Hashimoto2010,J.-Ph. Reid}\\

 &  & & $s$-wave & Nodal &ARPES & \cite{KFe2As2ARPES}\\

RbFe$_2$As$_2$ & 2.5 &4.55, 1.39 &$s+s$ & Nodeless & $\mu$SR& \cite{Shermadini1,Shermadini2}\\

& & &$d$ & Nodal & TC & \cite{Zhang6}\\

CsFe$_2$As$_2$ &1.8 & &$d$& Nodal & TC/HC & \cite{Hong2013},\cite{Wang5}\\

Ba$_{0.6}$K$_{0.4}$Fe$_2$As$_2$ &37 &7.5, 3.7 & $s$ & Nodeless, two gap &ARPES & \cite{Ding}\\

& 38 &7.3, 4.1 & $s$ & Nodeless, two gap &$\mu$SR & \cite{Hiraishi2009}\\

BaFe$_2$(As$_{0.7}$P$_{0.3}$)$_2$ &30 & & $s$ &Nodal &ARPES/TC & \cite{Zhang6,Yamashita2011}\\

Ba(Fe$_{0.77}$Ru$_{0.23}$)$_2$As$_2$ &17 & & & Nodal&TC & \cite{Qiu2012}\\

Ba(Fe$_{2-x}$Co$_x$)As$_2$& 22.1& 3.77, 1.57 & $s$ & Nodeless & $\mu$SR & \cite{Williams2009}\\

CaKFe$_4$As$_4$ &34.3 &5.82, 1.69 & $s+s$ & Nodeless, two gap & $\mu$SR/TDO/STM & \cite{Biswas,Cho}\\

KCa$_2$Fe$_4$As$_4$F$_2$ & 33.4 &7.03, 1.28& $s+d$ &Line nodes &$\mu$SR & \cite{Smidman1}\\

& & 10.14, 1.19&$d+d$  & Line nodes &$\mu$SR & \cite{Smidman1}\\

RbCa$_2$Fe$_4$As$_4$F$_2$ &29.2& 6.48, 0.70 & $s+s$& &$\mu$SR & \cite{Adroja2}\\

& &6.42, 0.73 & $s+d$ & Nodal &$\mu$SR & \cite{Adroja2}\\

CsCa$_2$Fe$_4$As$_4$F$_2$ &28.3 & 6.15, 1.24& $s+d$&Line nodes &$\mu$SR & \cite{Kirschner}\\

ThFeAsN &28.1 & 4.29, 0.25 & $s+d$, $s+s$&Nodal and nodeless &$\mu$SR/HC & \cite{Adroja3,Shiroka}\\

K$_2$Cr$_3$As$_3$ &5.8 &6.4 & $d$& Line nodes &$\mu$SR/TDO & \cite{Adroja1,Pang2}\\

Cs$_2$Cr$_3$As$_2$ & 2.1&6.0 & $d$ & Line nodes &$\mu$SR & \cite{Adroja2}\\

Rb$_2$Cr$_3$As$_3$ & 4.8 &4.2 & & Point nodes& NQR & \cite{Yang2015}\\

\end{tabular}
\end{table*}  

\noindent The nodal superconductivity observed  in KCa$_2$Fe$_4$As$_4$F$_2$ and CsCa$_2$Fe$_4$As$_4$F$_2$~\cite{Smidman1,Kirschner} is different to the case of the CaKFe$_4$As$_4$, where in the latter distinct signatures are found for a nodeless $s\pm$ pairing state~\cite{Biswas,Cho,Mou}, despite both systems being near optimal hole doping~\cite{Mou}. Moreover the electronic structure calculations by Wang {\it et al.}\cite{Wang7} are in line with many other Fe-based materials with hole pockets at the zone center and electron pockets at the edge, and therefore a nodeless $s\pm$ state might be expected. Consequently the change in superconducting gap structure between CaKFe$_4$As$_4$ and KCa$_2$Fe$_4$As$_4$F$_2$ is not similar to the case of (Ba$_{1-x}$K$_x$)Fe$_2$As$_2$, where the change from nodeless to nodal superconductivity is brought about by hole doping \cite{Ding,KFe2As2ARPES,Fukazawa2009, Dong2010}.  Guguchia {\it et al.}~\cite{Guguchia2015} proposed that pressure leads to a change from nodeless to nodal superconductivity in Ba$_{0.65}$Rb$_{0.35}$Fe$_2$As$_2$, while it was also suggested that nodal superconductivity occurs in Fe based materials when the  pnictogen is less than 1.33 \AA~above the Fe-layer \cite{Hashimoto2012}. Nevertheless, in KCa$_2$Fe$_4$As$_4$F$_2$ the As atoms are at heights of 1.40 and 1.44 \AA~\cite{Smidman1}, well above this upper limit and indeed above the values found for CaKFe$_4$As$_4$ (1.40 and 1.35 \AA )~\cite{Iyo2016}. This then suggests that this represents a different scenario for the occurence of   nodal superconductivity in FeSCs, and whether the asymmetry above and below  the Fe-layers which is present in both CaAFe$_4$As$_4$ and   $A$Ca$_2$Fe$_4$As$_4$F$_2$ plays a role in this, is yet to be determined. Meanwhile for RbCa$_2$Fe$_4$As$_4$F$_2$, Adroja {\it et al.}~\cite{Adroja2} find that the nodeless $s+s$- and nodal $s+d$-wave models can both fit the superfluid density well  ~\cite{Adroja2}.  Since this study could not distinguish between the nodeless and nodal models, additonal measurements are important to determine whether the gap structure is similar to the other materials in this family.

\noindent An outstanding puzzle in ThFeAsN is the origin of the  superconductivity in a stoichiometric compound in the apparent absence of a nearby ordered phase \cite{Shiroka}. Interestingly, band structure calculations predict a stripe antiferromagnetic ground state for   ThFeAsN which is not observed \cite{Singh,Wang5},  but instead ZF$-\mu$SR reveals the presence of magnetic fluctuations \cite{Adroja3,Shiroka}. Adroja {\it et al.}~\cite{Adroja3} determine that for ThFeAsN, $\sigma_{\rm sc}(T)$ can be modelled with a two-gap model with either an isotropic $s+s$ or a $s+d$ wave gap structure, instead of a single-gap isotropic $s$-wave, anisotropic $s$-wave, or $d$-wave models, which is in accordance with the common trend of multiband superconductivity in FeSCs. However, it was suggested that  the field-dependent heat capacity can discriminate between the two-gap nodal and nodeless scenarios, and from this it was concluded that the $(s + d)$-wave model is a better description of the gap function of ThFeAsN~\cite{Adroja3}.  Additionally, the gap value of 2$\Delta/k_BT_c$ = 4.29 obtained from the superfluid density analysis suggests strongly coupled superconductivity in ThFeAsN~\cite{Adroja3}. On the other hand, Shiroka {\it et al.} ~\cite{Shiroka} concluded that the TF-$\mu$SR results are best described by nodeless two-gap superconductivity. These results may be important for the proposal of theoretical models to explain the origin of unconventional superconductivity in ThFeAsN. It would be of particular interest to examine the nature of the superconductivity in the second superconducting dome in ThFeAsN$_{1-x}$O$_x$ ($x\sim$ 0.3) using $\mu$SR and neutron scattering measurements. 

\noindent The $\mu$SR investigations of both hole-doped and electron-doped ${\rm BaFe_2As_2}$ have been very useful for understanding the nature of the superconducting order parameter and  two-gap superconductivity is well supported in these doped 122 systems. $\mu$SR studies on hole doped Ba$_{1-x}$K$_{x}$Fe$_{2}$As$_{2}$ have shown two isotropic and nodeless superconducting gaps in optimally doped Ba$_{0.6}$K$_{0.4}$Fe$_2$As$_2$ with a strong coupling \cite{Hiraishi2009}. Evidence for  TRS breaking is also found in ion-irradiated Ba$_{0.27}$K$_{0.73}$Fe$_2$As$_2$  \cite{Grinenko2017}. Furthermore, a crossover from a nodeless  to nodal  gap is proposed to be induced by heavy hole doping near $x \approx 0.8$ \cite{Cho2016}, associated with the disappearance of the electron pockets in the Fermi surface at a Lifshitz transition in heavily hole doped materials. This kind of crossover was also seen in Ba$_{1-x}$Rb$_x$Fe$_2$As$_2$, which is induced by both heavy hole doping as well as by hydrostatic pressure \cite{Guguchia2011,Guguchia2016}. The $\mu$SR investigations on electron-doped  Ba(Fe$_{1-x}$Co$_x$)$_2$As$_2$ have also shown evidence for two-gap superconductivity, however, the superconductivity does not seem to be strongly coupled \cite{Williams2009, Williams2010}.

\noindent An important puzzle for the $A_2$Cr$_3$As$_3$ ($A$ = K, Rb, and Cs)  family of superconductors is the nature of the superconducting pairing symmetry, especially given the clear evidence for line nodes in the superconducting gap from $\mu$SR and tunnel diode oscillator based measurements \cite{Pang2, Adroja1,Adroja5}. One of the leading pairing symmetries predicted from theoretical proposals for a broad range of parameters is a  triplet $p_{z}$-wave pairing state arising due to ferromagnetic fluctuations \cite{Wu2,Zhou}. This state has a nodal plane at $k_z=0$, and is also able to account for the disappearance of paramagnetic limiting in the upper critical field perpendicular to the chain direction \cite{T. Kong, Balakirev}. On the other hand, the anisotropy of the paramagnetic limiting fields in $A_2$Cr$_3$As$_3$ may also be a consequence of the strong antisymmetric spin-orbit coupling arising due to the noncentrosymmetric crystal structure \cite{Smidman2017}. Another possibility is an $f$-wave pairing state which has three line nodes within the $ab$-plane \cite{Zhou}, which is compatible with the observed in-plane six fold modulation of the upper critical field \cite{Zuo2017}. Meanwhile the presence of spin fluctuations apparently modified below $T_c$ is revealed from ZF-$\mu$SR, pointing towards a link between these fluctuations and the superconductivity \cite{Adroja1,Adroja5}, in line with the above theoretical scenarios. With the findings  of low dimensional fluctuations compatible with a Tomonaga-Luttinger liquid \cite{Zhi,Watson}, it is of particular importance to understand any relation between these and the unconventional superconductivity.

\noindent Despite extensive efforts, there are still many unresolved questions related to the gap structure and pairing states of FeSCs. A thorough understanding is all the more important given the potential for novel applications, particularly those related to quantum computation~\cite{Dai1}. For instance, it is important to determine the origin of nodal superconductivity in some systems, and whether these are  manifestations of the same multigap sign-changing $s$-wave state, or if multiple pairing symmetries are realized. 
In addition to the various gap structure probes, inelastic neutron scattering investigations have been very important for elucidating the pairing states of FeSCs \cite{Dai1}, and therefore INS studies on all the  Fe-based and Cr-based materials presented in this review  are highly desirable. More generally, comprehensive studies using a broad variety of complementary methods are very important for a clear elucidation of the gap structures and pairing symmetries of FeSCs and related materials, as well as for the study of proximate magnetic and nematic phases and their relationship to superconductivity. To this end  $\mu$SR is an important tool for such studies, in particular being a sensitive local probe which can be used to determine the superconducting gap structure. Together with other methods such as thermodynamic probes, ARPES, inelastic neutron scattering, resonant inelastic X-ray scattering, and nuclear magnetic resonance, this can provide vital information for revealing the microscopic origin of high-$T_{\rm c}$ superconductivity.

\subsection*{Acknowledgements}{The work was completed at the ISIS pulsed muon and neutron source of the Rutherford Appleton Laboratory, UK. We are appreciative of several collaborators for providing us with materials, helpful discussions, and laboratory assistance at different $\mu$SR beamlines in the last few years. We would like to acknowledge H. Q. Luo, G.-H. Cao, J. Zhao, S. Blundell, A. M. Strydom,   F. Kirschner, A. D. Hillier, P. K. Biswas, M. R. Lees, C. Wang, H. Mao, F. Lang, P. Baker, M. Telling, Yu. Feng, B. Pan and F. L. Pratt for their valuable input in this work. MS acknowledges support of  the National Natural Science Foundation of China (Grant No. 11874320), the National Key R \& D Program of China (Grant No. 2017YFA0303100). DTA would like to thank the Royal Society of London for the UK-China Newton funding and CMPC-STFC, grant number CMPC-09108, for financial support. AB would like to acknowledge DST India, for Inspire Faculty Research Grant (DST/INSPIRE/04/2015/000169), and UK-India Newton funding for funding support.}

\subsection*{Conflicting Interests}{The authors declare that they have no conflict of interest.}


\begin{thebibliography}{99}


\bibitem{Onnes} H. K. Onnes: Comm. Phys. Lab. Univ. Leiden {\bf 120b}, {\bf  122b}, {\bf 124c}, (1911). 
\bibitem{Bardeen} J. Bardeen, L.N. Cooper, J.R. Schrieffer: Phys. Rev. {\bf 108}, 1175 (1957).
\bibitem{Cooper} L.N. Cooper: Phys. Rev. {\bf 104(4)}, 1189 (1956).  

\bibitem{Poole} C. P. Poole Jr., H.A. Farach, R.J. Creswick: {\it Superconductivity} (Academic Press, San Diego, 1995). 
\bibitem{McMillan} W. L. McMillan, Phys. Rev. {\bf 167}, 331 (1968).
\bibitem{Bednorz} J. G. Bednorz and K. A. M$\ddot{u}$ller, Z. Phys. B. {\bf 64(1)}, 189-193 (1986).
\bibitem{Wu1987} M. K. Wu, J. R. Ashburn, C. J. Torng, P. H. Hor, R. L. Meng, L. Gao, Z. J. Huang, Y. Q. Wang, and C. W. Chu, Phys. Rev. Lett. {\bf 58}, 908 (1987).
\bibitem{Schilling1993} A. Schilling, M. Cantoni, J. D. Guo and H. R. Ott, Nature \textbf{363}, 56–58 (1993).
\bibitem{Chu1993} C. W. Chu, L. Gao, F. Chen, Z. J. Huang, R. L. Meng and Y. Y. Xue, Nature \textbf{365}, 323–325 (1993).


\bibitem{Nicolas Doiron-Leyraud} N. D. Leyraud, C. Proust, D. LeBoeuf, J. Levallois, J. -B. Bonnemaison, R. Liang, D. A. Bonn, W. N. Hardy and L. Taillefer, Nature  {\bf 447}, 565-568 (2007).
\bibitem{T. Park} T. Park, V. A. Sidorov, F. Ronning, J. -X. Zhu, Y. Tokiwa, H. Lee, E. D. Bauer, R. Movshovich, J. L. Sarrao and J. D. Thompson, Nature {\bf 456}, 366-368 (2008). 
\bibitem{Y. Liu1} Y. Liu, D. L. Sun, J. T. Park, C. T. Lin, Physica C: Superconductivity and its Applications, {\bf 470}, S513-S515 (2010). 


\bibitem{Kamihara2}  Y. Kamihara, T. Watanabe, M. Hirano, and H. Hosono, J. Am. Chem. Soc. {\bf 130}, 3296 (2008).
\bibitem{Kamihara1}  Y. Kamihara, H. Hiramatsu, M. Hirano, R. Kawamura, H. Yanagi, T. Kamiya, and H. Hosono, J. Am. Chem. Soc. {\bf 128} 10012 (2006).
\bibitem{Chen2008} X. H. Chen, T. Wu, G. Wu, R. H. Liu, H. Chen and  D. F. Fang, Nature {\bf 453}, 761 (2008).
\bibitem{Takahashi2}  H. Takahashi, K. Igawa, K. Arii, Y. Kamihara, M. Hirano and H. Hosono, Nature {\bf 453}, 376 (2008).

\bibitem{Mazin} I. Mazin and J. Schmalian, Physica C {\bf 469}, 614 (2009).
\bibitem{Johnston}  D. C. Johnston, Adv. Phys. {\bf 59}, 803 (2010).
\bibitem{Paglione} J. Paglione and R. L. Greene, Nat. Phys. {\bf 6}, 645 (2010).
\bibitem{Wen} H.-H. Wen and S. Li, Ann. Rev. Condens. Matt. Phys. {\bf 2}, 121 (2011).
\bibitem{Stewart} G. R. Stewart, Rev. Mod. Phys. {\bf 83}, 1589 (2011).
\bibitem{Hirschfeld} P. J. Hirschfeld, M. M. Korshunov, and I. I. Mazin, Rep. Prog. Phys. {\bf 74}, 124508 (2011).
\bibitem{J. Schmalian1} J. Schmalian, D. Pines, and B. Stojkovic, Phys. Rev. Lett. {\bf 80}, 3839 (1998). 
\bibitem{Y. S. Lee1} Y. S. Lee, R. J. Birgeneau, M. A. Kastner, Y. Endoh, S. Wakimoto, K. Yamada, R. W. Erwin, S. -H. Lee, and G. Shirane, Phys. Rev. B {\bf 60}, 3643 (1999).

\bibitem{Fisher1} R Fisher, L. Degiorgi and Z. X. Shen  Rep. Prog. Phys. {\bf 74},  124506 (2011). 
\bibitem{Fernandes1} R. M. Fernandes, A. V. Chubukov and J. Schmalia, Nature Physics, {\bf 10}, 97  (2014).
\bibitem{Avci1} S. Avci, O. Chmaissem, J. M. Allred, S. Rosenkranz, I. Eremin, A. V. Chubukov, D. E. Bugaris, D. Y. Chung, M. G. Kanatzidis, J. -P. Castellan, J. A. Schlueter, H. Claus, D. D. Khalyavin, P. Manuel, A. Daoud-Aladine and R. Osborn,  Nature Communications {\bf 5}, 3845 (2014).
\bibitem{Fernandes2} R. M. Fernandes, L. H. VanBebber, S. Bhattacharya, P. Chandra, V. Keppens, D. Mandrus, M. A. McGuire, B. C. Sales, A. S. Sefat, and J. Schmalian, Phys. Rev. Lett. {\bf 105}, 157003 (2010). 
\bibitem{Fernandes3} R. M. Fernandes, E. Abrahams, and J. Schmalian, Phys. Rev. Lett. {\bf 107}, 217002 (2011).
\bibitem{Fang4} C. Fang, H. Yao, W. -F. Tsai, J. P. Hu, and S. A. Kivelson, Phys. Rev. B {\bf 77}, 224509 (2008). 
\bibitem{Xu1} C. Xu, M. M$\ddot{o}$ller, and S. Sachdev, Phys. Rev. B {\bf 78}, 020501(R) (2008). 
\bibitem{J. Dai} J. Dai, Q. Si, J.-X. Zhu, and E. Abrahams, Proc. Natl. Acad. Sci. {\bf 106}, 4118 (2009).
\bibitem{S. Onari} S. Onari and H. Kontani, Phys. Rev. Lett. {\bf 109}, 137001 (2012). 
\bibitem{S. Onari1} S. Onari, Y. Yamakawa, and H. Kontani, Phys. Rev. Lett. {\bf 112}, 187001 (2014). 
\bibitem{S. Kumar} F. Kr$\ddot{u}$ger, S. Kumar, J. Zaanen, and J. van den Brink, Phys. Rev. B {\bf 79}, 054504 (2009). 
\bibitem{W. Lv} W. Lv, J. Wu, and P. Phillips, Phys. Rev. B {\bf 80}, 224506 (2009). 
\bibitem{C.-C. Lee} C.-C. Lee, W.-G. Yin, and W. Ku, Phys. Rev. Lett. {\bf 103}, 267001 (2009).
\bibitem{Yamakawa1} Y. Yamakawa, S. Onari, and H. Kontani, Phys. Rev. X 6, 021032 (2016).
\bibitem{Moon2012} E.-G. Moon and S. Sachdev, Phys. Rev. B {\bf 85}, 184511 (2012).
\bibitem{R. M. Fernandes1} R. M. Fernandes,  and A. J. Millis,  Phys. Rev. Lett. {\bf 111}, 127001 (2013).
\bibitem{Fan Yang1} Fan Yang, Fa Wang, and Dung-Hai Lee, Phys. Rev. B 88, 100504 (2013). 


\bibitem{F Steglichhfs} F. Steglich, O. Stockert, S. Wirth, C. Geibel, H. Q. Yuan, S. Kirchner and Q. Si, Journal of Physics: Conference Series, {\bf 449}, 012028 (2013).
\bibitem{F. Steglich} F. Steglich, J. Aarts, C. D. Bredl, W. Lieke, D. Meschede, W. Franz, and H. Schiefer Phys. Rev. Lett. {\bf 43}, 1892 (1979). 
\bibitem{C Petrovic1} C. Petrovic, P. G. Pagliuso, M. F. Hundley, R. Movshovich, J. L. Sarrao, J. D. Thompson, Z. Fisk, and P. Monthoux, Journal of Physics: Condensed Matter {\bf 13}, L337 (2001). 
\bibitem{Bauer} E. Bauer, G. Hilscher, H. Michor, Ch. Paul, E. W. Scheidt, A. Gribanov, Yu. Seropegin, H. No$\ddot{e}$l, M. Sigrist, and P. Rogl, Phys. Rev. Lett. {\bf 92}, 027003 (2004). 
\bibitem{G. R. Stewart1} G. R. Stewart, Z. Fisk, J. O. Willis, and J. L. Smith, Phys. Rev. Lett. {\bf 52}, 679 (1984).
\bibitem{Premala Chandra} P. Chandra, P. Coleman and  R. Flint, Nature  {\bf 493}, 621-626 (2013).



\bibitem{Hardy1993} W. N. Hardy, D. A. Bonn, D. C. Morgan, Ruixing Liang, and Kuan Zhang, Phys. Rev. Lett. {\bf 70}, 3999 (1993).
\bibitem{Annett1990} J. F. Annett, Advances in Physics {\bf 39}, 83-126 (1990).
\bibitem{Scalapino2012} D. J. Scalapino, Rev. Mod. Phys. {\bf 84}, 1383  (2012).
\bibitem{Mazin2008} I. I. Mazin, D. J. Singh, M. D. Johannes, and M. H. Du, Phys. Rev. Lett. {\bf 101}, 057003 (2008).
\bibitem{Dai1} P. Dai, J. Hu and E. Dagotto, Nature Physics {\bf 8}, 709-718 (2012); P. Dai, Rev. Mod. Phys.  {\bf 87}, 855 (2015).
\bibitem{Sato1} T. Sato, K. Nakayama, Y. Sekiba, P. Richard, Y. -M. Xu, S. Souma, T. Takahashi, G. F. Chen, J. L. Luo, N. L. Wang, and H. Ding, Phys. Rev. Lett. {\bf 10}, 047002 (2009).
\bibitem{Qian2011} T. Qian, X.-P. Wang, W.-C. Jin, P. Zhang, P. Richard, G. Xu, X. Dai, Z. Fang, J.-G. Guo, X.-L. Chen, and H. Ding, Phys. Rev. Lett. {\bf 106}, 187001 (2011).
\bibitem{Niu2015} X. H. Niu {\ et al.}  Phys. Rev. B {\bf 92}, 060504 (2015).
\bibitem{Zhao2016} L. Zhao {\ et al.} Nat. Commun. {\bf 7}, 10608 (2016).
\bibitem{J.-Ph. Reid} J.-Ph. Reid, M. A. Tanatar, A. Juneau-Fecteau, R. T. Gordon, S. Rene de Cotret, N. Doiron-Leyraud, T. Saito, H. Fukazawa, Y. Kohori, K. Kihou, C. H. Lee, A. Iyo, H. Eisaki, R. Prozorov, and Louis Taillefer, Phys. Rev. Lett. {\bf 109}, 087001 (2012).
\bibitem{Maiti2011} S. Maiti, M. M. Korshunov, T. A. Maier, P. J. Hirschfeld, and A. V. Chubukov, Phys. Rev. Lett. {\bf 107}, 147002 (2011).
\bibitem{Reid2012}  J.-Ph. Reid, A. Juneau-Fecteau, R. T. Gordon, S. Rene de Cotret, N. Doiron-Leyraud, X. G. Luo, H. Shakeripour, J. Chang, M. A. Tanatar, H. Kim, R. Prozorov, T. Saito, H. Fukazawa, Y. Kohori, K. Kihou, C. H. Lee, A. Iyo, H. Eisaki, B. Shen, H.-H. Wen, and Louis Taillefer, Supercond. Sci. Technol. {\bf 25}, 084013 (2012).
\bibitem{Thomale2011} Ronny Thomale, Christian Platt, Werner Hanke, Jiangping Hu, and B. Andrei Bernevig, Phys. Rev. Lett. {\bf 107}, 117001 (2011).
\bibitem{Lee2009} Wei-Cheng Lee, Shou-Cheng Zhang, and Congjun Wu, Phys. Rev. Lett. {\bf 102}, 217002 (2009). I. I. Mazin, Phys. Rev. B {\bf 84}, 024529 (2011).
\bibitem{Cho2016}  K. Cho, M. Kończykowski, S. Teknowijoyo, M. A. Tanatar, Y. Liu, T. A. Lograsso, W. E. Straszheim, V. Mishra, S. Maiti, P. J. Hirschfeld, R. Prozorov, Sci. Adv. {\bf 2}, e1600807 (2016).
\bibitem{KFe2As2ARPES} K. Okazaki {\ et al.} Science {\bf 337}, 1314-1317 (2012).

\bibitem{Khodas2012} M. Khodas and A. V. Chubukov, Phys. Rev. Lett. {\bf 108}, 247003 (2012).
\bibitem{Nica2017} E. M. Nica, R. Yu, and Q. Si, npj Quantum Mat. {\bf 2}, 24 (2017).




\bibitem{Luke1998} G. M. Luke, Y. Fudamoto, K. M. Kojima, M. I. Larkin, J. Merrin, B. Nachumi, Y. J. Uemura, Y. Maeno, Z. Q. Mao, Y. Mori, H. Nakamura and M. Sigrist. Nature {\bf 394},  558–561 (1998).
\bibitem{Luke1993} G. M. Luke, A. Keren, L. P. Le, W. D. Wu, Y. J. Uemura, D. A. Bonn, L. Taillefer, and J. D. Garrett, Phys. Rev. Lett. {\bf 71}, 1466 (1993).
\bibitem{Xia2006} J. Xia, Y. Maeno, P. T. Beyersdorf, M. M. Fejer, and A. Kapitulnik,  Phys. Rev. Lett. {\bf 97}, 167002 (2006).
\bibitem{Schemm2014} E. R. Schemm, W. J. Gannon, C. M. Wishne, W. P. Halperin, and A. Kapitulnik, Science {\bf 345}, 190-193 (2014). 
\bibitem{Reotier1995} P. D. de R\'eotier, A. Huxley, A.Yaouanc, J. Flouquet, P. Bonville, P. Imbert, P. Pari, P. C. M. Gubbens, A. M. Mulders, Phys.  Lett. A {\bf 205}, 239-243 (1995).
\bibitem{Mackenzie2003} A. P. Mackenzie and Y. Maeno, Rev. Mod. Phys. {\bf 75}, 657 (2003).
\bibitem{Joynt2002} R. Joynt and L. Taillefer, Rev. Mod. Phys. {\bf 74}, 235 (2002).
\bibitem{Hillier2009} A. D. Hillier, J. Quintanilla, and R. Cywinski, Phys. Rev. Lett. {\bf 102}, 117007 (2009).
\bibitem{Hillier2012} A. D. Hillier, J. Quintanilla, B. Mazidian, J. F. Annett, and R. Cywinski, Phys. Rev. Lett. {\bf 109}, 097001 (2012).
\bibitem{Weng2016} Z. F. Weng, J. L. Zhang, M. Smidman, T. Shang, J. Quintanilla, J. F. Annett, M. Nicklas, G. M. Pang, L. Jiao, W. B. Jiang, Y. Chen, F. Steglich and H. Q. Yuan, Phys. Rev. Lett. {\bf 117}, 027001 (2016). 
\bibitem{Sigrist2005} M. Sigrist, AIP Conf. Proc. {\bf 789}, 165 (2005).
\bibitem{Ghosh2018} S. K. Ghosh, J. F. Annett, J. Quintanilla, arXiv:1803.02618.
\bibitem{Mahyari2017}  Z. Lotfi Mahyari, A. Cannell, C. Gomez, S. Tezok, A. Zelati, E. V. L. de Mello, J.-Q. Yan, D. G. Mandrus, and J. E. Sonier, Phys. Rev. B {\bf 89}, 020502(R) (2014).
\bibitem{Grinenko2017}  V. Grinenko, P. Materne, R. Sarkar, H. Luetkens, K. Kihou, C. H. Lee, S. Akhmadaliev, D. V. Efremov, S.-L. Drechsler, and H.-H. Klauss Phys. Rev. B {\bf 95}, 214511 (2017).

\bibitem{CrAsSC1} W. Wu, J. Cheng, K. Matsubayashi, . Kong, F. Lin, C. Jin, N. Wang, Y. Uwatoko, and J. Luo, Nat. Commun. \textbf{5}, 5508 (2014).
\bibitem{CrAsSC2} H. Kotegawa, S. Nakahara, H. Tou, and H. Sugawara, J. Phys. Soc. Jpn. \textbf{83}, 093702 (2014).
\bibitem{J. Bao} J. -Ke Bao, J. -Yong Liu, C. -Wei Ma, Z. -Hao Meng, Z. -Tu Tang, Y. -Lei Sun, H. -Fei Zhai, H. Jiang, H. Bai, C. -Mu Feng, Z. -An Xu, and G. -Han Cao, Phys. Rev. X {\bf 5}, 011013 (2015). 
\bibitem{Z. Tang1} Z. -Tu Tang, J. -Ke Bao, Y. Liu, Y. -Lei Sun, A. Ablimit, H. -Fei Zhai, H. Jiang, C. -Mu Feng, Z. -An Xu, and G. -Han Cao,  Phys. Rev. B {\bf 91}, 020506(R), (2015).
\bibitem{Z. Tang} Z. -Tu Tang, J. -Ke Bao, Z. Wang, H. Bai, H. Jiang, Y. Liu, H. -Fei Zhai, C. -Mu Feng, Z. -An Xu, G. -Han Cao, Science China Materials, {\bf 58}, 16-10 (2015). 
\bibitem{Mu1} Q. G. Mu, B. B. Ruan,  B. J. Pan,  T. Liu,  J. Yu,  K. Zhao,  G. F. Chen and Z. A. Ren, arXiv preprint arXiv:1801.01010.
\bibitem{Mu}  Q. G. Mu, B. B. Ruan,  K. Zhao, B. J. Pan, T. Liu,  L. Shan,  G.F. Chen, and Z. A. Ren, arXiv preprint arXiv:1805.05257.
\bibitem{Zhao2}  K. Zhao, Q.G. Mu, T. Liu, B.J. Pan, B. B. Ruan, L. Shan, G. F.  Chen, and Z. A. Ren, arXiv preprint arXiv:1805.11577.
\bibitem{T. Kong} T. Kong, S. L. Budko, and P. C. Canfield, Phys. Rev. B {\bf 91}, 020507(R) (2015).
\bibitem{Balakirev} F. F. Balakirev, T. Kong, M. Jaime, R. D. McDonald, C. H. Mielke, A. Gurevich, P. C. Canfield, and S. L. Budko, Phys. Rev. B {\bf 91}, 220505(R) (2015).
\bibitem{Pang2} G. Pang, M. Smidman, W. Jiang, J. Bao, Z. Weng, Y. Wang, L. Jiao, J. Zhang, G. Cao and H. Yuan,  Phys. Rev. B {\bf 91}, 220502 (2015).
\bibitem{X. Wu} X. Wu, C. Le, J. Yuan, H. Fan, J. Hu, Chin. Phys. Lett. {\bf 32}, 057401 (2015).
\bibitem{Zhou} Y. Zhou, C. Cao, and F.-C. Zhang, Science Bulletin {\bf 62}, 208 (2017).
\bibitem{Wu2} X. Wu, F. Yang, C. Le, H. Fan, and J. Hu, Phys. Rev. B {\bf 92}, 104511 (2015).
\bibitem{H. Zhong} H. Zhong, X. -Yong Feng, H. Chen, J. Dai,  arXiv1503.08965.
\bibitem{Zhi} H. Z. Zhi, T. Imai, F. L. Ning, Jin-Ke Bao, and Guang-Han Cao, Phys. Rev. Lett. {\bf 114}, 147004 (2015).
\bibitem{Adroja1} D. T. Adroja, A. Bhattacharyya, M. Telling, Y. Feng, M. Smidman, B. Pan, J. Zhao, A. D. Hillier, F. Pratt and A. Strydom, Phys. Rev. B {\bf 92}, 134505 (2015).

\bibitem{Watson} M. D. Watson, Y. Feng, C. W. Nicholson, C. Monney, J. M. Riley, H. Iwasawa, K. Refson, V. Sacksteder, D. T. Adroja, J. Zhao, and M. Hoesch, Phys. Rev. Lett. {\bf 118}, 097002 (2017).

\bibitem{Zhi3} H. Zhi, D. Lee, T. Imai, Z. Tang, Y. Liu, and G. Cao, Phys. Rev. B {\bf 93}, 174508 (2016). 
\bibitem{Yang2015} J. Yang, Z. T. Tang, G. H. Cao, and G. Zheng, Phys. Rev. Lett. {\bf 115}, 147002 (2015).



\bibitem{Adroja5} D. T. Adroja, A. Bhattacharyya, M. Smidman, A. D. Hillier, Y. Feng, B. Pan, J. Zhao, M. R. Lees, A. Strydom, and P. K. Biswas, J. Phys. Soc. Jpn. {\bf 86}, 044710 (2017). 

\bibitem{Khasanov2007} R. Khasanov, S. Str\"{a}ssle, D. Di Castro, T. Masui, S. Miyasaka, S. Tajima, A. Bussmann-Holder, and H. Keller, Phys. Rev. Lett. {\bf 99}, 237601 (2007).
\bibitem{Ohishi2012} K. Ohishi, Y. Ishii, I. Watanabe, H. Fukazawa, T. Saito, Y. Kohori, K. Kihou, C.-H. Lee, H. Kito, A. Iyo, and H. Eisaki, J. Phys. Soc. Jpn. {\bf 81}, SB046 (2012).
\bibitem{Amato} A. Amato: Rev. Mod. Phys. {\bf 69(4)}, 1119 (1997).
\bibitem{Blundell} S. J. Blundell: Contemporary Physics {\bf 40(3)}, 175 (1999). (see also cond-mat/0207699).
\bibitem{Feyerherm} R. Feyerherm, A. Amato, F. N. Gygax, A. Schenck, C. Geibel, F. Steglich, N. Sato, T. Komatsubara: Phys. Rev. Lett. {\bf 73(13)}, 1849 (1994).
\bibitem{Sonier} J. E. Sonier, J. H. Brewer, R. F. Kiefl, Rev. Mod. Phys. {\bf 72(3)}, 769 (2000). 
\bibitem{McKenzie1} Iain McKenzie, Annu. Rep. Prog. Chem., Sect. C: Phys. Chem.{\bf 109}, 65--112, (2013).
\bibitem{Khasanov} R. Khasanov and Z. Guguchia, Supercond. Sci. Technol. {\bf 28} 034003 (2015).




\bibitem{Adroja2} D. T. Adroja, F. K. K. Kirschner, F. Lang, M. Smidman, A. D. Hillier, Z. -Cheng Wang, G. -Han Cao, G. B. G. Stenning, and S. J. Blundell, arXiv:1802.07334. 


\bibitem{Tegel} M. Tegel, Iron pnictide superconductors, Dissertation, Ludiwig-Maximilians-Universit$\ddot{o}$t 2011.



\bibitem{Sonier1} J. E. Sonier, Rep. Prog. Phys. {\bf 70}, 1717-1755 (2007).
\bibitem{Y. Jie1} Y. Jie, Y. Hua,S. YuJun, Z. BeiYi and J. Kim,  Science China, Physcs, Mechanics and Astronomy, {\bf 58}, 107401 (2015). 
\bibitem{PCarretta} P. Carretta, R. De Renzi, G. Prando and S. Sanna, Physica Script, {\bf 88}, 068504 (2013).
\bibitem{Y. Bang} Y. Bang and G. R. Stewart, J. Phys.: Condens. Matter {\bf 29}, 123003 ( (2017).
\bibitem{H. Hosono} H. Hosono, K. Tanabe, E. Takayama-Muromachi, H. Kageyama, S. Yamanaka, H. Kumakura, M. Nohara, H. Hiramatsu, and S. Fujitsu, Sci. Technol. Adv. Mater. {\bf 16}, 033503 (2015).


\bibitem{W. Barford1} W. Barford and J. M. F. Gunn, Physica C {\bf 156}, 515 (1988). 
\bibitem{E. H. Brandt} E. H. Brandt, Phys. Rev. B {\bf 68}, 054506 (2003). 



\bibitem{Prozorov2006} R. Prozorov and R. W. Giannetta, Supercond. Sci. Technol. {\bf 19}, R41–R67 (2006). 
\bibitem{Bhattacharyya1} A. Bhattacharyya, D. T. Adroja, J. Quintanilla, A. D. Hillier, N. Kase, A. M. Strydom and J. Akimitsu, Phys. Rev. B {\bf 91} 060503 (2015). 
\bibitem{Bhattacharyya2} A. Bhattacharyya, D. T. Adroja, N. Kase, A. D. Hillier, J. Akimitsu and A. M. Strydom,  Scientific reports {\bf 5}, (2015). 

\bibitem{Annett2} J. F. Annett,  Advances in Physics {\bf 39}, 83-126 (1990). 
\bibitem{Carrington2003} A. Carrington and F. Manzano, Physica C {\bf 385}, 205 (2003).
\bibitem{Lin2011} J.-Y. Lin, Y. S. Hsieh, D. A. Chareev, A. N. Vasiliev, Y. Parsons, and H. D. Yang, Phys. Rev. B {\bf 84}, 220507(R) (2011). 
\bibitem{Smidman2017} M. Smidman, M. B. Salamon, H Q Yuan, and D. F. Agterberg, Rep. Prog. Phys. {\bf 80}, 036501 (2017).


\bibitem{Wang1} Z. -C. Wang, C. -Y. He, S. -Q. Wu, Z. -T. Tang, Y. Liu, A. Ablimit, C.-M. Feng, and G.-H. Cao, J. Am. Chem. Soc. {\bf 138}, 7856 (2016).
\bibitem{Wang2} Z. Wang, C. He, Z. Tang, S. Wu, and G. Cao, Sci. China Mater. {\bf 60}, 83 (2017).
\bibitem{Johnston2010} D. C. Johnston, Adv. Phys. {\bf 59}, 803 (2010).
\bibitem{Ishida} J. Ishida, S. Iimura and H. Hosono, Phys. Rev. B {\bf 96}, 174522 (2017). 
\bibitem{Wang7} G. Wang, Z. Wang, and X. Shi, Europhys. Lett. {\bf 116}, 37003 (2016).
\bibitem{Smidman1} M. Smidman, F. K. K. Kirschner,  D. T. Adroja, A. D. Hillier, F. Lang, Z. C. Wang, G. H. Cao, and S. J. Blundell, Phys. Rev. B {\bf 97}, 060509(R) (2018).
\bibitem{Zhang6n} Z. Zhang, A. F. Wang, X. C. Hong, J. Zhang, B. Y. Pan, J. Pan, Y. Xu, X. G. Luo, X. H. Chen, and S. Y. Li, Phys. Rev. B {\bf 91}, 024502 (2015).
\bibitem{Cui} J. Cui, Q.-P. Ding, W. R. Meier, A. E. Bohmer, T. Kong V. Borisov, Y. Lee, S. L. Budko, R. Valent, P. C. Canfield, and Y. Furukawa, Phys. Rev. B {\bf 96}, 104512 (2017).
\bibitem{Kirschner} F. K. K. Kirschner, D. T. Adroja, Z.-Cheng Wang, F. Lang, M. Smidman, P. J. Baker, G.-Han Cao, and S. J. Blundell, Phys. Rev. B {\bf 97}, 060506(R) (2018).
\bibitem{Biswas} P. K. Biswas, A. Iyo, Y. Yoshida, H. Eisaki, K. Kawashima and A. D. Hillier, Phys. Rev. B {\bf 95}, 140505 (2017).
\bibitem{Cho} K. Cho, A. Fente, S. Teknowijoyo, M. A. Tanatar, K. R. Joshi, N. M. Nusran, T. Kong, W. R. Meier, U. Kaluarachchi, I. Guillamon, H. Suderow, S. L. Budko, P. C. Canfield and R. Prozorov, Phys. Rev. B {\bf 95}, 100502 (2017).
\bibitem{Mou} D. Mou, T. Kong, W. R. Meier, F. Lochner, L. -L. Wang, Q. Lin, Y. Wu, S. L. Budko, I. Eremin, D. D. Johnson, P. C. Canfield and A. Kaminski, Phys. Rev. Lett. {\bf 117}, 277001 (2016).
\bibitem{Zhang} W. -L. Zhang, W. R. Meier, T. Kong, P. C. Canfield,  and G. Blumberg, arXiv:1804.06963v1.
 

\bibitem{Wang3} C. Wang, Z. C. Wang,Y. M. Mei, Y. K. Li, L. Li, Z. T. Tang, Y. Liu, P. Zhang, H. F. Zhai, Z. A. Xu, and G. H. Cao, J. Am. Chem. Soc. {\bf 138}, 2170 (2016).
\bibitem{Wang4} Z. C. Wang, C. Y. He, Z. T. Tang, S. Q. Wu, and G. H. Cao, Sci. China Mater. {\bf 60}, 83 (2017).
\bibitem{Singh} D. J. Singh, J. Alloys Compd. {\bf 687}, 786 (2016).
\bibitem{Wang5} G. Wang and X. Shi, Europhys. Lett. {\bf 113}, 67006 (2016).
\bibitem{Mao} H. Mao, C. Wang, H. E. Maynard-Casely, Q. Huang, Z. Wang, G. Cao, S. Li, and H. Luo, Europhys. Lett. {\bf 117}, 57005 (2017).
\bibitem{Albedah} M. A. Albedah, F. Nejadsattari, Z. M.Stadnik, Z. -Cheng Wang, C. Wang, G. -Han Cao, J. Alloys Compd. {\bf 695}, 1128 (2017).
\bibitem{Bai-Zhuo Li} Bai-Zhuo Li, Zhi-Cheng Wang, Jia-Lu Wang, Fu-Xiang Zhang, Dong-Ze Wang, Feng-Yuan Zhang, Yu-Ping Sun , Qiang Jing , Hua-Fu Zhang, Shu-Gang Tan, Yu-Ke Li, Chun-Mu Feng, Yu-Xue Mei,, Cao Wang and Guang-Han Cao, J. Phys.: Condens. Matter {\bf 30}, 255602 (2018).
\bibitem{Barbero} N. Barbero, S. Holenstein, T. Shang, Z. Shermadini, F. Lochner, I. Eremin, C. Wang, G.-H. Cao, R. Khasanov, H.-R. Ott, J. Mesot, and T. Shiroka, Phys. Rev. B {\bf 97}, 140506(R) (2018).
\bibitem{Adroja3} D. T. Adroja, A. Bhattacharyya, P. K. Biswas, M. Smidman, A. D. Hillier, H. Mao, H. Luo, G. -Han Cao, Z. Wang, and C. Wang, Phys. Rev. B {\bf 96}, 144502 (2017).


\bibitem{Shiroka} T. Shiroka, T. Shang, C. Wang, G.-H. Cao, I. Eremin, H.-R. Ott and J. Mesot, Nature Communications {\bf 8}, 156 (2017).
\bibitem{Sanna2010} S. Sanna, R. De Renzi, T. Shiroka, G. Lamura, G. Prando, P. Carretta, M. Putti, A. Martinelli, M. R. Cimberle, M. Tropeano, and A. Palenzona, Phys. Rev. B {\bf 82}, 060508(R) (2010).

\bibitem{Yuan2003}  H. Q. Yuan, F. M. Grosche, M. Deppe, C. Geibel, G. Sparn, F. Steglich, Science {\bf 302}, 2104-2107 (2003).
\bibitem{Jun Zhang1} Jun Zhang, Feng-Liang Liu, Tian-Ping Ying , Na-Na Li , Yang Xu , Lan-Po He, Xiao-Chen Hong, Yun-Jie Yu , Ming-Xiang Wang, Jian Shen, Wen-Ge Yang and Shi-Yan Li. npj Quantum Materials {\bf 2}, 49 (2017).

\bibitem{Medvedev1} S. Medvedev, T. M. McQueen, I. A. Troyan, T. Palasyuk, M. I. Eremets, R. J. Cava, S. Naghavi, F. Casper, V. Ksenofontov, G. Wortmann and C. Felser, Nature Materials. {\bf 8}, 630, (2009).


\bibitem{Shen C} C. Shen, B. Si, C. Cao, X. Yang, J. Bao, Q. Tao, Y. Li, G. Cao, and  Zhu-An Xu, J. Appl. Phys. {\bf 119}, 083903 (2016).
\bibitem{Mukuda H} H. Mukuda, F. Engetsu, T. Shiota, K. To Lai, M. Yashima, Y. Kitaoka, S. Miyasaka, and S. Tajima, J. Phys. Soc. Jpn. {\bf 83}, 083702 (2014). 
\bibitem{Soshi Iimura1} Soshi Iimura, Satoru Matsuishi, Hikaru Sato, Taku Hanna, Yoshinori Muraba, Sung Wng Kim, Jung Eun Kim, Masaki Takata and Hideo Hosono, Nature Communications {\bf 3}, 943 (2012). 
\bibitem{Yang J}  J. Yang, R. Zhou, L. L. Wei, H. X. Yang, J. Q. Li, Z. X. Zhao, Guo-qing Zheng, Chin. Phys. Lett. {\bf 32}, 107401 (2015). 
\bibitem{Rustem Khasanov} R. Khasanov, H. Luetkens  A. Amato H. -Henning Klauss, Z. -An Ren, J. Yang, W. Lu, and Z. -Xian Zhao, Phys. Rev. {\bf 78}, 092506 (2008).



\bibitem{Canfield2010} P. C. Canfield and S. L. Bud'ko, Annu. Rev. Condens. Matter Phys. {\bf 1}, 27--50 (2010).
\bibitem{Mandrus2010} D. Mandrus, A. S. Sefat, M. A. McGuire, and B. C. Sales, Chem. Mater. {\bf 22}, 715 (2010).
\bibitem{Stewart2011} G. R. Stewart, Rev. Mod. Phys. {\bf 83}, 1589 (2011).
\bibitem{Damascelli2003} A. Damascelli, Z. Hussain, and Z.-X. Shen, Rev. Mod. Phys. {\bf 75}, 473 (2003).
\bibitem{Lee2006} P. A. Lee, N. Nagaosa, and X.-G. Wen, Rev. Mod. Phys. {\bf 78}, 17 (2006).
\bibitem{Goko2009} T. Goko, A. A. Aczel, E. Baggio-Saitovitch, S. L. Bu\'dko, P. C. Canfield, J. P. Carlo, G. F. Chen, P. Dai, A. C. Hamann, W. Z.
Hu, H. Kageyama, G. M. Luke, J. L. Luo, B. Nachumi, N. Ni, D. Reznik, D. R. Sanchez-Candela, A. T. Savici, K. J. Sikes, N. L. Wang, C. R. Wiebe, T. J. Williams, T. Yamamoto, W. Yu, and Y. J. Uemura, Phys. Rev. B {\bf 80}, 024508 (2009).
\bibitem{Ding} H. Ding, P. Richard, K. Nakayama, K. Sugawara, T. Arakane, Y. Sekiba, A. Takayama, S. Souma, T. Sato, T. Takahashi, Z. Wang, X. Dai, Z. Fang, G. F. Chen, J. L. Luo and N. L. Wang, Europhys. Lett. {\bf 83}, 47001 (2008).
\bibitem{Hiraishi2009} M. Hiraishi, R. Kadono, S. Takeshita, M. Miyazaki, A. Koda, H. Okabe, and J. Akimitsu, J. Phys. Soc. Jpn. {\bf 78}, 023710 (2009). 
\bibitem{Zhao2008}  L. Zhao, H. Liu, W. Zhang, J. Meng, X. Jia, G. Liu, X. Dong, G. F. Chen, J. L. Luo, N. L. Wang, G. Wang, Y. Zhou, Y. Zhu, X. Wang, Z. Zhao, Z. Xu, C. Chen, X. J. Zhou, Chin. Phys. Lett. {\bf 25}, 4402 (2008).
\bibitem{Nakayama2009}  K. Nakayama, T. Sato, P. Richard, Y.-M. Xu, Y. Sekiba, S. Souma, G. F. Chen, J. L. Luo, N. L. Wang, H. Ding et al. Europhys. Lett. {\bf 85}, 67002 (2009).
\bibitem{Guguchia2011} Z. Guguchia, Z. Shermadini, A. Amato, A. Maisuradze, A. Shengelaya, Z. Bukowski, H. Luetkens, R. Khasanov, J. Karpinski, and H. Keller, Phys. Rev. B {\bf 84}, 094513 (2011).
\bibitem{Williams2009} T. J. Williams, A. A. Aczel, E. Baggio-Saitovitch, S. L. Bu\'dko, P. C. Canfield, J. P. Carlo, T. Goko, J. Munevar, N. Ni, Y. J.
Uemura, W. Yu, and G. M. Luke, Phys. Rev. B {\bf 80}, 094501 (2009).
\bibitem{Williams2010} T. J. Williams, A. A. Aczel, E. Baggio-Saitovitch, S. L. Bu\'dko, P. C. Canfield, J. P. Carlo, T. Goko, H. Kageyama, A. Kitado, J. Munevar, N. Ni, S. R. Saha, K. Kirschenbaum, J. Paglione, D. R. Sanchez-Candela, Y. J. Uemura, G.M. Luke, Phys. Rev. B {\bf 82}, 094512 (2010).
\bibitem{Hashimoto2010} K. Hashimoto, M. Yamashita, S. Kasahara, Y. Senshu, N. Nakata, S. Tonegawa, K. Ikada, A. Serafin, A. Carrington, T. Terashima, H. Ikeda, T. Shibauchi, and Y. Matsuda, Phys. Rev. B {\bf 81}, 220501(R) (2010).
\bibitem{Yamashita2011} M. Yamashita, Y. Senshu, T. Shibauchi, S. Kasahara, K. Hashimoto, D. Watanabe, H. Ikeda, T. Terashima, I. Vekhter, A. B. Vorontsov, and Y. Matsuda, Phys. Rev. B {\bf 84}, 060507(R)  (2011).
\bibitem{Zhang6} Y. Zhang, R. Ye, Q. Q. Ge, F. Chen, Juan Jiang, M. Xu, B. P. Xie, and D. L. Feng, Nature Phys. {\bf 8}, 371 (2012).
\bibitem{Qiu2012} X. Qiu, S. Y. Zhou, H. Zhang, B. Y. Pan, X. C. Hong, Y. F. Dai, Man Jin Eom, Jun Sung Kim, Z. R. Ye, Y. Zhang, D. L. Feng, and S. Y. Li, Phys. Rev. X {\bf 2}, 011010 (2012).
\bibitem{Guguchia2016}  Z. Guguchia, R. Khasanov, Z. Bukowski, F. von Rohr, M. Medarde, P. K. Biswas, H. Luetkens, A. Amato, and E. Morenzoni, Phys. Rev. B {\bf 93}, 094513 (2016).
\bibitem{Xu2013}  N. Xu, P. Richard, X. Shi, A. van Roekeghem, T. Qian, E. Razzoli, E. Rienks, G.-F. Chen, E. Ieki, K. Nakayama, T. Sato, T. Takahashi, M. Shi, H. Ding, Phys. Rev. B {\bf 88}, 220508(R) (2013).
\bibitem{Hodovanets2014}  H. Hodovanets, Y. Liu, A. Jesche, S. Ran, E. D. Mun, T. A. Lograsso, S. L. Bu\'dko, P. C. Canfield, Phys. Rev. B {\bf 89}, 224517 (2014).
\bibitem{Rotter1} M. Rotter, M. Tegel, D. Johrendt, I. Schellenberg, W. Hermes, and R. P\"{o}ttgen, Phys. Rev. B {\bf 78}, 020503(R) (2008).
\bibitem{Huang2008} Q. Huang, Y. Qiu, Wei Bao, M. A. Green, J. W. Lynn, Y. C. Gasparovic, T. Wu, G. Wu, and X. H. Chen, Phys. Rev. Lett. {\bf 101}, 257003 (2008).
\bibitem{Rotter} M. Rotter, M. Tegel, and D. Johrendt, Phys. Rev. Lett. {\bf 101}, 107006 (2008).
\bibitem{Aczel2008} A. A. Aczel, E. Baggio-Saitovitch, S. L. Budko, P. C. Canfield, J. P. Carlo, G. F. Chen, P. Dai, T. Goko,W. Z. Hu, G.M. Luke, J.
L. Luo, N. Ni, D. R. Sanchez-Candela, F. F. Tafti, N. L.Wang, T. J. Williams, W. Yu, and Y. J. Uemura, Phys. Rev. B {\bf 78}, 214503 (2008).
\bibitem{Wiesenmayer2011} E. Wiesenmayer, H. Luetkens, G. Pascua, R. Khasanov, A. Amato, H. Potts, B. Banusch, H.-H. Klauss, and D. Johrendt,
Phys. Rev. Lett. {\bf 107}, 237001 (2011).
\bibitem{Mallett2017}  B. P. P. Mallett, C. N. Wang, P. Marsik, E. Sheveleva, M. Yazdi-Rizi, J. L. Tallon, P. Adelmann, Th. Wolf, and C. Bernhard
Phys. Rev. B {\bf 95}, 054512 (2017).

\bibitem{Guguchia2015}  Z. Guguchia, A. Amato, J. Kang, H. Luetkens, P. K. Biswas, G. Prando, F. von Rohr, Z. Bukowski, A. Shengelaya, H. Keller, E. Morenzoni, Rafael M. Fernandes and R. Khasanov Nat. Commun. {\bf 6}, 8863 (2015).
\bibitem{Sefat2008} A. S. Sefat, R. Jin, M. A. McGuire, B. C. Sales, D. J. Singh, and D. Mandrus, Phys. Rev. Lett. {\bf 101}, 117004 (2008).
\bibitem{Ni2008} N. Ni, M. E. Tillman, J.-Q. Yan, A. Kracher, S. T. Hannahs, S. L. Bu\'dko, and P. C. Canfield, Phys. Rev. B {\bf 78}, 214515 (2008).
\bibitem{Wang2009} X. F. Wang, T. Wu, G. Wu, R. H. Liu, H. Chen, Y. L. Xie and X. H. Chen, New J. Phys. {\bf 11}, 045003 (2009).
\bibitem{Bernhard2009} C. Bernhard, A. J. Drew, L. Schulz, V. K. Malik, M. Rössle, C. Niedermayer, T. Wolf, G. D. Varma, G. Mu, H.-H. Wen, H. Liu, G. Wu, and X. H. Chen, New J. Phys. {\bf 11}, 055050 (2009). 
\bibitem{Marsik2010} P. Marsik, K. W. Kim, A. Dubroka, M. Roessle, V. K. Malik, L. Schulz, C. N. Wang, Ch. Niedermayer, A. J. Drew, M. Willis, T. Wolf, C. Bernhard, Phys. Rev. Lett. {\bf 105}, 057001 (2010).
\bibitem{Sonier2011} J. E. Sonier, W. Huang, C. V. Kaiser, C. Cochrane, V. Pacradouni, S. A. Sabok-Sayr, M. D. Lumsden, B. C. Sales, M. A. McGuire, A. S. Sefat, and D. Mandrus Phys. Rev. Lett. {\bf 106}, 127002 (2011).
\bibitem{Bernhard2012} C. Bernhard, C. N. Wang, L. Nuccio, L. Schulz, O. Zaharko, J. Larsen, C. Aristizabal, M. Willis, A. J. Drew, G. D. Varma,
T. Wolf, and Ch. Niedermayer, Phys. Rev. B {\bf 86}, 184509 (2012).
\bibitem{Larsen2015} J. Larsen, B. Mencia Uranga, G. Stieper, S. L. Holm, C. Bernhard, T. Wolf, K. Lefmann, B. M. Andersen, and C. Niedermayer
Phys. Rev. B {\bf 91}, 024504 (2015).
\bibitem{Khasanov2009} R. Khasanov, A. Maisuradze, H. Maeter, A. Kwadrin, H. Luetkens, A. Amato, W. Schnelle, H. Rosner, A. Leithe-Jasper, and H.-H. Klauss, Phys. Rev. Lett. {\bf 103}, 067010 (2009).






\bibitem{Uemura1} Y. J. Uemura, G. M. Loke, B. J. Sternlieb, J. H. Brewer, J. F. Carolan, W. N. Hardy, R. Kadono, J. R. Kempton, R. F. Kiefl, S. R. Kreitzman, P. Mulhern, T. M. Riseman, D. Ll. Williams, B. X. Yang, S. Uchida, H. Takagi, J. Gopalkrishnan, A. W. Sleight, M. A. Subramanian, C. L. Chien, M. Z. Cieplak, G. Xiao, V. Y. Lee, B. W. Statt, C. E. Stronach, W. J. Kossler, and X. H. Yu, Phys. Rev. Lett. {\bf 62}, 2317 (1989).
\bibitem{Hillier1} A. D. Hillier and R. Cywinski, Appl. Magn. Reson. {\bf 13}, 95 (1997).


\bibitem{Hashimoto2012} K. Hashimoto, S. Kasahara, R. Katsumata, Y. Mizukami, M. Yamashita, H. Ikeda, T. Terashima, A. Carrington, Y. Matsuda, and T. Shibauchi, Phys. Rev. Lett. {\bf 108}, 047003 (2012).
\bibitem{Fukazawa2009} H. Fukazawa, Y. Yamada, K. Kondo, T. Saito, Y. Kohori, K. Kuga, Y. Matsumoto, S.Nakatsuji, H. Kito, P. M. Shirage, K. Kihou, N. Takeshita, C.-H. Lee, A. Iyo, and H. Eisaki, J. Phys. Soc. Jpn {\bf 78}, 083712 (2009).
\bibitem{Dong2010} J. K. Dong, S. Y. Zhou, T. Y. Guan, H. Zhang, Y. F. Dai,
X. Qiu, X. F. Wang, Y. He, X. H. Chen, and S. Y. Li, Phys. Rev. Lett. {\bf 104}, 087005 (2010).
\bibitem{Iyo2016} A. Iyo, K. Kawashima, T. Kinjo, T. Nishio, S. Ishida, H. Fujihisa, Y. Gotoh, K. Kihou, H. Eisaki, and Y. Yoshida,  J. Am. Chem. Soc. 138, 3410-3415 (2016).
\bibitem{Smidman2017} M. Smidman, M. B. Salamon, H. Q. Yuan, D. F. Agterberg, Rep. Prog. Phys. {\bf 80}, 036501 (2017).
\bibitem{Zuo2017}Huakun Zuo, Jin-Ke Bao, Yi Liu, Jinhua Wang, Zhao Jin, Zhengcai Xia, Liang Li, Zhuan Xu, Jian Kang, Zengwei Zhu, and Guang-Han Cao, Phys. Rev. B {\bf 95}, 014502 (2017).


\bibitem{J. D. Fletcher} J. D. Fletcher, A. Serafin, L. Malone, J. G. Analytis, J. -H. Chu, A. S. Erickson, I. R. Fisher, and A. Carrington, Phys. Rev. Lett. {\bf 102}, 147001 (2009).
\bibitem{Mike Sutherland} M. Sutherland, J. Dunn, W. H. Toews, E. O. Farrell, J. Analytis, I. Fisher, and R. W. Hill, Phys. Rev. B {\bf 85}, 014517 (2012).
\bibitem{Allan2012} M. P. Allan, A. W. Rost, A. P. Mackenzie, Yang Xie, J. C. Davis, K. Kihou, C. H. Lee, A. Iyo, H. Eisaki, T.-M. Chuang, Science {\bf 336}, 563 (2012).
\bibitem{K. Umezawa} K. Umezawa, Y. Li, H. Miao, K. Nakayama, Z.-H. Liu, P. Richard, T. Sato, J. B. He, D.-M. Wang, G. F. Chen, H. Ding, T. Takahashi, and S.-C. Wang, Phys. Rev. Lett. {\bf 108}, 037002 (2012).
\bibitem{Takeshita2009} S. Takeshita and R. Kadono, New J. Phys. {\bf 11}, 035006 (2009).
\bibitem{FeSCTan} C. Tan, T. P. Ying, Z. F. Ding, J. Zhang, D. E. MacLaughlin, O. O. Bernal, P. C. Ho, K. Huang, I. Watanabe, S. Y. Li, and L. Shu, Phys. Rev B {\bf 97}, 174524 (2018).
\bibitem{Hsu}  F. -Chi Hsu, J. -Yong Luo, K. -Wei Yeh, T. -Kun Chen, T. -Wen Huang, P. M. Wu, Y. -Chi Lee, Y. -Lin Huang, Y. -Yi Chu, D. -Chung Yan, and M. -Kuen Wu. Proc. Natl Acad. Sci. {\bf 105}, 14262 (2008).
\bibitem{Khasanov2008} R. Khasanov, K. Conder, E. Pomjakushina, A. Amato, C. Baines, Z. Bukowski, J. Karpinski, S. Katrych, H.-H. Klauss, H. Luetkens, A. Shengelaya, and N. D. Zhigadlo, Phys. Rev. B {\bf 78}, 220510(R) (2008).
\bibitem{Miao} H. Miao, P. Richard, Y. Tanaka, K. Nakayama, T. Qian, K. Umezawa, T. Sato, Y. -M. Xu, Y. -B. Shi, N. Xu, X. -P. Wang, P. Zhang, H. -B. Yang, Z. -J. Xu, J. S. Wen, G. -D. Gu, X. Dai, J.-P. Hu, T. Takahashi, H. Ding, Phys. Rev. B {\bf 85}, 094506 (2012).
\bibitem{Shermadini1} Z. Shermadini, J. Kanter, C. Baines, M. Bendele, Z. Bukowski, R. Khasanov, H.-H. Klauss, H. Luetkens, H. Maeter, G. Pascua, B. Batlogg, and A. Amato, Phys. Rev. B {\bf 82}, 144527 (2010).
\bibitem{Shermadini2} Z. Shermadini, H. Luetkens, A. Maisuradze, R. Khasanov, Z. Bukowski, H. H. Klauss, and A. Amato, Phys. Rev. B {\bf 86}, 174516 (2012).
\bibitem{Hong2013} X. C. Hong, X. L. Li, B. Y. Pan, L. P. He, A. F. Wang, X. G. Luo, X. H. Chen, and S. Y. Li, Phys. Rev. B {\bf 87}, 144502 (2013).


\end{thebibliography}
\end{document}